\documentclass[fleqn,usenatbib]{mnras}

\usepackage{newtxtext,newtxmath}

\usepackage[T1]{fontenc}
\usepackage{ae,aecompl}

\usepackage{graphicx}	
\usepackage{amsmath}	
\usepackage{amssymb}	

\usepackage{subfigure}

\title[The WWL pathfinder]{The Wendelstein Weak Lensing (WWL) pathfinder: Accurate weak lensing masses for \emph{Planck} clusters}

\author[R. L. Rehmann et al.]{
R. L. Rehmann,$^{1,2}$\thanks{E-mail: rehmann@usm.uni-muenchen.de}
D. Gruen,$^{3,4,5}$
S. Seitz,$^{1,2}$ 
R. Bender,$^{1,2}$ 
A. Riffeser,$^{1,2}$ 
M. Kluge,$^{1,2}$ 
\newauthor 
C. Goessl,$^{1,2}$ 
U. Hopp,$^{1,2}$ 
A. Mana,$^{1,2}$ 
C. Ries,$^{1}$ 
M. Schmidt$^{1}$ 
\\
$^{1}$University-Observatory, Ludwig-Maximilians-University, Scheinerstrasse 1, D-81679 Munich, Germany\\
$^{2}$Max Planck Institute for Extraterrestrial Physics, Giessenbachstrasse, D-85748 Garching, Germany\\
$^{3}$SLAC National Accelerator Laboratory, Menlo Park, CA 94025, USA\\
$^{4}$KIPAC, Physics Department, Stanford University, Stanford, CA 94305, USA\\
$^{5}$Einstein Fellow
}

\date{Accepted XXX. Received YYY; in original form ZZZ}

\pubyear{2017}

\begin{document}
\label{firstpage}
\pagerange{\pageref{firstpage}--\pageref{lastpage}}
\maketitle

\begin{abstract}

We present results from the Wendelstein Weak Lensing (WWL) pathfinder project,
in which we have observed three intermediate redshift \emph{Planck} clusters of galaxies with the new 30'${\times 30}$' wide field imager at the 2m~\emph{Fraunhofer} Telescope at Wendelstein Observatory.
We investigate the presence of biases in our shear catalogues and estimate their impact on our weak lensing mass estimates. The overall calibration uncertainty depends on the cluster redshift and is below 8.1--15~per~cent for $z\approx$0.27--0.77. It will decrease with improvements on the background sample selection and the multiplicative shear bias calibration.

We present the first weak lensing mass estimates for PSZ1~G109.88+27.94 and PSZ1~G139.61+24.20,  two SZ-selected cluster candidates. 
Based on Wendelstein colors and SDSS photometry, we find that the redshift of PSZ1~G109.88+27.94 has to be corrected to $z\approx0.77$.
We investigate the influence of line-of-sight structures on the weak lensing mass estimates and find upper limits for two groups in each of the fields of PSZ1~G109.88+27.94 and PSZ1~G186.98+38.66.
We compare our results to SZ and dynamical mass estimates from the literature, and in the case of PSZ1~G186.98+38.66 to previous weak lensing mass estimates.
We conclude that our pathfinder project demonstrates that weak lensing cluster masses can be accurately measured with the 2m~\emph{Fraunhofer} Telescope.

\end{abstract}




\section{Introduction}\label{sec:one:intro}

With masses above ${~10^{14}\ \mathrm{M_\odot}}$, 
clusters of galaxies are the massive end of the distribution of collapsed structures in the Universe.
By measuring the abundance of clusters as a function of mass and redshift, the evolution of structure formation can be studied. 
Halo abundance experiments \citep[for a review, cf.][]{allen2011cosmological} 
let us examine the underlying dark matter density field, as well as the growth rate of structure.

In order to exploit their full potential as cosmological probes, masses have to be determined accurately. 
Large surveys rely on inexpensive observables, 
such as richness, X-ray luminosity and SZ Compton parameter,
which do not provide an absolute mass scale.
The Mass-Observable Relation (MOR) relates the cosmology dependent theoretical 
cluster mass function to observables. Using samples of galaxy clusters for which weak lensing mass estimates are available, the 
MOR can be calibrated \citep[e.g.][]{hoekstra2012canadian,marrone2012locuss,gruen2014weak, mantz2016weighing, melchior2017weak}. 

In this work, we show that the Wendelstein Wide Field Imager (WWFI) installed at the 2m~\emph{Fraunhofer}~telescope \citep{hopp2008improving, hopp2014commissioning, kosyra201464} can be used to conduct cluster weak lensing studies and add to the list of clusters with accurate mass estimates. 
The paper is organized as follows.
In Section~\ref{sec:two:data}, we introduce our data and describe the reduction procedure
and the photometry.
We elaborate on the shear measurement procedure in 
Section~\ref{sec:three:shear} and
address the possibility of systematic bias in our shape catalogues.
Section~\ref{sec:four:photoz} provides a description of our background sample selection procedure and 
a discussion of the impact of uncertainties 
in the redshift estimation on the derived cluster masses.
We explain the 
weak lensing analysis of our clusters in Section~\ref{sec:five:wlanalysis}.
Our results are presented in Section~\ref{sec:six:results} and compared to SZ and X-ray studies in Section~\ref{sec:seven:compare}. 
Concluding remarks are given in Section~\ref{sec:eight:conclude}. 
Throughout, we use a flat standard $\Lambda$CDM model with $\Omega_\mathrm{M} = 0.27$ and $H_0 = 72~\mathrm{km\ s^{-1} Mpc^{-1}}$.

\section{Observations and data reduction}\label{sec:two:data}

\subsection{Instruments}\label{sec:instruments}
We have observed our targets with the 2.0m \emph{Fraunhofer} Telescope \citep[see][]{hopp2014commissioning} using the Wendelstein Wide Field Imager \citep[see][]{kosyra201464}.
The WWL pathfinder project has been among the first projects to provide science verification during telescope commissioning using the WWFI as the scientific first light instrument since 2014.
The camera consists of a $2\times 2$ mosaic of (4k)$^2$ pixel 15\ $\mu$m e2v CCDs. Each of the four CCDs has four readout ports. The field of view is 27.6' $\times$ 29.0' $\approx 0.22$\ deg$^2$ with a pixel scale of $0.2$\ arcsec$/$pixel. The filter wheels are equipped with five optical SDSS-like broad-band filters $ugriz$. 
For this project, we have used three-band photometric $gri$ data only. 
With sub-arcsecond median seeing at the telescope site and a design that aims to reduce the amount of ghost images, WWFI data are suitable for cluster weak lensing studies.

\subsection{Cluster sample}\label{sec:fields}

For this work we have selected three \emph{Planck} clusters of galaxies \citep{ade2014b} in order to increase the MOR calibration sample. 
Thanks to the location of the telescope, clusters far up in the northern hemisphere can be included in the WWL sample. 
We are the first to target PSZ1~G109.88+27.94 and PSZ1~G139.61+24.20 for weak lensing studies.
The large number of field stars allows for an exquisite test for how well PSF anisotropies of the camera can be modeled. 
This strategy can turn out to be problematic, however,
when stars become too many or bright stars cause over-saturation and bleeding effects.
We show the PSF modelling in Section~\ref{sec:psf}.

Table~\ref{tab:sample} provides more information on our targets. Spectroscopic redshifts are available for all of these clusters. We note that the spectroscopic redshift of PSZ1~G109.88+27.94 that is referenced in \citet{ade2014b} could not be confirmed. Our data, as well as SDSS photometry \citep[DR-14][]{abolfathi2017fourteenth} suggests, that the true redshift of this cluster is likely much larger ($z\approx0.8$). More details on the redshift estimation of this object are given in Sections~\ref{sec:four:photoz}~and~\ref{sec:109}.

We have obtained three band photometric data with 
the WWFI SDSS-like $g,r,i$~filters. 
We use the $r$~band as our lensing band. The good seeing in this band and the long integration~times (up to 10.5 hours) make it the most useful for weak lensing analyses.   
\begin{table*}
	\centering
	\caption{The cluster sample investigated in this paper and the reference field W-EGS that is needed for the background sample selection (cf. Section~\ref{sec:EGS}). Columns from left
        to right: object name, right ascension, declination and spectroscopic
        redshift as given in the PSZ1 catalogue, WWFI exposure time and 
        PSF FWHM in the $g,r,i$ stacks respectively. The referenced redshifts are from (a)~Photo-$z$ from \citet[DR-14][]{abolfathi2017fourteenth}, (b)~\citet{ade2015b} and (c)~\citet{piffaretti2011vizier}.}
	\label{tab:sample}
	\begin{tabular}{lcccccr} 
		\hline
		Object & RA (J2000) & Dec (J2000) & $z$ & $t_\mathrm{exp}$ (h) & $\mathrm{FWHM ('')}$ & WWFI filter\\
		\hline
		\hline
                      &            &             &      &  4.15  & 0.92 &$g$\\
		PSZ1~G109.88+27.94 & 18:23:15.1 & +78:24:27 & $0.77^\mathrm{(a)}$  &8.68 & 0.91 &$r$ \\  
                      &            &             &       &  3.08 & 0.79 &$i$\\
                                            \hline                   

                      &            &             &      &  0.70 & 1.19 &$g$\\   
		PSZ1~G139.61+24.20 & 06:22:13.9 & +74:41:39   & $0.267^\mathrm{(b)}$  &10.5 & 0.91 &$r$\\
                      &            &             &       & 2.45 & 1.03  &$i$\\
                                            \hline              

                      &            &             &       & 1.48 & 1.10 &$g$\\        
		PSZ1~G186.98+38.66 & 08:50:12.0 & +36:03:36 & $0.378^\mathrm{(c)}$   & 5.55 & 0.89 &$r$\\
                      &            &             &       & 1.36 & 1.08  &$i$\\	
                                            \hline
	
                      &            &             &           & 4.29 & 1.05 &$g$ \\
		W-EGS                & 14:19:48.0   & +52:54:36 &         & 6.60 & 0.85 &$r$\\  
                      &            &             &            & 3.98 &  1.22 &$i$\\
		\hline
                      
	\end{tabular}
\end{table*}

\subsection{Reference field}\label{sec:EGS}

\begin{figure}
	\includegraphics[width=\columnwidth]{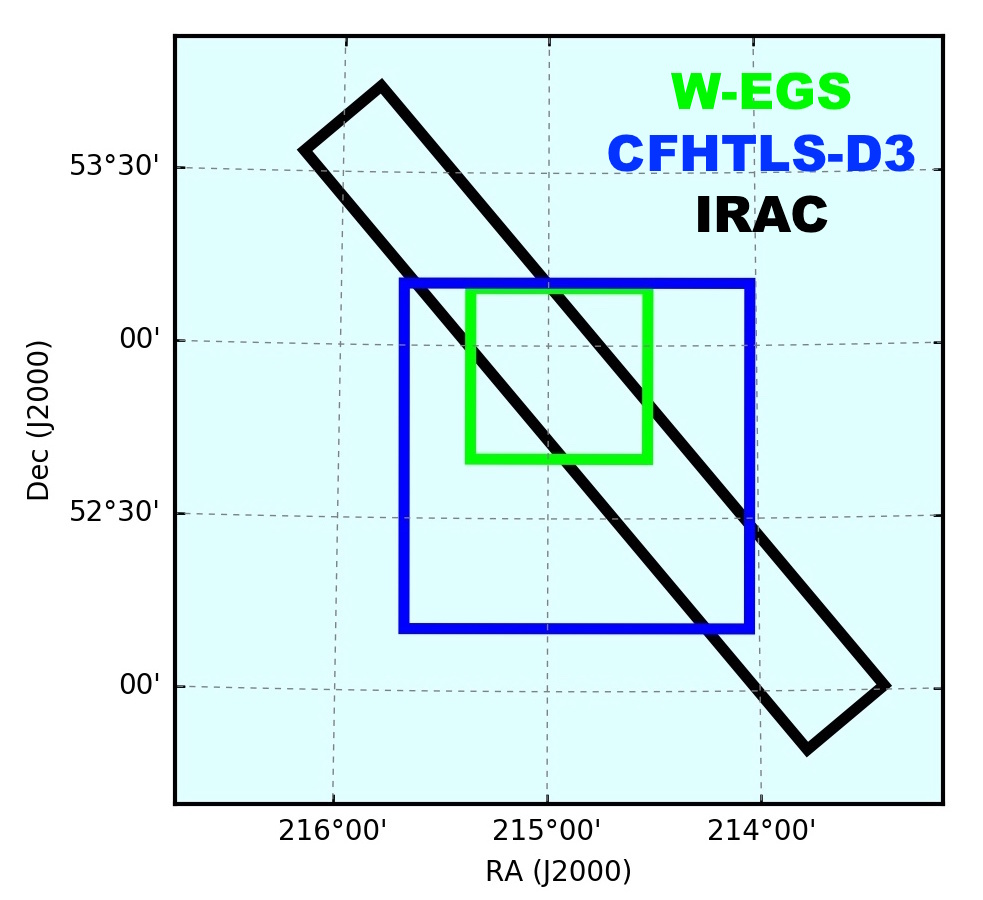}
    \caption{Footprints of W-EGS (green), CFHTLS-D3 field (blue) and EGS covered with IRAC (black).} 
    \label{fig:fields}
\end{figure}

In order to get a clean background sample selection for the weak lensing analysis, 
we need reliable redshift estimates.
This can be achieved by comparing the flux of the observed galaxies in all available filters
to galaxies with known redshifts \citep{gruen2014weak, gruen2017selection}. Those reference galaxies have to be observed with the same set of filters as has been used 
for the cluster fields. We call this reference field W-EGS, since  
we have chosen a subregion of the extended groth strip, an extremely well studied patch on the sky, for our analysis. 
It overlaps with 
CFHTLS-D3 (the Canada-France-Hawaii Telescope Legacy Survey-Deep3) 
\citep{davis2007all}, which provides extremely deep $u^*,g',r',i',z'$ data, and is also covered in the near-infrared with $J,H,K$
\citep{bielby2012wircam}. Using this data set, we use a template-fitting approach to get good photometric redshift estimates for the field. 
In accordance to the \emph{Planck} cluster fields, we have observed W-EGS 
in $g,r,i$ and obtained a photometric catalogue including approximately 25~000 galaxies.   
Fig.~\ref{fig:fields} shows the footprints of W-EGS and CFHTLS-D3, as well as the region of the extended groth strip observed with \emph{Spitzer}/IRAC in the mid-infrared \citep{barmby2008catalog}.
The background sample selection is described in more detail in Section~\ref{sec:four:photoz}.
\subsection{Data reduction}\label{sec:reduction}

We perform de-biasing, flat-fielding and
masking of cosmic rays \citep{goessl2002image} and 
charge persistences 
in the 
raw image frames.
In a first step, the background subtraction, final astrometry and co-addition of the resampled
images is done using \texttt{SCAMP} \citep{bertin2006automatic} and \texttt{SWarp} \citep{bertin2002terapix}.
We exclude frames with too large PSF size and too low sky transparency (cf.~equation~\ref{eq:trans}). 

In order to select the appropriate cut on the seeing, we consider the distribution of the
FWHM of the PSF modeled as a polynomial function in the single frames via \texttt{SExtractor}
\citep{bertin1996sextractor} and \texttt{PSFEx}. We select the maximum
FWHM in such a way, that we do not lose too much depth. 

We define the transparency of a frame $i$ as 
\begin{equation}T_{\mathrm{f},i} = \frac{t_i\times F_{\mathtt{SCAMP},i}}{{\max}
\big\{t_i\times F_{\mathtt{SCAMP},i}\big\}},\label{eq:trans}\end{equation} 
where $t_i$ is the exposure time and $F_{\mathtt{SCAMP},i}$ the flux scale of image $i$ as calculated by \texttt{SCAMP}.
The transparency threshold has been set to $T_{\mathrm{f},i} \geq 50$~per~cent of the ideal transparency 
of all nights (${\max}
\{t_i\times F_{\mathtt{SCAMP},i}\}$). 
This quantity is a measure for the amount of absorption in the atmosphere, 
i.e. ${T_{\mathrm{f},i}\approx100}$~per~cent in a cloudless night.
Although a stricter cut would be preferable in order to ensure an unbiased photometry, 
a 50~per~cent-cut has turned out to sufficiently exclude any noisy exposures while not removing too many frames for the stacking.  
We do not rely on a constant photometric solution ${T_{\mathrm{f},i}=100}$, as long as the transparency is comparable in the cluster fields and the W-EGS stacks.

\subsection{Photometry}\label{sec:photometry}

Due to a lack of standard star observations,
we fix the $r$~band zero-points by calibrating the fluxes of the field stars
relative to the Pan-STARRS PV3 (Panoramic Survey Telescope And Rapid Response System Processing Version 3)
catalogue \citep[cf.][]{flewelling2016pan}.  
We use \texttt{SExtractor} 
in dual image mode to detect objects in the $r$-band stacks 
and extract the flux 
from all filter bands.
We use \texttt{IRAF}\footnote[1]{\url{http://iraf.noao.edu/}} 
to convolve all images of an individual field to the same PSF and measure AB magnitudes in an aperture with a diameter of 8 pixels (1.6 arcsec).  
We perform Stellar Locus Regression (SLR) in order to find the zero-point offsets
in the remaining $g$ and $i$~bands \citep[for more details cf.~e.g.][]{brimioulle2013dark}. The minimization of the residuals in colour-colour diagrams is
done with respect to the stellar library of \citet{pickles1998stellar}.  
Our zero-points have an uncertainty of only 1~per~cent.

\section{PSF modeling and shear measurement}\label{sec:three:shear}

\subsection{PSF modeling}\label{sec:psf}

\begin{figure}
   	\subfigure[PSZ1~G109.88+27.94]{\includegraphics[width = 2.5in]{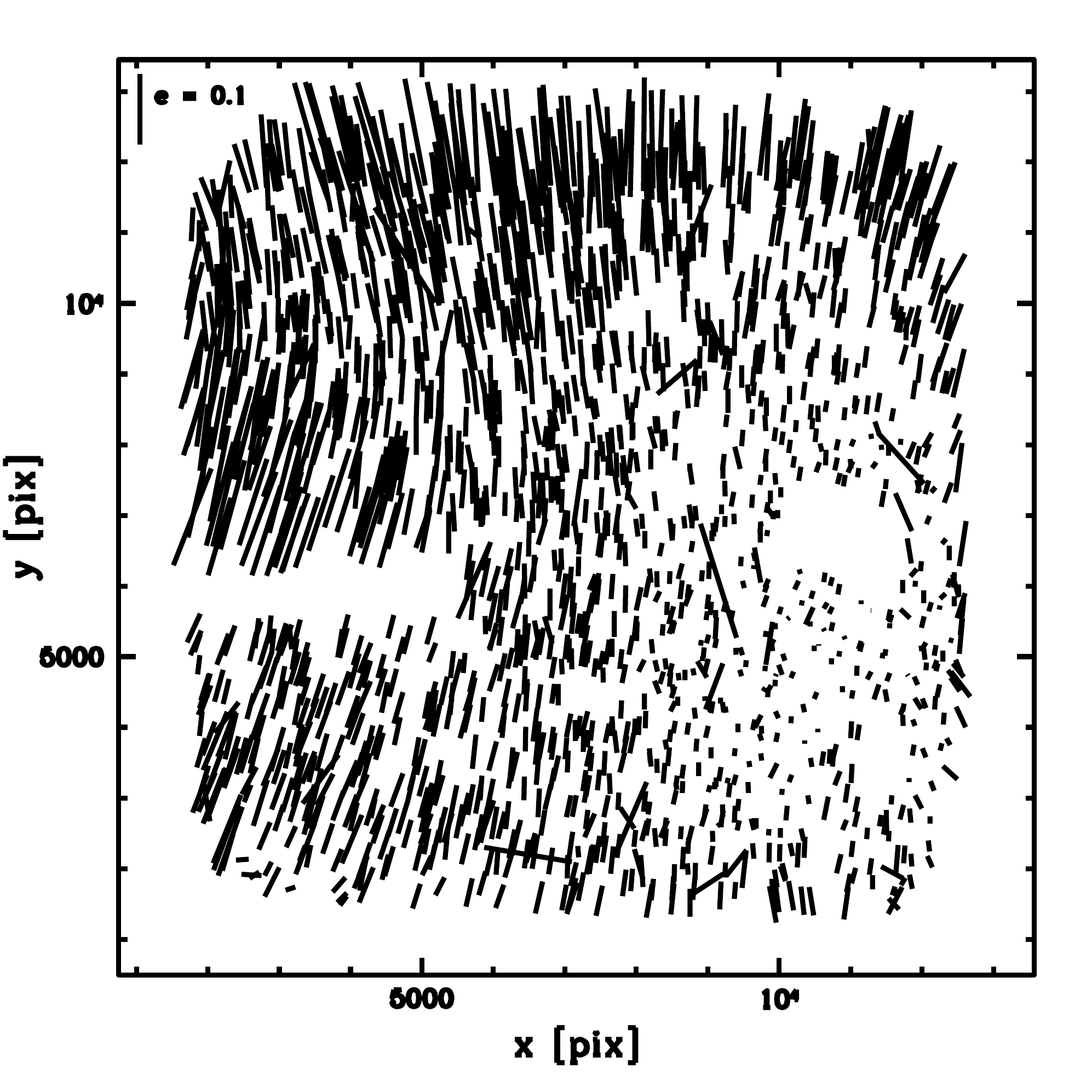}}\\  
	\subfigure[PSZ1~G139.61+24.20]{\includegraphics[width = 2.5in]{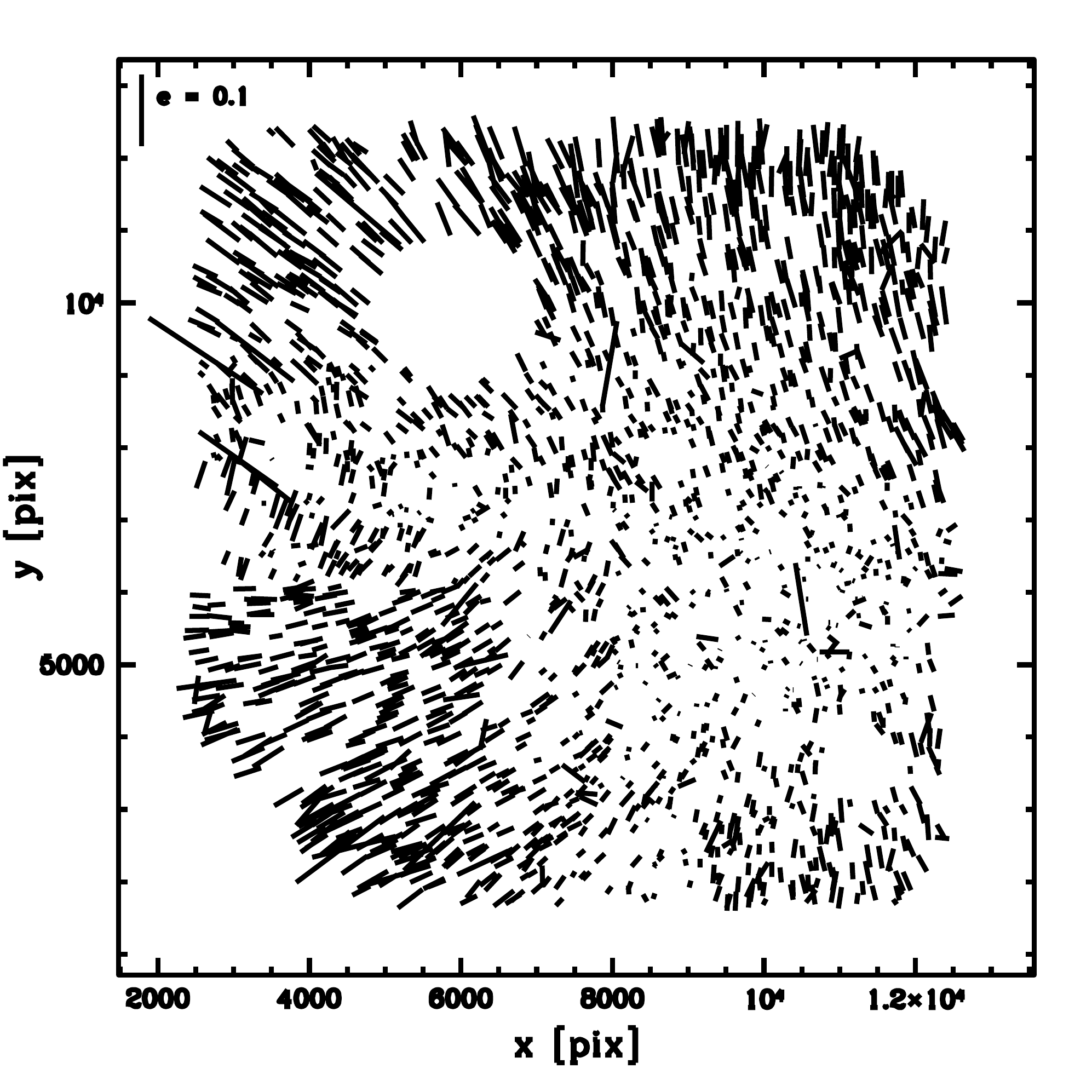}}\\
	\subfigure[PSZ1~G186.98+38.66]{\includegraphics[width = 2.5in]{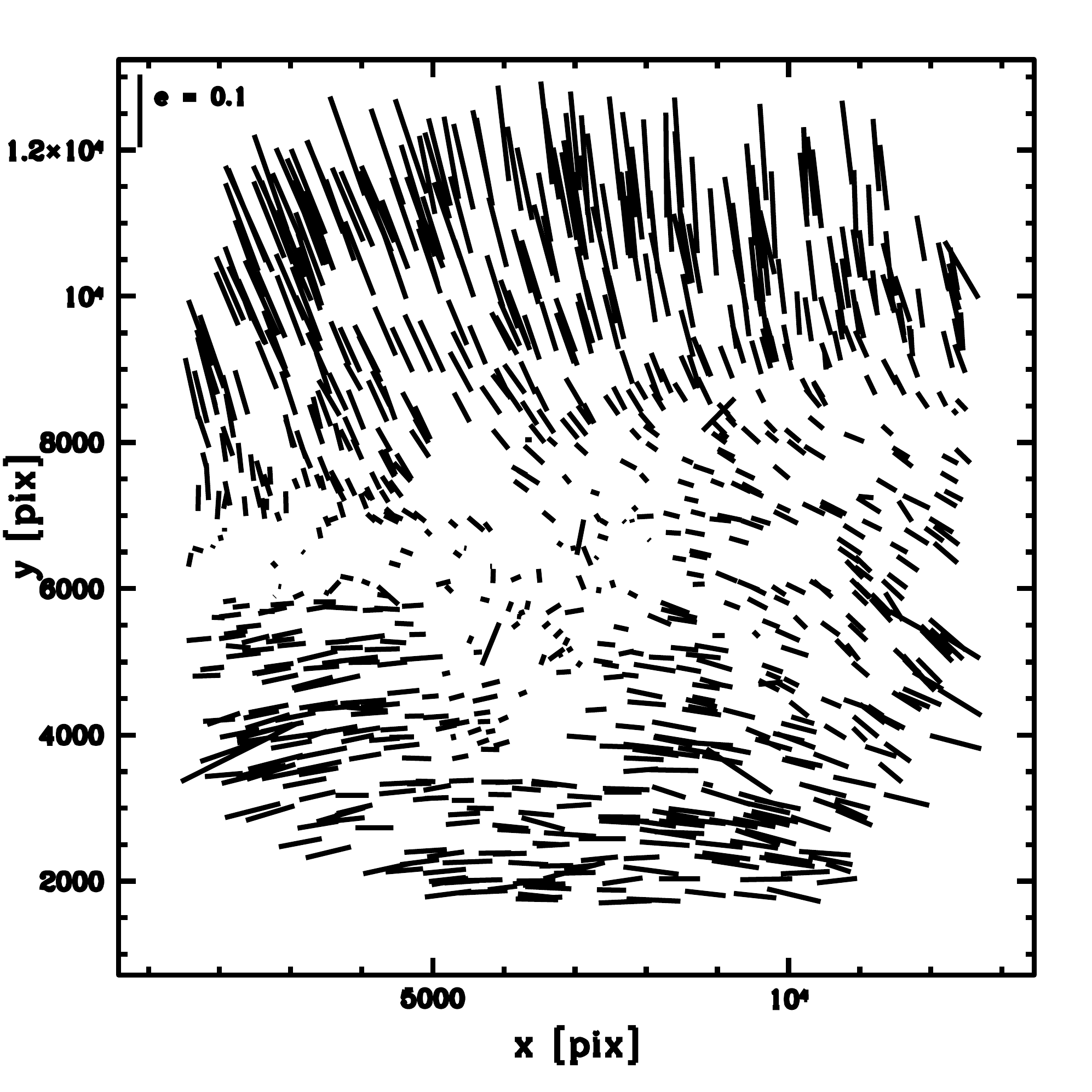}}
   \caption[Whisker plots of the stellar ellipticities]{The pattern of the point spread function illustrated as whisker plots of the stellar ellipticities of the $r$~band stacked images.  In the upper left corner of each image, we show a 10 per~cent ellipticity for comparison. Calibration experiments conducted with the WWFI during the period of observations are the main reason why the PSF pattern is unstable over time. Bright stars have made it necessary for us to mask part of the field, which explains the regions of "missing data" especially for PSZ1~G109.88+27.94 and PSZ1~G139.61+24.20.}
    \label{fig:whisker}
\end{figure}

The surface brightness profiles of galaxies and stars alike are modified by the atmosphere, the telescope and the CCD. Only pre-seeing
galaxy shapes can be used to perform a weak lensing analysis.
The PSF is defined as the response of the \emph{Fraunhofer} Telescope optics to a point source.
The field stars can therefore be used to measure the PSF at the corresponding image positions. 
We show the ellipticity of the field stars in our stacked $r$-band images in Fig.~\ref{fig:whisker}. 
We verify that the selected stars are homogeneously distributed across the field of view. Gaps in the pattern shown in Fig.~\ref{fig:whisker} are the results of masking. 
Usually the affected areas are centred on very bright stars, as the increased photon noise affects the photometry and the shape measurement.
In the case of PSZ1~G109.88+27.94 however, part of the chip gaps had to be masked as well, since, as a result of the dithering of the camera, the stacks were much shallower in this region and the PSF model could not fit the data well.

We use the values of the measured stellar ellipticity in order to model the overall PSF distribution as a function of pixel coordinates. We filter the photometric catalogues according to the \texttt{SExtractor} star classifier, \texttt{FLAG}=0 and size larger than the stellar flux radius. 
We then use \texttt{PSFEx} \citep{bertin2011automated} to model the spatial variation of the PSF as a 
polynomial function of position in the image plane. 
We find that the splitting of the stacked images into 16 subregions of equal area and running \texttt{PSFEx} on each field improves the quality of the modeled PSF ellipticities.

\begin{figure}
   	\subfigure[PSZ1~G109.88+27.94]{\includegraphics[width = 2.5in]{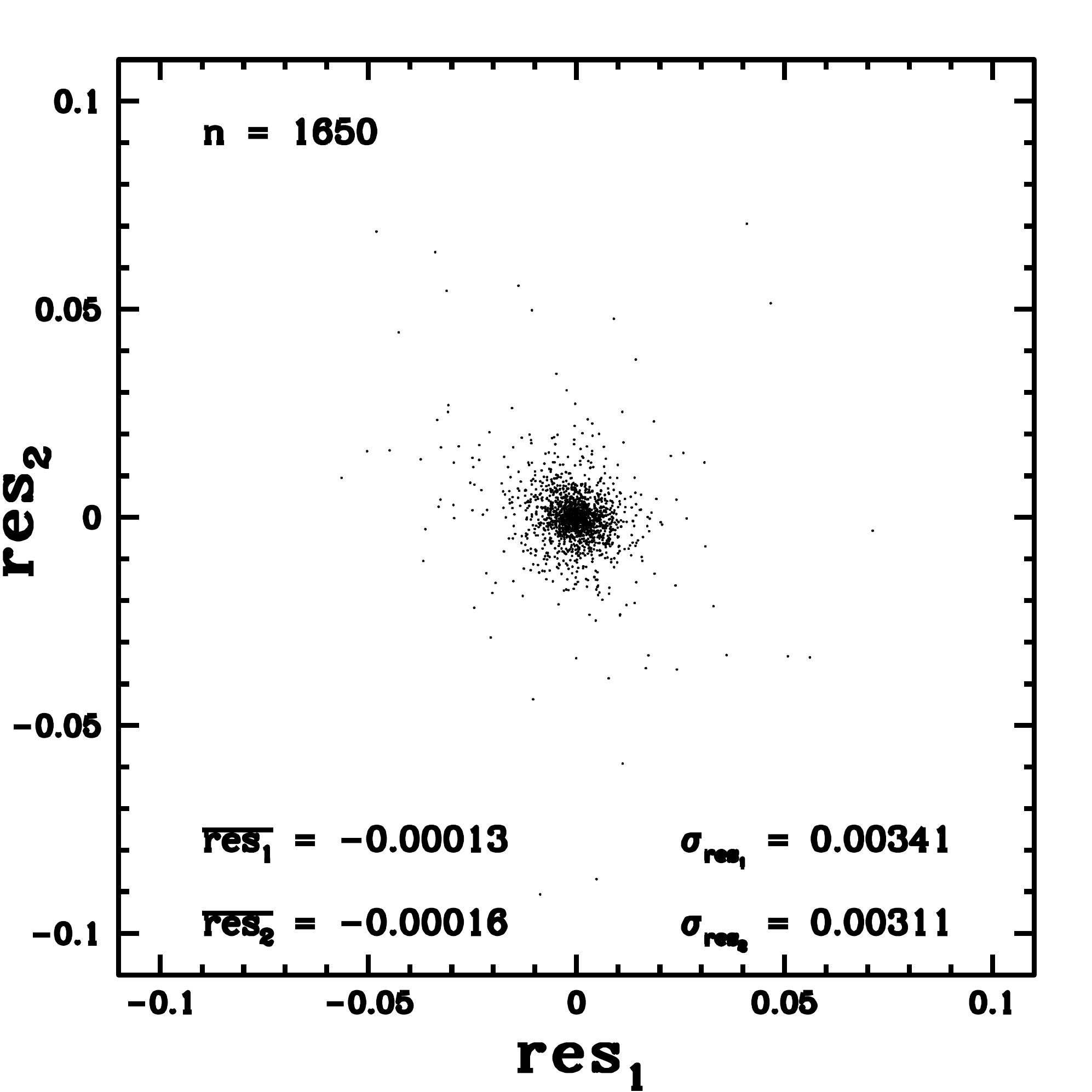}}\\
	\subfigure[PSZ1~G139.61+24.20]{\includegraphics[width = 2.5in]{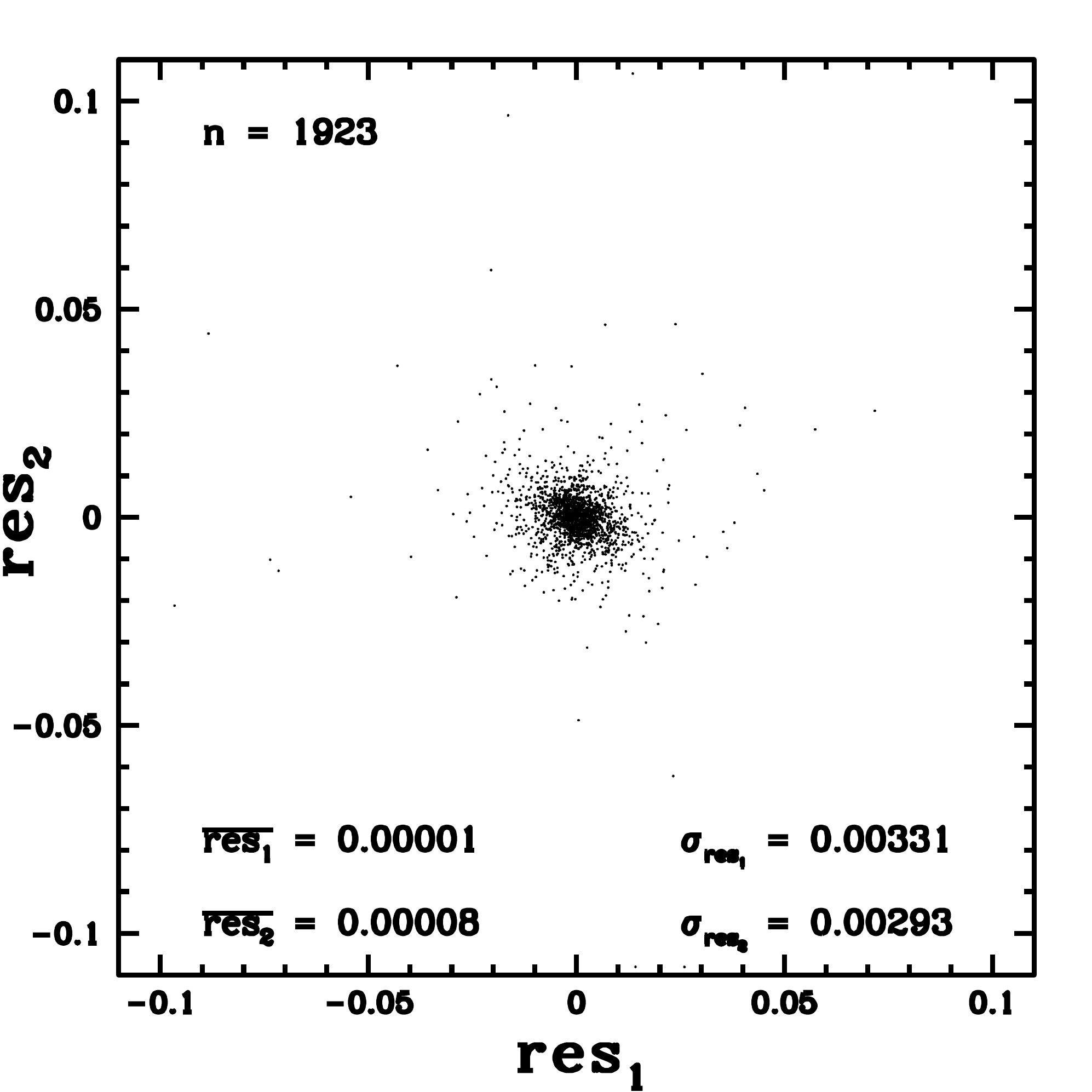}}\\
	\subfigure[PSZ1~G186.98+38.66]{\includegraphics[width = 2.5in]{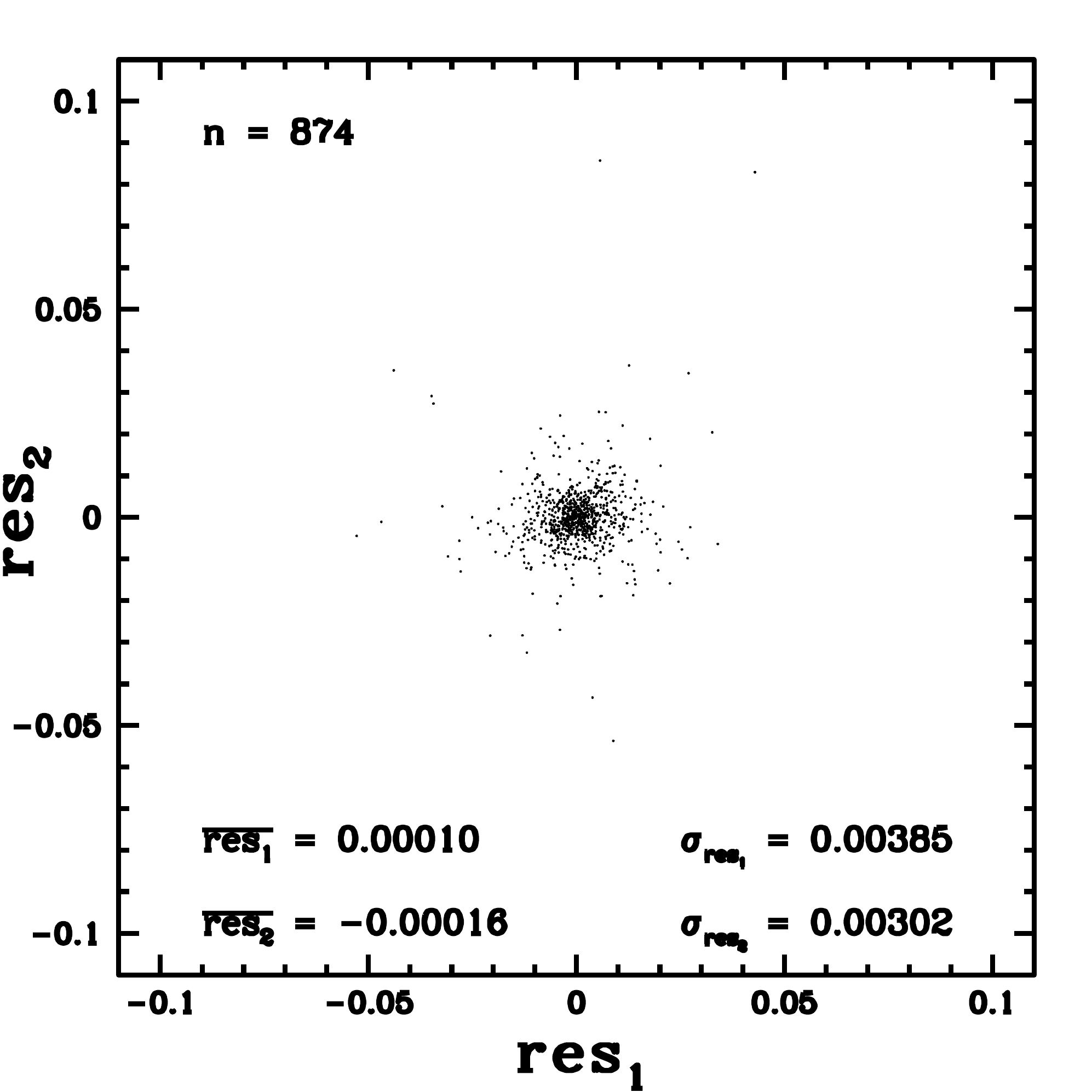}}
\caption[Residual ellipticities of stellar objects and modeled stars]{The residuals of measured and modeled ellipticities of the field stars in the $r$~band stacked images. The residuals  ${\boldsymbol{\mathrm{res}}} = (\boldsymbol{e^\star}-\boldsymbol{e}^\mathrm{PSF})$ scatter around zero and are nearly symmetrically distributed. Mean values for the components of ${\boldsymbol{\mathrm{res}}}$, standard deviations $\sigma_{\boldsymbol{\mathrm{res}}}$ and the number of stars used for the fit $n$ are given for the individual fields.}
       \label{fig:psfres}
\end{figure}

\begin{figure}
   	\subfigure[PSZ1~G109.88+27.94]{\includegraphics[width = 2.5in]{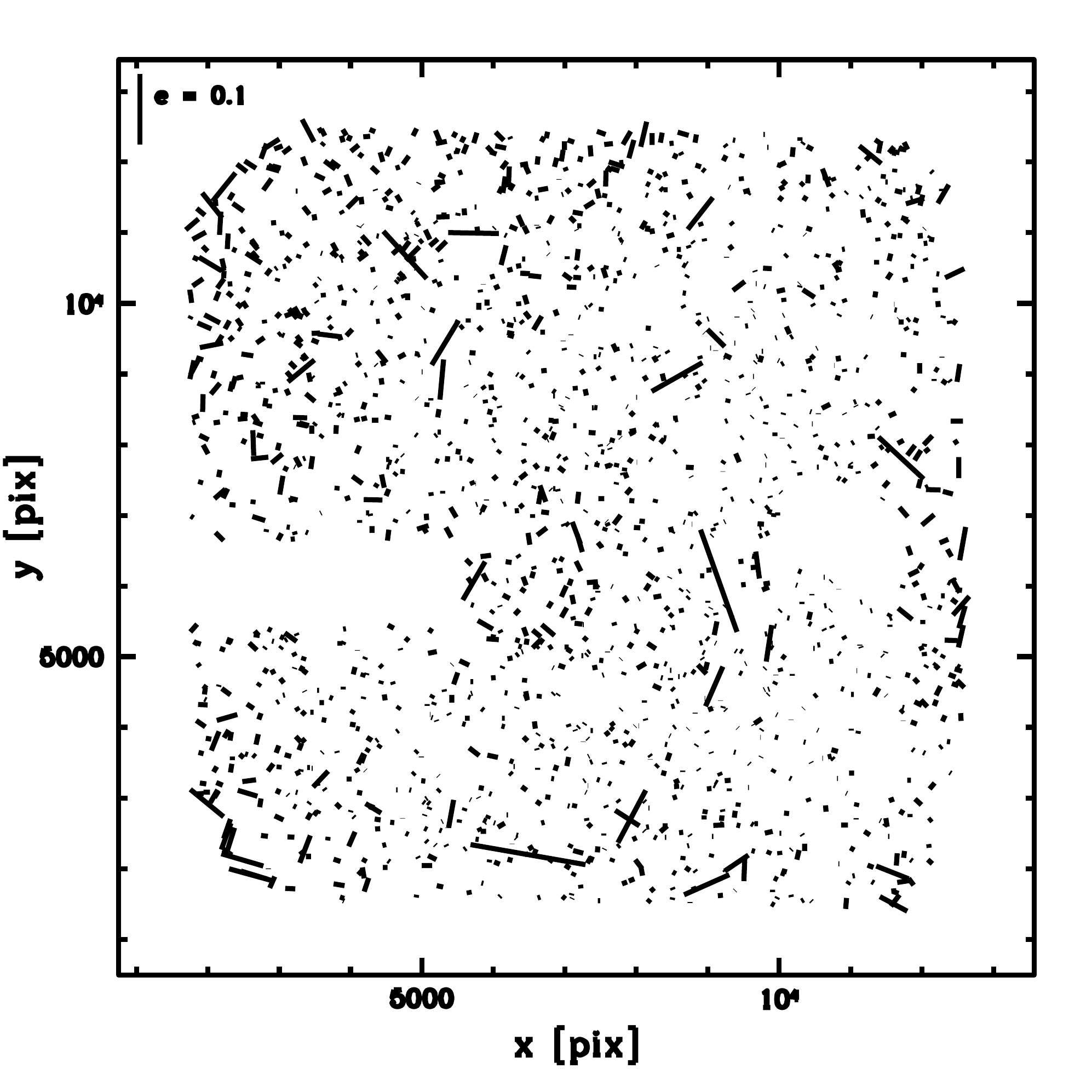}}\\
	\subfigure[PSZ1~G139.61+24.20]{\includegraphics[width = 2.5in]{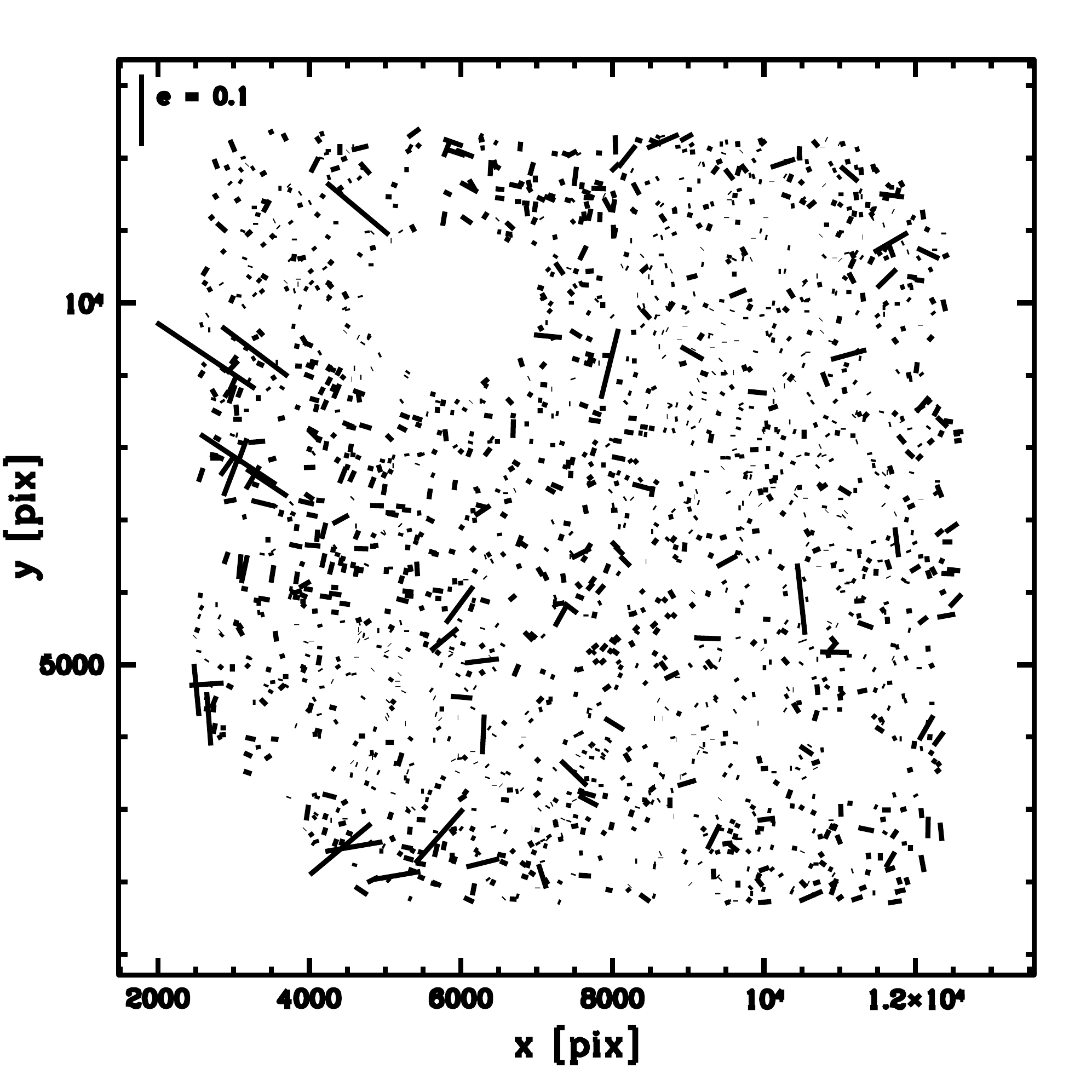}}\\
	\subfigure[PSZ1~G186.98+38.66]{\includegraphics[width = 2.5in]{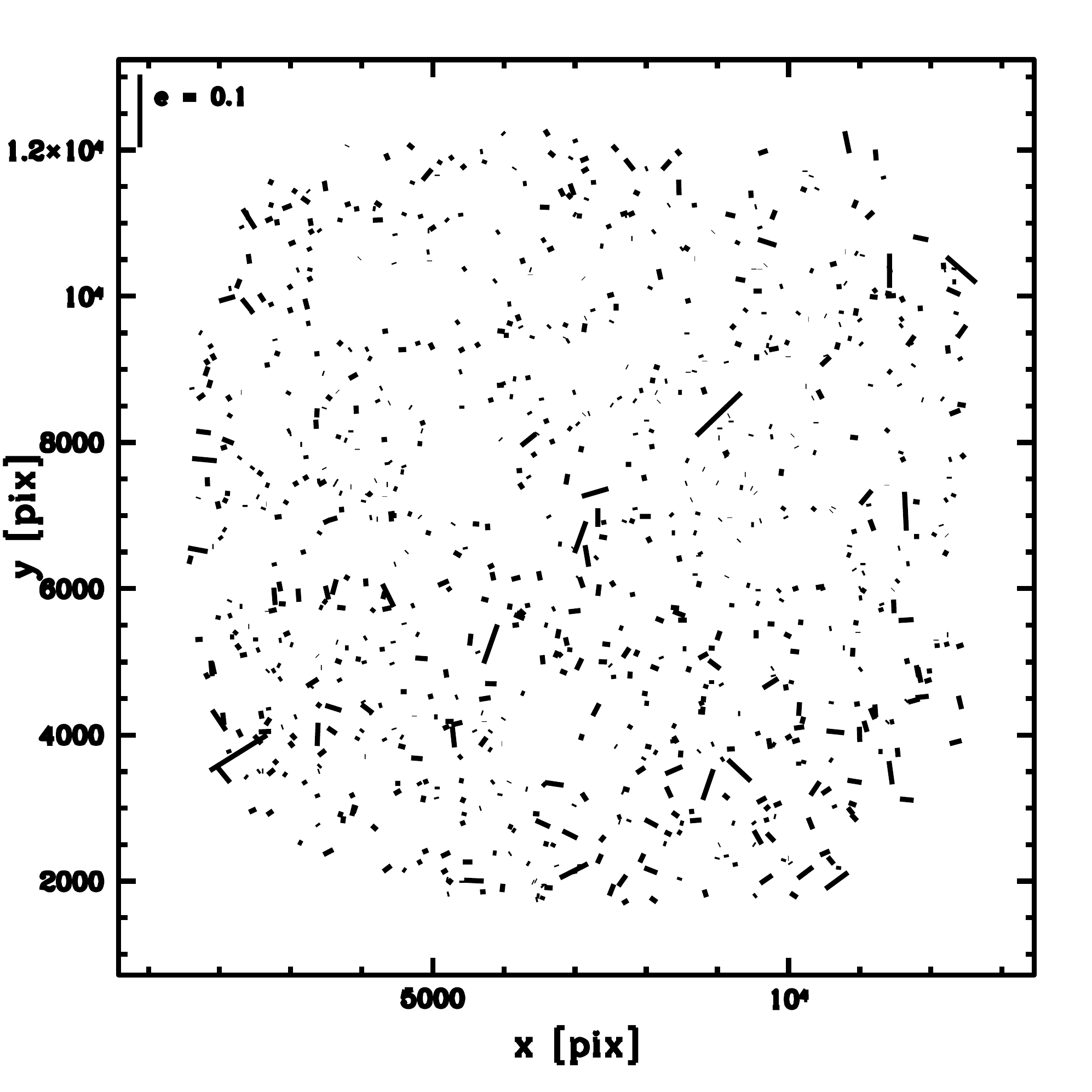}}
	\caption[Whisker plots of the residual ellipticities]{Whisker plots of the residuals of measured and modeled ellipticities of the field stars in the $r$~band stacked images. 
In the upper left corner of each image, we show a 10 per~cent ellipticity for comparison.
Large residuals could indicate wrong shape measurements of individual stars or an imperfect reconstruction of the local point spread function. However, unresolved double stars or a contamination of the sample by non-stellar objects can explain most of the few per~cent level ellipticity residuals. 
}
    \label{fig:whiskerres}
\end{figure}

The order of the polynomial has to be chosen carefully.
We use our PSF models to create mock star catalogues and visualize how well we can reproduce the PSF of our telescope optics.
We consider the residuals of the field and modeled stars (Fig~\ref{fig:psfres}) 
and make whisker plots of the ellipticity residuals (Fig.~\ref{fig:whiskerres}).
Although the WWFI~PSF can be rather elliptical, our modeled stellar ellipticities fit the measured data well, with absolute value of the mean residual ellipticities between $10^{-5}$ and $1.6\times10^{-4}$. The error of the mean residual ellipticities is well below 0.004. Comparing these results to previous weak lensing studies using instruments like MegaCam \citep{brimioulle2013dark}, the WFI on MPG/ESO \citep{gruen2013weak} and SuprimeCam \citep{von2014weighing}, we conclude that the WWFI PSF is well behaved and that we can model its spatial behaviour very accurately.

\begin{figure}
  	{\includegraphics[width = \columnwidth]{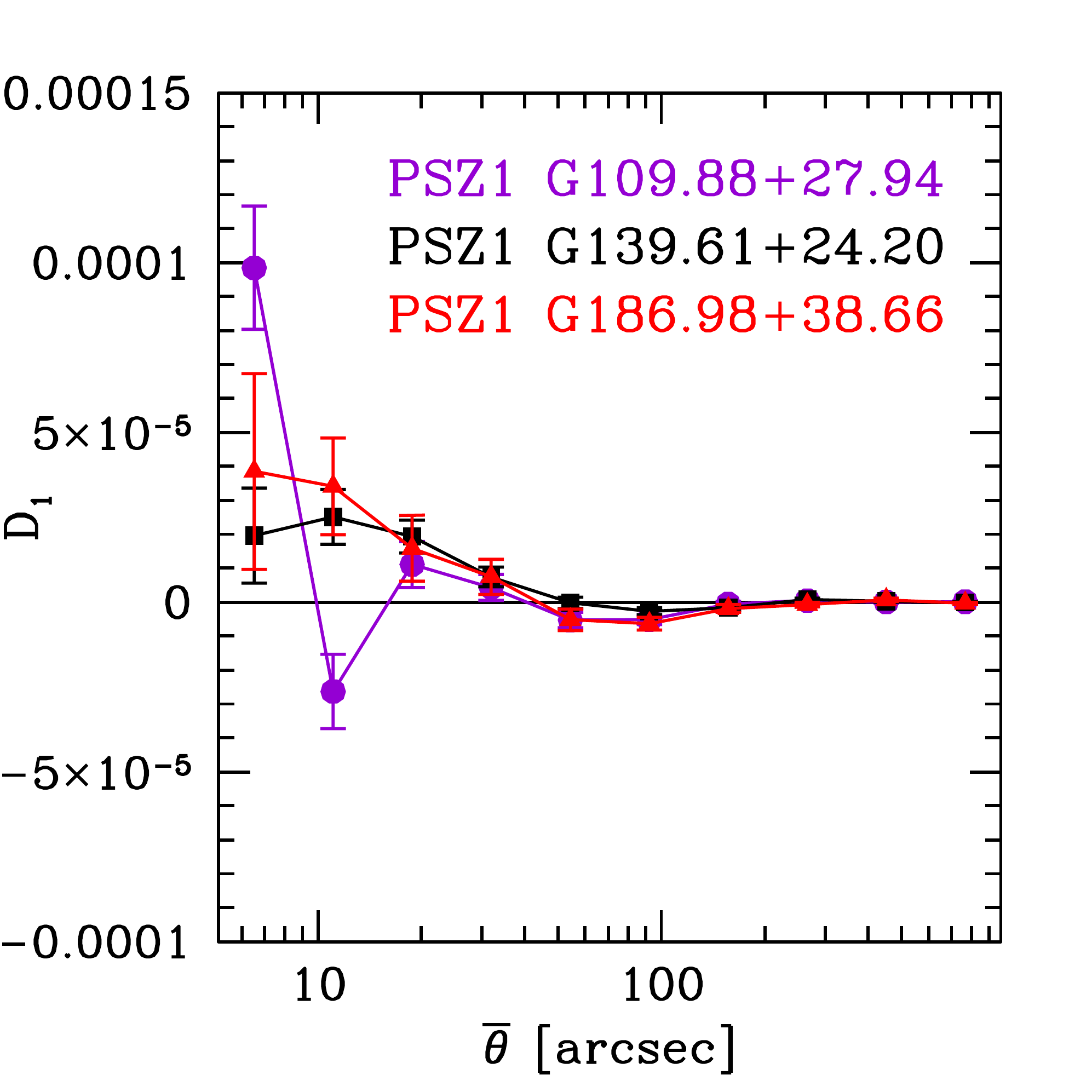}}\\
	{\includegraphics[width = \columnwidth]{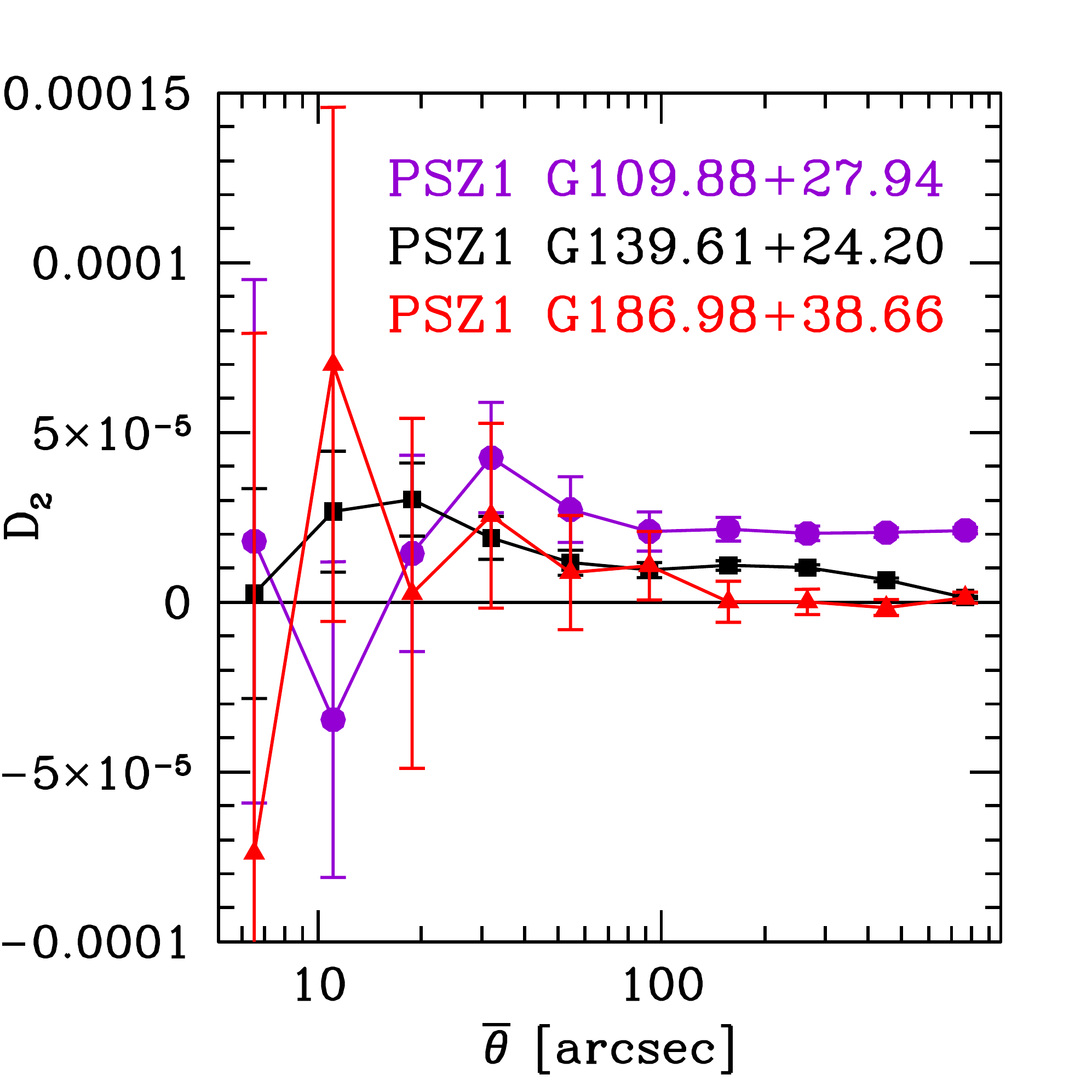}}
        \caption{The auto-correlation function of the ellipticity residuals $D_1$ and the cross-correlation function of the residuals 
          and the measured ellipticities $D_2$ of the field stars in the $r$-band, as defined in equation~(\ref{d1d2}) \citep[cf.][]{rowe2010improving}. The angular separation of the objects is denoted by $\bar\theta$.
          The \emph{Rowe} statistics for PSZ1~G109.88+27.94, PSZ1~G139.61+24.20 and PSZ1~G186.98+38.66 are presented in purple (circles), 
          black (squares) and red (triangles) respectively. On large scales, $D_1$ is consistent with zero for all clusters. 
          With the exception of the innermost datapoint of PSZ1~G109.88+27.94, all values of $D_1$ are smaller than 
          $5\times 10^{-5}$. None of the correlation functions show signs of overfitting, 
          as the cross-correlation function of the residuals and the measured ellipticities is mostly non-negative on large scales. 
          For small angular separations, the error bars grow increasingly large, yet $D_2$ is consistent with zero.
          }
       \label{fig:rowe}
\end{figure}

A metric to identify the best fitting PSF model has been proposed 
by \cite{rowe2010improving}. 
The ellipticity residual auto-correlation function $D_1$ and the cross-correlation function of 
residuals and and measured ellipticities $D_2$, defined as
\begin{equation} 
\begin{split}
D_1
\left(
\bar{\theta}
\right) 
&\equiv 
\left\langle 
\left( 
\boldsymbol{e^\star} - \boldsymbol{e}^\mathrm{PSF}
\right)*
\left(
\boldsymbol{e^\star} - \boldsymbol{e}^\mathrm{PSF}
\right)
\right\rangle  
\left(
\bar{\theta}
\right)\\
D_2
\left(
\bar{\theta}
\right) & \equiv 
\left\langle  
\boldsymbol{e^\star}*
\left( 
\boldsymbol{e^\star} - \boldsymbol{e}^\mathrm{PSF}
\right)
+
\left( 
\boldsymbol{e^\star} - \boldsymbol{e}^\mathrm{PSF}
\right)
*\boldsymbol{e^\star}
\right\rangle 
\left(
\bar{\theta}
\right) ,
\label{d1d2}
\end{split}
\end{equation} can be used as a measure of the quality of the fit.
Naturally, a perfect model means $D_{1,2} = 0$ on all scales. 
The closer to zero $D_1(\bar\theta)$ and $D_2(\bar\theta)$ are, the better the reconstructed PSF pattern. 
For our weak lensing analysis, separation angles of ${\bar\theta \gtrsim 60\ \mathrm{arcsec}}$ are of interest. 
We reject PSF models, for which $\lvert D_{1}(\bar\theta\geq1\ \mathrm{arcmin})\rvert~>~10^{-5}$.
For a perfect PSF model both, $D_{1}$ and $D_{2}$, are equal to zero.
For each cluster field, we model the PSF as polynomials with increasing order. 
We then compare the corresponding $D_{1,2}(\bar\theta)$ and select the model being closest to zero while not showing signs of overfitting. 
The best fitting \emph{Rowe}~statistics for our cluster sample are presented in Fig.~\ref{fig:rowe}. The best-fitting PSF orders are $5,6,6$ for PSZ1~G109.88+27.94, PSZ1~G139.61+24.20 and PSZ1~G186.98+38.66, respectively. 

Again comparing to literature, we find our \emph{Rowe}~statistics to be very similar to the findings of \citet{gruen2014weak}, both in amplitude of $D_{1,2}$ and general trend in that the offset of $D_2$ from zero can be as large as $3\times 10^{-5}$. They also find $D_1$ to be much better behaved than $D_2$ and show a consistency with zero on even the smallest scales for their cluster sample. \citet{gruen2014weak} do not see an offset of $D_2$ to zero in their data. 

Fig.~\ref{fig:rowe} gives the first indication, that our PSF model for the $r$-band stacked image of PSZ1~G109.88+27.94 might be biased, as $D_2(\bar\theta \gtrsim 1\ \mathrm{arcmin}) \approx 2\times10^{-5}$. We investigate this possibility and the implications on the quality of our mass estimate for PSZ1~G109.88+27.94 further in Section~\ref{sec:classicleak} and Section~\ref{sec:leakage}.

\subsection{Shape measurement}\label{sec:ksb}

We prepare the galaxy catalogue for the actual shape measurement procedure by preselecting 
unsaturated sources with flux radii larger than the stellar flux radius.
Using our \texttt{PSFEx} model and an implementation of the \texttt{KSB+} 
\citep{kaiser1994method, luppino1997detection, hoekstra1998weak} 
pipeline of \cite{gruen2013weak}, we calculate pre-seeing galaxy shapes. 

We run \texttt{KSB+} on the prepared postage stamps and the \mbox{\texttt{PSFEx}} PSF model simultaneously.  
Polarizations \citep{kaiser1994method} are calculated from the second moments of the surface brightness distribution $I(\boldsymbol\theta)$ 
of the galaxies. These are measured within an aperture weighted with a Gaussian weight function $w(\lvert  \theta \rvert)$ 
centred on the galaxy centroid.  
We scale $w(\lvert  \theta \rvert)$ with the measured half-light radius of the observed galaxy \citep{gruen2013weak}. 

Given weighted second moments of an object with surface brightness distribution $I(\boldsymbol\theta)$
\begin{equation} Q_{ij} = \int \mathrm{d}^2\theta\ w(\lvert  \theta \rvert) I(\boldsymbol\theta)\ \theta_i \theta_j ,\end{equation} 
the polarization is defined as   
\begin{equation} \boldsymbol{e} = \frac{1}{Q_{11} + Q_{22}} \begin{pmatrix}
Q_{11} - Q_{22} \\
 2 Q_{12} \\
 \end{pmatrix} . \end{equation} 
The component of the observed polarization that comes purely from the intrinsic orientation of the galaxy is called $\boldsymbol{e}^\mathrm{int}$. For our measurement of how weak gravitational lensing distorts the observed galaxy images,
we have to assume that the galaxies are randomly oriented in the sky, i.e. $\langle\boldsymbol{e}_\mathrm{int}\rangle = 0$. 
Strictly speaking, this is not true. 
A preferential orientation of source galaxies could be caused by e.g. intrinsic alignments \citep[for a recent review, cf.][]{joachimi2015galaxy}.
However, this effect has no impact on weak lensing cluster masses.

\subsection{Shear measurement}\label{sec:shear}

Polarizations are defined as weighted second moments of the image intensities. The observed post-seeing polarization $\boldsymbol{e}^\mathrm{obs}$ does not only depend on $\boldsymbol{e}^\mathrm{int}$ but 
also on the polarization of the PSF image $\boldsymbol{p}$ and the reduced shear $\boldsymbol{g}$ distorting the intrinsic galaxy shape. 
Introducing
tensors to describe the linear response of $\boldsymbol{e}^\mathrm{obs}$ to  $\boldsymbol{p}$ and $\boldsymbol{g}$, 
the observed polarization can be expressed as
\begin{equation} \boldsymbol{e}^\mathrm{obs} = \boldsymbol{e}^\mathrm{int} +\hat{\boldsymbol{P}}^\mathrm{sm} p 
+ \hat{\boldsymbol{P}}^{\gamma}\boldsymbol{g}, \label{pol}\end{equation}
 where $\hat{\boldsymbol{P}}^\mathrm{sm}$ is the smear polarizability tensor and $\hat{\boldsymbol{P}}^{\gamma}$ denotes the shear responsivity tensor.
Point-like sources have ${\boldsymbol{e}^\mathrm{int}=\boldsymbol{g}=0}$, which means that, in the absence of PSF distortions, 
the ellipticity of a star is zero. The atmospheric seeing is described as a large circularly symmetric disc convolved with a small, 
highly anisotropic distortion. The PSF anisotropy corrected ellipticity is thus given by 
\begin{equation}\boldsymbol{e}^\mathrm{cor}= \boldsymbol{e}^\mathrm{obs} - \hat{\boldsymbol{P}}^\mathrm{sm} \boldsymbol{p} , \end{equation} 
where the position dependent vector $\boldsymbol{p}$ is estimated from the PSF image \citep{luppino1997detection,hoekstra1998weak}.

Now taking the response of the ellipticity to the shear into account, 
while assuming the galaxies to be randomly oriented, we obtain an estimate for the ensemble reduced shear
\begin{equation}\langle \boldsymbol{g} \rangle= \langle \boldsymbol{\epsilon} \rangle=
\left\langle \frac{2}{\mathrm{tr}{\hat{\boldsymbol{P}}^{\gamma}}} \boldsymbol{e}^\mathrm{cor} \right\rangle,\label{eq:ge}\end{equation} 
where $\boldsymbol\epsilon$ is the measured galaxy ellipticity. In equation~\ref{eq:ge}, we have  approximated the inverted shear responsivity tensor as 
${(\hat{\boldsymbol{P}}^{\gamma})^{-1} \approx 2 / \mathrm{tr}{\hat{\boldsymbol{P}}}}$.
In accordance to \cite{gruen2013weak}, we only include objects with successful \texttt{KSB+} shape measurements 
with $\mathrm{tr}{\hat{\boldsymbol{P}}} \geq 0.1$ in our shape catalogues.

The reduced shear $g$ combines the effects on the galaxy images induced by the convergence $\kappa$ and the shear $\gamma$ 
\begin{equation}
g = \frac{\gamma}{1-\kappa}. 
\label{eq:g}
\end{equation}
In polar coordinates, where the lens is at the origin, the tangential and cross component of the gravitational shear can be written in terms of the polar angle $\varphi$,  
\begin{equation}
\begin{split}
g_\mathrm{t} &= -\left[g_1\left(\boldsymbol\theta\right)\cos\left(2\varphi\right) + g_2\left(\boldsymbol\theta\right)\sin\left(2\varphi\right)\right]\\
g_\mathrm{x} &= -g_1\left(\boldsymbol\theta\right)\sin\left(2\varphi\right) + g_2\left(\boldsymbol\theta\right)\cos\left(2\varphi\right)
.
\label{eq:gtgx}
\end{split}
\end{equation}

\subsection{Impact of biases on the WWL shapes} 

The FWHM of the PSF in our lensing band is in the sub-asrcsecond regime and
the PSF ellipticity can be as large as ${\sim15}$~per~cent in some areas of the stacked images 
(cf.~Fig.~\ref{fig:whisker}). The question arises whether a weak lensing measurement is actually feasible with our current WWFI data.
In order to verify the quality of our shape catalogues, we have to test whether our implementation of \texttt{KSB+} recovers unbiased galaxy shapes in the presence of large PSF ellipticity, or if our PSF models are an insufficient description of the true PSF in our stacks. 

Calibration bias is caused by a poor correction for the effects of atmospheric seeing on galaxy shapes.
\cite{heymans2006shear} describe the effect on the measured shear as a multiplicative and additive bias, 
\begin{equation} \langle\boldsymbol\epsilon^\mathrm{obs} -\boldsymbol\epsilon^\mathrm{true}\rangle = m \times \langle\boldsymbol\epsilon^\mathrm{true}\rangle + \boldsymbol{c}.
\label{eq:shearbias}
\end{equation}
In the presence of a calibration bias, $m$ is expected to be non-zero, while PSF systematics 
also imply $\boldsymbol{c} \neq 0$.
If the response of $\boldsymbol\epsilon^\mathrm{obs}$ to shear is non-linear, a third term has to be introduced \citep[their~equation~11]{heymans2006shear}, 
which we do not consider in our analysis.

\subsubsection{Multiplicative shear bias}\label{sec:noisebias}

The multiplicative shape measurement bias $m$ influences the amplitude of the measured ellipticities, which translates into a change of the shear amplitude. 
As the mass measurement scales with the shear, $m$ biases weak lensing cluster mass estimates.
We determine the multiplicative bias by repeating the shape measurement on simulated galaxy images.
We correct our shape catalogues by applying a S/N-calibration of the galaxy ellipticities. 
\cite{gruen2013weak} determined 
the dependency of $m$ in our pipeline on the signal-to-noise ratio of a galaxy by fitting a functional form of $m$,
\begin{equation} m \approx -0.025 -0.17 \exp\left(-\frac{(\mathrm{S}/\mathrm{N})_\mathrm{gal}}{17}\right),\end{equation} with a minimum (S/N)$_\mathrm{gal}$ of 10. 
They define $m$ as the deviation of the ratio of mean shapes as measured on simulated cluster fields using their implementation of \texttt{KSB+} 
to the true shapes from 1. As expected, the absolute value of the multiplicative bias decreases as a function of signal-to-noise until it reaches an almost 
constant value for large (S/N)$_\mathrm{gal}$. 

The simulations that have been used are not a perfect representation of the WWFI data. One of the main problems is likely the large PSF ellipticity. A dependency of $m$ on the profiles of the galaxies and the distribution of their sizes and ellipticities has also not been taken into account. For these reasons we expect a \emph{residual multiplicative shear bias}. 
We conservatively estimate this bias to be up to 5~per~cent.
In order to get a feeling for how realistic this value might be, we make use of the fact that one cluster in our sample (PSZ1~G186.98+38.66) has also been observed with Subaru and is part of the WtG project. We match the shape catalogues and calculate the tangential shear signals using the different data sets. Our mean tangential shear in 5 radial bins from the cluster centre is proportional to the shear signal calibrated by WtG with a proportionality constant that is consistent with 1, i.e. $\sigma_\mathrm{m} = 0.03 \pm 0.07$. Selection effects and biases in the WtG galaxy shape catalogue aside, this test shows that $\sigma_\mathrm{m} = 0.05$ is an adequate budget for the residual multiplicative shear bias.

\subsubsection{Additive shear biases}

We consider three different kinds of additive systematics in our shape catalogues, one of which is constant over the whole fields and two of which are related to the PSF.
We investigate their impact on our cluster masses in Section~\ref{sec:meane},~\ref{sec:modelbias}~and \ref{sec:leakage1}, respectively. 

\paragraph{Mean ellipticity}\label{sec:meane}

\begin{figure*}
	\centering
	{\includegraphics[width = 2.2in]{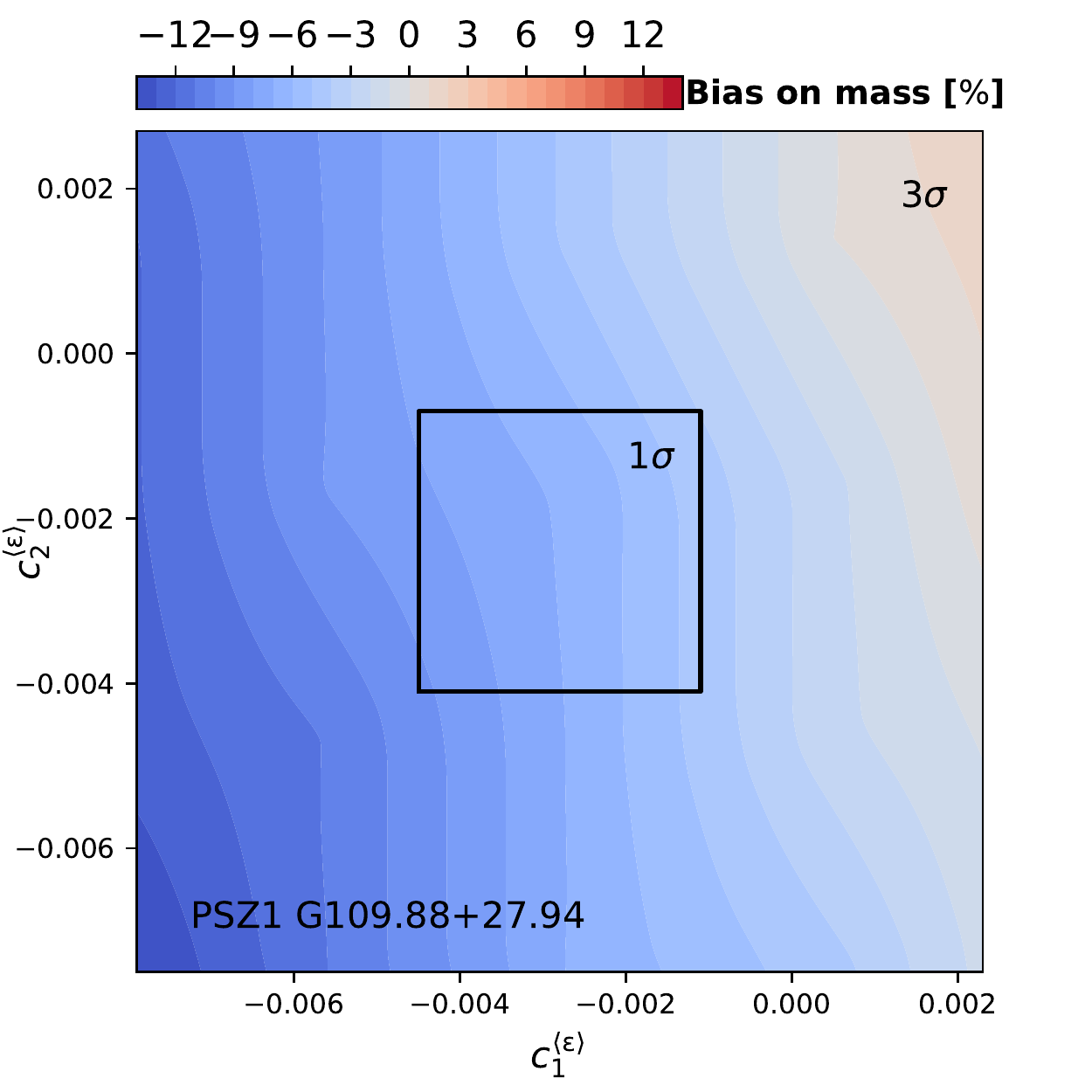}}  
	{\includegraphics[width = 2.2in]{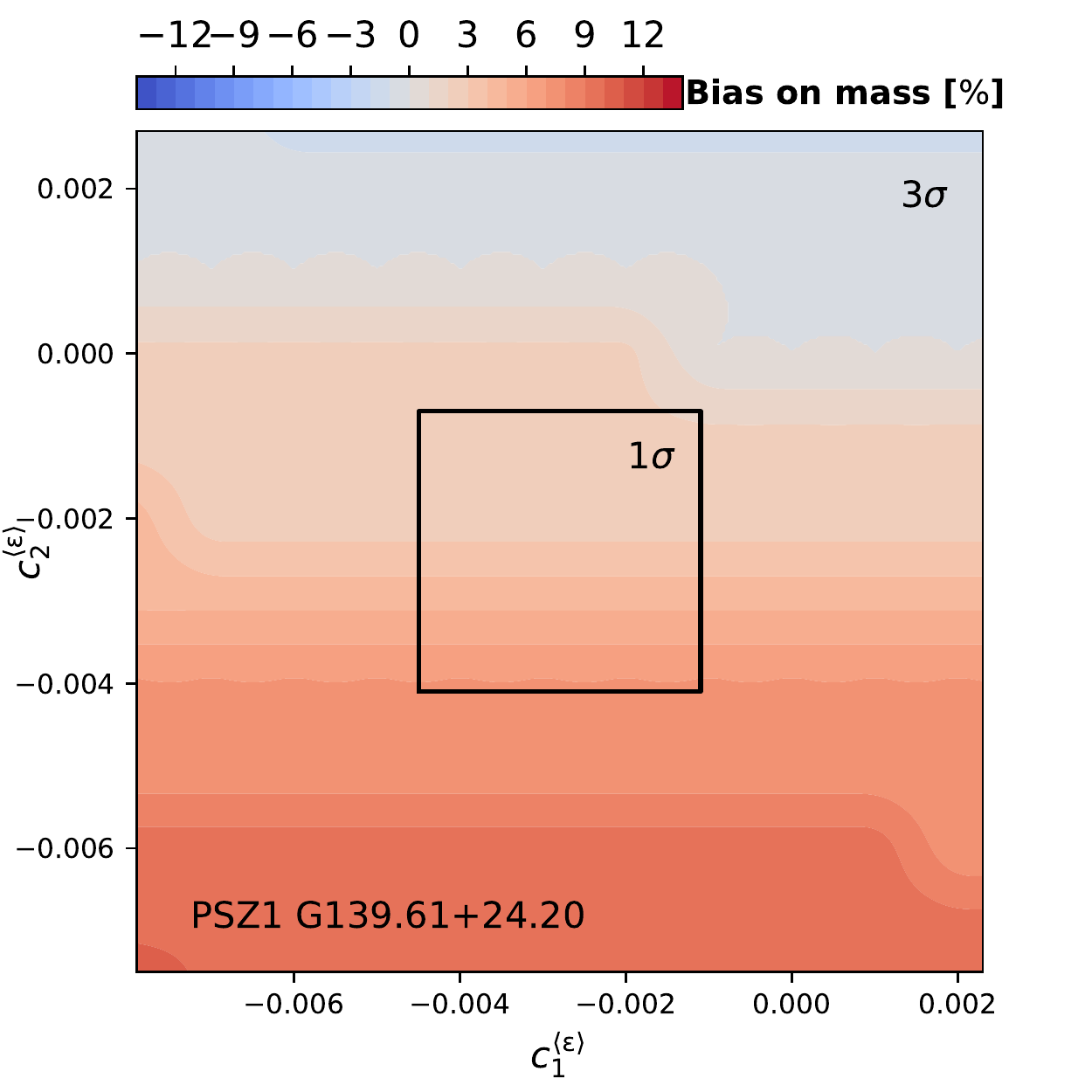}}
	{\includegraphics[width = 2.2in]{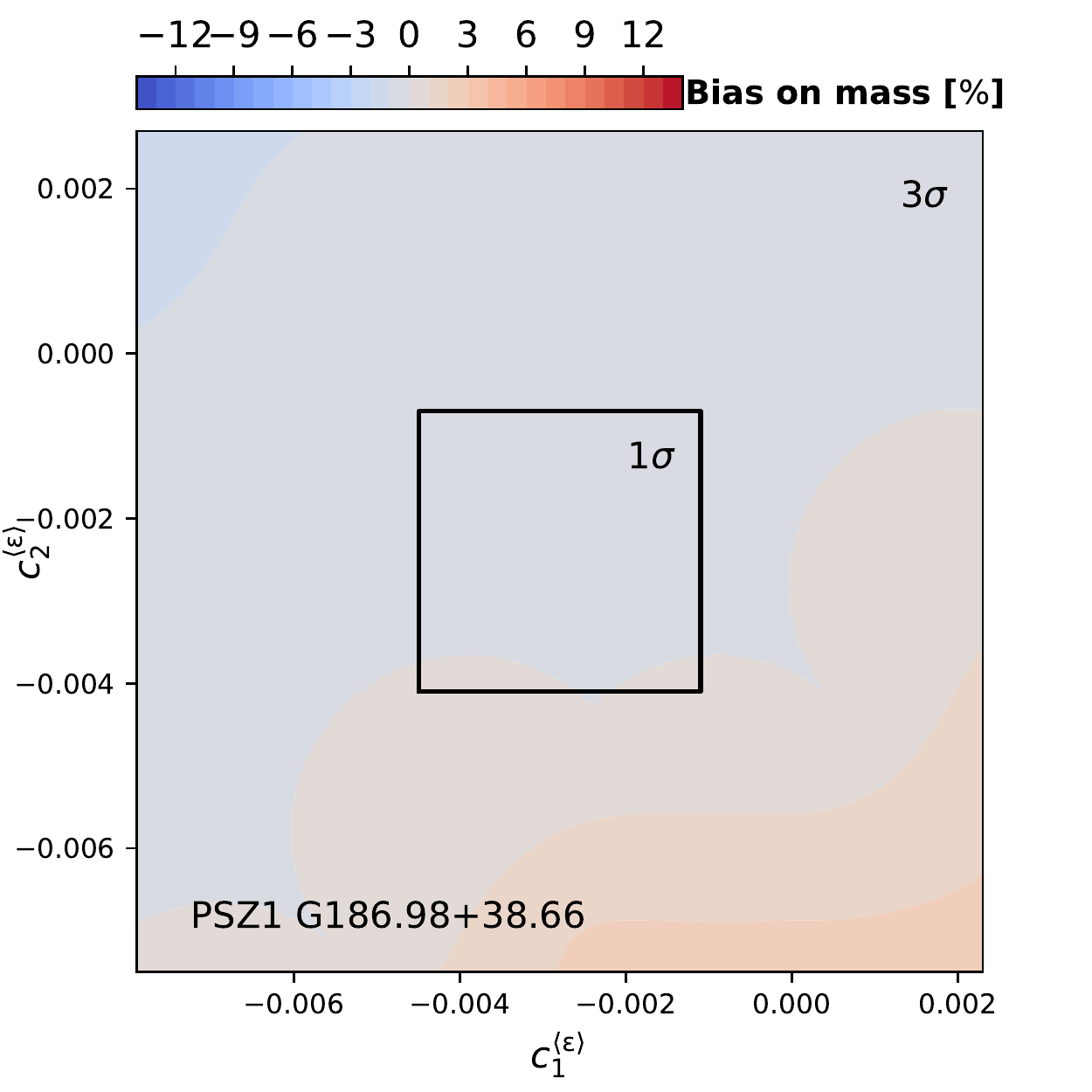}}
\caption{The added uncertainty on the mass at given components of the additive shear bias $c^\mathrm{\langle\epsilon\rangle}_{1,2}$ varies between the cluster fields. The plots show the $3\sigma$ intervals of the measured mean ellipticities in the survey. The solid black box encompasses all values of $c^\mathrm{\langle\epsilon\rangle}_{1,2}$ within $1\sigma$ 
of the best fitting value of $\boldsymbol c^\mathrm{\langle\epsilon\rangle}$.}
       \label{fig:massbias}
\end{figure*}

We find a spatially \emph{constant mean ellipticity} of ${{c^\mathrm{\langle\epsilon\rangle}_{1,2} = (-2.8,-2.4) \pm (1.7,1.7) \times 10^{-3}}}$ in our cluster fields. 
In the presence of a radially non-symmetric mask of random orientation, this causes a mean tangential shear component 
that affects the recovered weak lensing cluster mass. 
The impact of mean ellipticity bias on the mass depends on the applied mask in the individual field. Consequently, it is really a "statistical" uncertainty and will decrease with increasing WWL sample size.
We estimate the uncertainty on the cluster mass by adding a constant offset of ${\boldsymbol {c}^\mathrm{\langle\epsilon\rangle}}$ to the observed galaxy ellipticity, running our two-parameter NFW fitting code (cf.~Section~\ref{sec:nfw}) and comparing the results to the mass profile that best fits our original data. 
In this way we find the statistical uncertainty on the cluster mass caused by the mean ellipticity in the data to be smaller than ${\sigma^\mathrm{\langle\epsilon\rangle} \lesssim 9.0\%,7.2\%,0.3\%}$ for PSZ1~G109.88+27.94, PSZ1~G139.61+24.20 and PSZ1~G186.98+38.66, respectively. 
We show the impact of a hypothetical additive shear bias within $3\sigma$ from the measured mean ellipticity in Fig.~\ref{fig:massbias}. 
As expected, the conservative masking in the stacks of PSZ1~G109.88+27.94 and PSZ1~G139.61+24.20 causes an impact of a non-zero $c^\mathrm{\langle\epsilon\rangle}$ on the cluster mass. 
The weak lensing mass of PSZ1~G186.98+38.66, however, changes even less significantly. Even if we assume an additive shape bias that is $3\sigma$ larger than $\boldsymbol c^\mathrm{\langle\epsilon\rangle}$, the uncertainty due to this on the cluster mass is less than 2~per~cent for this field.

\citet{heymans2012cfhtlens} find a mean ellipticity of ${c_{1,2} = (0.1,2.0) \pm (0.1,0.1) \times 10^{-3}}$ in the CFHTLenS shape catalogues. The value scales with galaxy size and signal-to-noise.
Even surveys as large as the Dark Energy Survey (DES) still find a non-zero mean ellipticity. 
\citet{jarvis2015science} find mean ellipticities of $\langle e_{1,2}\rangle = (0.1,6.8)\times10^{-4}$ in their \texttt{IM3SHAPE} and ${\langle e_{1,2}\rangle = (-0.4,10.2)\times10^{-4}}$ in their \texttt{NGMIX} shape catalogue for the Science Verification data. The DES Year 1 survey shows a mean ellipticity of  ${\langle e_{1,2}\rangle = (3.5,2.8)\times10^{-4}}$ and ${\langle e_{1,2}\rangle = (0.4,2.9)\times10^{-4}}$ using \texttt{METACALIBRATION} and  \texttt{IM3SHAPE}, respectively \citep{zuntz2017dark}.
This shows that the mean ellipticity can be different if another shape measurement technique is applied to the same data and that ${\mid\langle e \rangle\mid}$ might decrease if the survey area increases. 

The reason for this phenomenon is not known but does not depend on the PSF model. The linear model introduced in Section~\ref{sec:classicleak} disentangles the PSF dependent from the spatially constant part of the galaxy ellipticities. 

\paragraph{Model bias}\label{sec:modelbias}

An insufficiently modeled PSF can cause a position dependent additive shape bias, which we call the \emph{PSF~model bias}. Our PSF models were designed to minimize the PSF~model bias (cf.~Section~\ref{sec:psf}).
We use the \emph{Rowe} statistics as a diagnostics to verify the presence of PSF~model bias in our weak lensing shear catalogues. 
$D_1$ is consistent with zero for angular separations $\gtrsim30''$ for all fields, and also $D_2$ is small (cf.~Fig.~\ref{fig:rowe}). 
We take the cross-correlation of the stellar and the residuals of stellar and model ellipticity as the maximum variance of our PSF model, i.e. ${{|c^\mathrm{mPSF}|} \lesssim \sqrt{\langle D_2(\bar{\theta}>1') \rangle}}$.   
We find ${|c^\mathrm{mPSF}| \lesssim (4,3,1)\times10^{-3}}$ for PSZ1~G109.88+27.94, PSZ1~G139.61+24.20 and PSZ1~G186.98+38.66, respectively. We find upper limits of PSF~model~bias on the cluster masses equal to $\sigma_\mathrm{c^\mathrm{mPSF}} \lesssim 8.5\%,3.5\%,0.5\%$ by adding the maximum PSF~model~bias to the measured galaxy shapes before fitting for $M_\mathrm{200m}$.
Compared to the spatially constant mean ellipticity, this bias is subdominant for PSZ1~G186.98+38.66. In the field of PSZ1~G109.88+27.94, the effect of PSF~model bias can be as large as the mean ellipticity bias. As $c^\mathrm{mPSF}$ could be negative, these two effects could also cancel out and leave our mass constraints unaffected by these additive biases.

\paragraph{PSF leakage}\label{sec:leakage1}

\begin{figure}
   	{\includegraphics[width = \columnwidth]{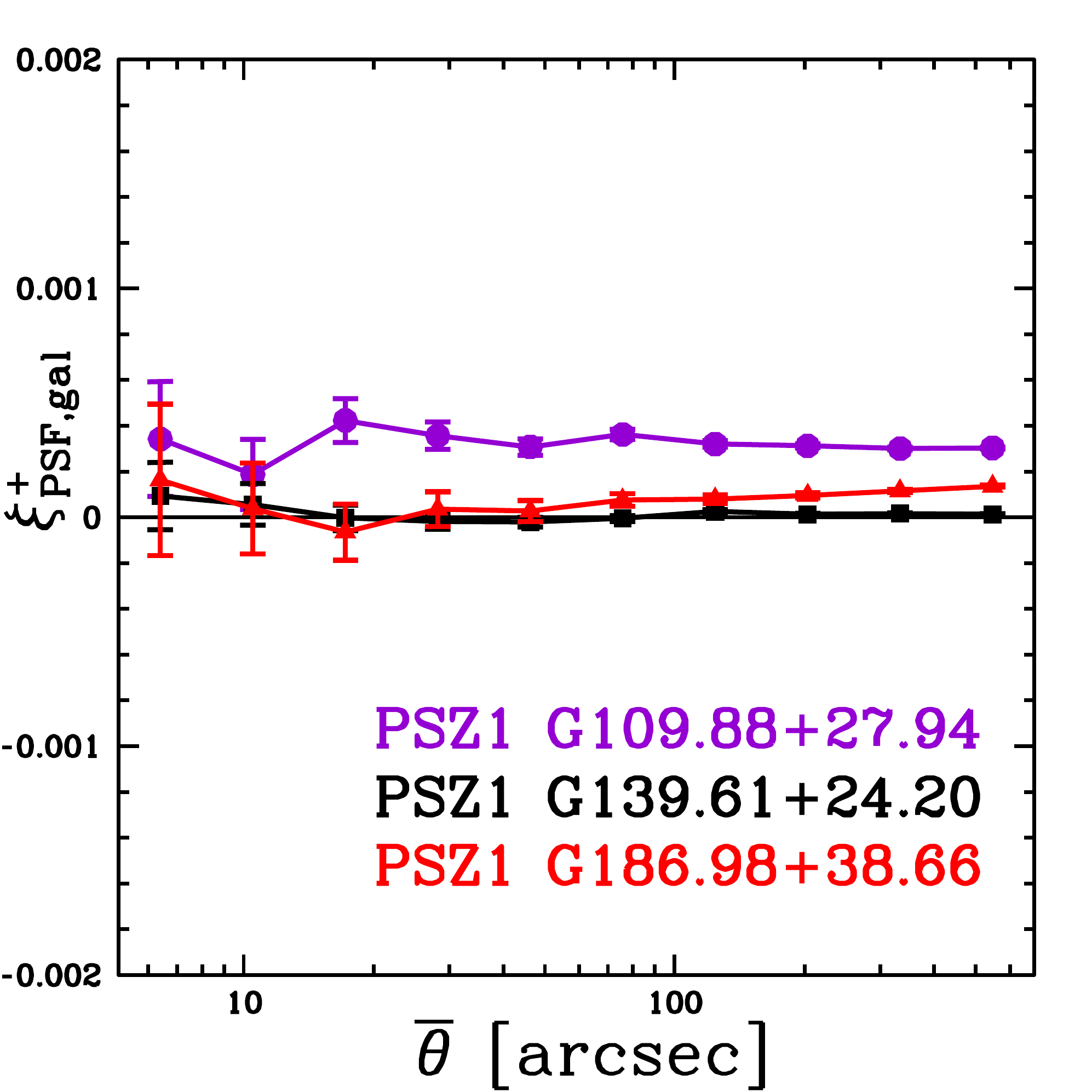}}
\caption{The cross-correlation function of the ellipticities of the PSF and the galaxies as a function of angular 
separation for PSZ1~G109.88+27.94 (violet circles), PSZ1~G139.61+24.20 (black squares) and PSZ1~G186.98+38.66 (red triangles).
}
       \label{fig:xipg}
       \end{figure}

Even if there is only a small PSF~model bias in our shape catalogues, there is a second type of position dependent additive bias. 
\emph{PSF~leakage} occurs, if the PSF is not deconvolved properly from the source images 
or if selection depends on the alignment of PSF and galaxy ellipticity. 
This causes the observed gravitational shear and the PSF~ellipticity to be correlated. 
We can use this effect to confirm the presence of some leakage in our shape catalogues by calculating 
the cross-correlation function of PSF model and galaxy ellipticity
\begin{equation}
{\xi^+_{\mathrm{PSF,gal}}} = \left\langle 
\boldsymbol{e}^\mathrm{PSF}
*
 \boldsymbol\epsilon^\mathrm{gal}
\right\rangle
 +
 \left\langle
 \boldsymbol\epsilon^\mathrm{gal}
*
\boldsymbol{e}^\mathrm{PSF}
\right\rangle.
\label{eq:xipsfgal}
\end{equation}
In the equation above, $\boldsymbol\epsilon^\mathrm{gal}$ denotes the galaxy ellipticities, as measured by our \texttt{KSB+} pipeline. At each galaxy position, we also have an estimate for the value of the PSF ellipticity predicted by our \texttt{PSFEx}~model $\boldsymbol{e}^\mathrm{PSF}$.
As can be seen in Fig.~\ref{fig:xipg}, the correlation between galaxy and PSF ellipticities is small ({$\lesssim 0.0006$}) and strongest for PSZ1~G109.88+27.94.  
On large scales, PSZ1~G186.98+38.66, too, shows some correlation ${({\xi^+_{\mathrm{PSF,gal}}} \sim 0.0001)}$, while the PSZ1~G139.61+24.20 galaxy shapes seem to be unbiased by PSF leakage.

\subparagraph{Linear fit: \emph{Classical approach}}\label{sec:classicleak}

The amount of leakage is generally described as a linear dependency of the galaxy shapes on the PSF ellipticities. We can then rewrite equation~\ref{eq:shearbias} in the following way: 
\begin{equation}
\langle \boldsymbol\epsilon^\mathrm{obs} - (1-m) \boldsymbol\epsilon^\mathrm{true}\rangle = \boldsymbol{c}
= \boldsymbol{\alpha}\langle \boldsymbol{e}^\mathrm{PSF} \rangle + \boldsymbol{b}.
\label{eq:leak1}
\end{equation}  
Under the assumption that the galaxies are randomly oriented (i.e.~${\langle\boldsymbol{\epsilon}^\mathrm{true} \rangle = 0}$), we can find the \emph{linear leakage factor} $\boldsymbol\alpha$
and the \emph{additive leakage bias} $\boldsymbol{b}$ by assuming a linear model for the observed mean galaxy ellipticities $\left\langle\boldsymbol\epsilon^\mathrm{obs}\right\rangle$, 
\begin{equation}
\left\langle\boldsymbol\epsilon^\mathrm{obs}\right\rangle = \boldsymbol\alpha\cdot\left\langle\boldsymbol{e}^\mathrm{PSF}\right\rangle + \boldsymbol b ,
\label{eq:fit1}
\end{equation}      
where we average over galaxies with similar PSF~ellipticities.
The additive leakage bias $\boldsymbol{b}$ is a noisy measure for the mean ellipticity $\boldsymbol{c^\mathrm{\langle\epsilon\rangle}}$ (cf.~Section~\ref{sec:meane}), as it is constant over the whole field.

\begin{figure}
   	{\includegraphics[width = \columnwidth]{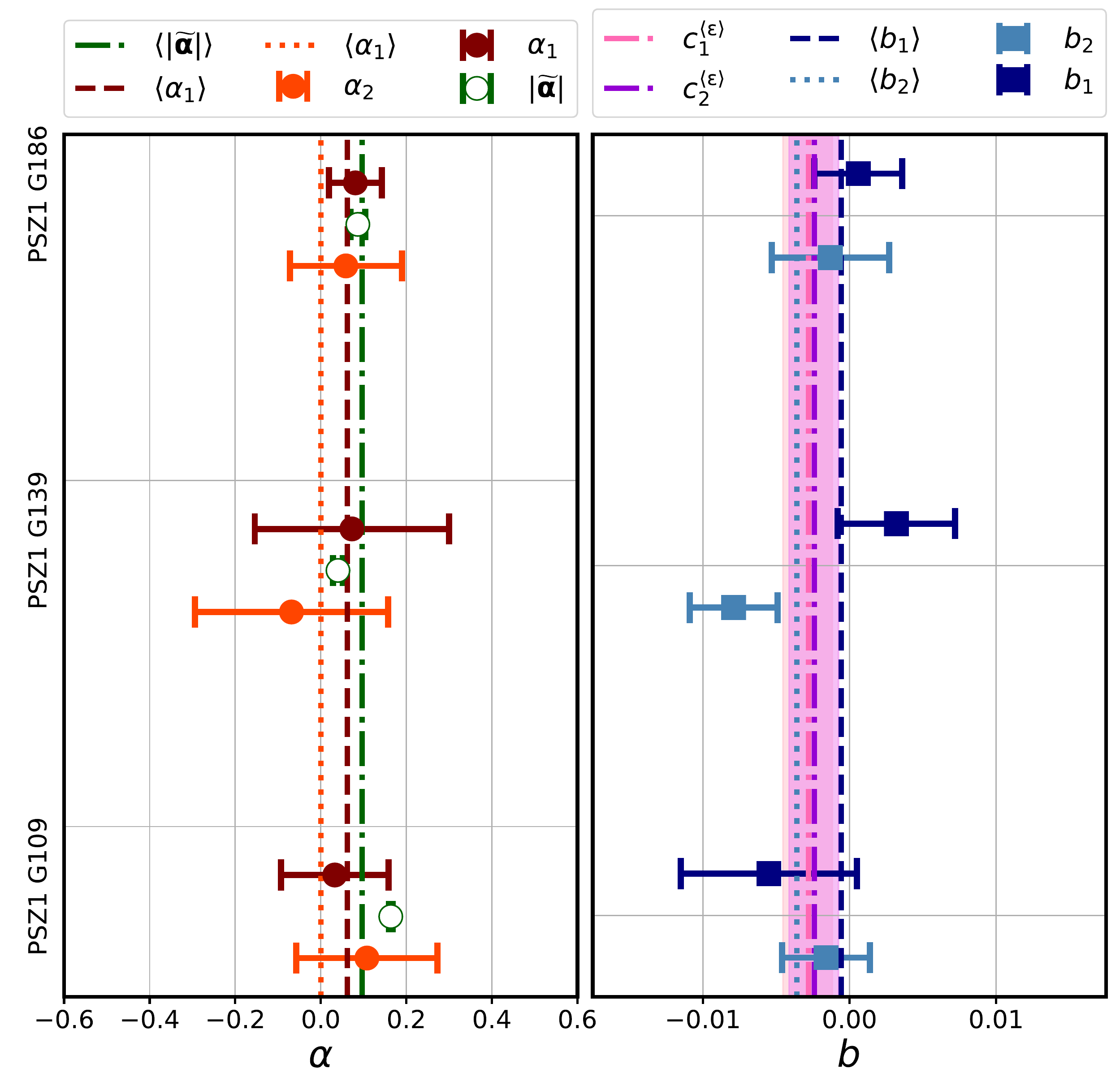}}
\caption{We perform a linear regression and fit equation~\ref{eq:fit1} to our data. \emph{Left panel}: The first component of the leakage factor $\alpha_1$ can be seen in red and the second component $\alpha_2$ in orange. 
The open green symbols show the results using our method from Section~\ref{sec:leakage} to estimate the absolute value of the vector $\tilde\alpha$ from equation~\ref{etsys}.
The same colour scheme is used to show the mean value of $\alpha_1,\alpha_2$ and $\tilde\alpha$ (dashed, dotted and dotted-dashed line, respectively).
\emph{Right panel}: The dark blue symbols show the values of the first component of the additive leakage bias $b_1$, the second component $b_2$ is shown in light blue. The dark (light) blue dashed (dotted) line shows the mean value of $b_1$ ($b_2$). The mean ellipticities and their $1\sigma$ intervals are shown in pink ($c^\mathrm{\langle\epsilon\rangle}_1$) and purple ($c^\mathrm{\langle\epsilon\rangle}_2$).
}
       \label{fig:fitparamsleak}
\end{figure}

The $\xi^+_{\mathrm{PSF,gal}}$ of the cluster fields suggests that, out of all cluster fields in our WWL~pathfinder~sample, the linear leakage factor $\alpha$ should be largest for PSZ1~G109.88+27.94 (cf.~Fig~\ref{fig:xipg}). We fit our model of $\left\langle\boldsymbol\epsilon^\mathrm{obs}\right\rangle$ (cf. equation~\ref{eq:fit1}) to the data to find estimates on $\boldsymbol\alpha$ and $\boldsymbol b$.
The results of our linear fit can be seen in Fig,~\ref{fig:fitparamsleak}.
Our estimated components of the linear leakage factor $\alpha_{1,2}$ are all consistent with zero but have large uncertainty due to shape noise. We can neither confirm that $\alpha_1 = \alpha_2$, nor can we make any statements on the variation of $\boldsymbol\alpha$ in the different fields.
The same holds for the additive leakage bias. The error bars are much larger than the error on the mean of the galaxy ellipticities in the fields $\sigma^\mathrm{c^{\langle\epsilon\rangle}}_{1,2}$. Though there is no indication that $c^\mathrm{\langle\epsilon\rangle}_1 \neq c^\mathrm{\langle\epsilon\rangle}_2$, we find a $2\sigma$ evidence that $b_1 \neq b_2$ for PSZ1~G139.61+24.20.  

If we average $b_1$ and $b_2$ over all fields, we find consistent results compared to the mean ellipticity. 
The mean leakage factor in the fields is $\alpha_{1,2} = (0.06,0.03) \pm (0.02,0.06)$. 
The strategy to fit a linear relation between $\boldsymbol\epsilon$ and $\boldsymbol{e}^\mathrm{PSF}$ does not yield results we can use to correct our ellipticities. 
Below, we present a new approach to correct for the effect of the additive shear bias $\alpha e^\mathrm{PSF}$ on cluster weak lensing mass estimates despite the low number of galaxies in our catalogues.

\subparagraph{Leakage correction for small shape catalogues}\label{sec:leakage}

\begin{figure}
	{\includegraphics[width = \columnwidth]{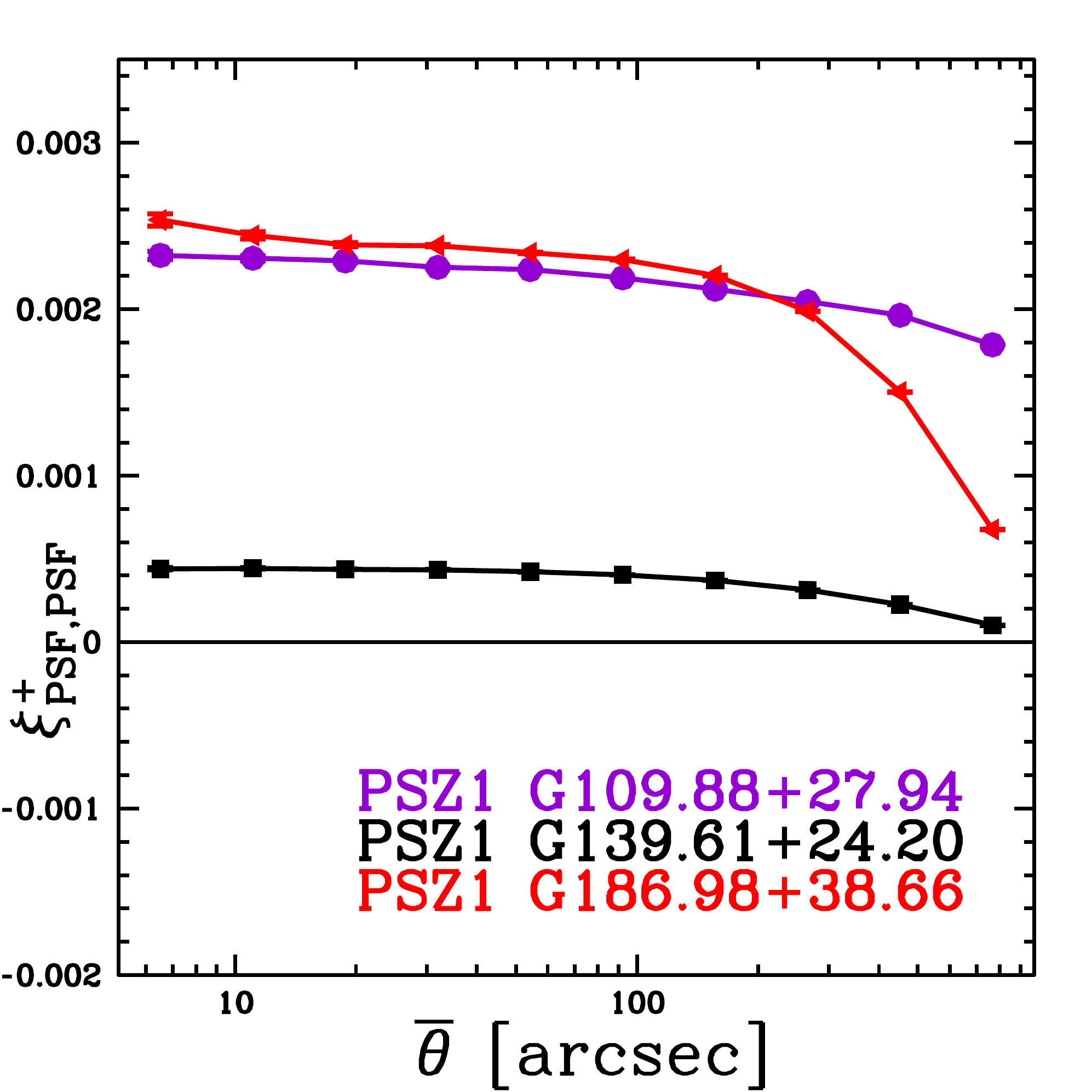}}
\caption{The auto-correlation function of the ellipticities of the PSF as a function of angular 
separation for PSZ1~G109.88+27.94 (violet circles), PSZ1~G139.61+24.20 (black squares) 
and PSZ1~G186.98+38.66 (red triangles).}
       \label{fig:xipp}
\end{figure}

For our cluster weak lensing analysis, we are only interested in the azimuthally averaged tangential component of the additive shape bias $\langle \boldsymbol c \rangle$. We introduce a method to model the additive bias on the mean tangential shear averaged in circular annuli around the cluster centre $\langle \boldsymbol c \rangle \coloneqq e_\mathrm{t}^\mathrm{sys}$. 
We make the assumption that the additive shear bias caused by leakage is linearly dependent on the PSF ellipticity and that a constant contribution is negligibly small (i.e.~${\langle \boldsymbol b \rangle = 0}$).
We further assume that ${\alpha_1 \approx \alpha_2}$ 
to write the observed tangential shear signal measured in radial bins as 
\begin{equation}
\begin{split}
\left\langle e_\mathrm{t}^\mathrm{obs} \right\rangle &= 
\left\langle -\left( e_\mathrm{1}^\mathrm{true} + {\alpha} e_\mathrm{1}^\mathrm{PSF}\right) \cos(2\boldsymbol\theta)
- \left( e_\mathrm{2}^\mathrm{true} + {\alpha} e_\mathrm{2}^\mathrm{PSF}\right) \sin(2\boldsymbol\theta) \right\rangle \\
&=
\left\langle e_\mathrm{t}^\mathrm{true} \right\rangle + \left\langle {\alpha} e_\mathrm{t}^\mathrm{PSF} \right\rangle :=
\left\langle e_\mathrm{t}^\mathrm{true} \right\rangle + \underbrace{\tilde{{\alpha}}\left\langle e_\mathrm{t}^\mathrm{PSF} \right\rangle}_{\left\langle e_\mathrm{t}^\mathrm{sys}\right\rangle} \ ,
\label{eq:alpha2}
\end{split}
\end{equation}
where we have defined the \emph{tangential~PSF~leakage~factor~$\tilde{\alpha}$}.
In order to calculate the \emph{systematic tangential shear} profile $\langle e_\mathrm{t}^\mathrm{sys}\rangle$,
we first have to estimate $\tilde{\alpha}$ and then measure the tangential shear profile of our model~PSF~ellipticities.
Using our definition of $\tilde\alpha$, we can write
$\left\langle e_i * e^\mathrm{PSF}_i \right\rangle = \tilde\alpha \left\langle e^\mathrm{PSF}_i * e^\mathrm{PSF}_i \right\rangle$, where we can identify $\left\langle e_i * e^\mathrm{PSF}_i \right\rangle$ as ${\xi^+_{\mathrm{PSF},\mathrm{gal}}}(\bar{\boldsymbol\theta})$ (cf.~Fig.~\ref{fig:xipg}). 
$\left\langle e^\mathrm{PSF}_i * e^\mathrm{PSF}_i \right\rangle$ is the autocorrelation~function of the PSF ellipticity ${\xi^+_{\mathrm{PSF},\mathrm{PSF}}}(\bar{\boldsymbol\theta})$ and is shown in Fig.~\ref{fig:xipp} for the cluster fields. The angular galaxy separation is once again denoted by $\bar{\boldsymbol\theta}$.
We calculate ${\xi^+_{\mathrm{PSF},\mathrm{gal}}}(\bar{\boldsymbol\theta})$ and $\xi^+_{\mathrm{PSF},\mathrm{PSF}}(\bar{\boldsymbol\theta})$ in 10 bins of angular galaxy separation and 
fit the linear leakage factor $\tilde{\alpha}$ to the functional ${\xi^+_{\mathrm{PSF},\mathrm{gal}}}(\bar{\boldsymbol\theta}) = \tilde{\alpha}\xi^+_{\mathrm{PSF},\mathrm{PSF}}(\bar{\boldsymbol\theta})$. The standard errors are estimated using bootstrapping.
We can now compare the results of the fit with the values for the components of $\boldsymbol\alpha$ we have obtained in Section~\ref{sec:classicleak} (cf.~Table~\ref{tab:alpha}).
\begin{table*}
	\centering
	\caption{Field name, the leakage factor $\tilde{\alpha}$ from equation~\ref{etsys} and its bootstrap error, the first and second component of the leakage factor $\boldsymbol\alpha$ (cf.~equation~\ref{eq:fit1}) and their $1\sigma$ errors.}
	\label{tab:alpha}
	\begin{tabular}{lcccccr} 
		\hline
		Object &$\tilde{\alpha}$&$\sigma_{\tilde{\alpha}}$ & $\alpha_1$ & $\sigma_{\alpha_1}$ & $\alpha_2$& $\sigma_{\alpha_2}$\\
		\hline
				\hline

		PSZ1~G109.88+27.94 	&  0.15	& 0.01 	& 0.03	&  0.13 & 0.11  &   0.17 \\
		PSZ1~G139.61+24.20	& 0.03 	& 0.02 	& 0.07	&  0.23 & -0.07  &   0.23 \\
		PSZ1~G186.98+38.66 	& 0.03 	& 0.01  	& 0.08   &  0.06 & 0.06   &   0.13 \\		
		\hline                      
	\end{tabular}
\end{table*}
$\xi^+_{\mathrm{PSF,gal}}$ and ${\xi^+_{\mathrm{PSF},\mathrm{PSF}}}$ are not proportional in all cases. This is due to varying patterns of the PSF in the fields. While the increase of $\xi^+_{\mathrm{PSF,gal}}$ is proportional to the amount of PSF leakage into the galaxy shapes, ${\xi^+_{\mathrm{PSF},\mathrm{PSF}}}$ is simply a measure for the spatial variation of the PSF pattern. The PSF patterns of  PSZ1~G109.88+27.94 and PSZ1~G186.98+38.66  are very homogenous in large parts of the images (cf.~Fig~\ref{fig:whisker}), which means that $\xi^+_{\mathrm{PSF},\mathrm{PSF}}(\bar{\boldsymbol\theta})$ is large on small and intermediate scales and then decreases faster for PSZ1~G186.98+38.66 than for PSZ1~G109.88+27.94. The whisker plot of PSZ1~G139.61+24.20, however, shows that the PSF ellipticities are not correlated with each other to the same degree and not at all on the largest scales. More complex PSF patterns are more difficult to correct, and consequently the amount of PSF leakage increases with decreasing $\xi^+_{\mathrm{PSF},\mathrm{PSF}}(\bar{\boldsymbol\theta})$.  

Using the classical approach to estimate $\alpha_{1,2}$, the components of the galaxy shapes $e_{1,2}$ could already be corrected directly but the errors are usually $\sim$~2-4 times larger than the absolute values of $\alpha_{1,2}$.
With our new model for the leakage factor, we get $\tilde{\alpha} \approx (15 \pm 1 , 3 \pm 2 , 3 \pm 1)\times10^{-2}$ for PSZ1~G109.88+27.94, PSZ1~G139.61+24.20 and PSZ1~G186.98+38.66 respectively.
Confirming our prediction from Section~\ref{sec:classicleak}, we find that the tangential leakage factor is largest for PSZ1~G109.88+27.94. It is more than four times larger than the
estimated $\tilde\alpha$ in the field of PSZ1~G139.61+24.20. This strong field dependency of $\tilde\alpha$ could be explained by the choice of our model, i.e. on the underlying assumption ${\boldsymbol{b}=0}$.

We cannot disentangle any contribution from a possible PSF~model~bias from $\tilde\alpha$, which could affect our estimates for the tangential leakage factor.
This effect varies locally in each field, since the PSF patterns are not the same and the quality of our PSF~models might be very different. 
As mentioned in Section~\ref{sec:modelbias}, the \emph{Rowe} statistics are very sensitive to the amount of PSF~model~bias in our stacks. While the PSF~model bias is negligibly small for PSZ1~G186.98+38.66, it might become important for PSZ1~G109.88+27.94 and PSZ1~G139.61+24.20.
Indeed, the residuals ${\tilde\alpha-\alpha_1}$, ${\tilde\alpha-\alpha_2}$, ${\alpha_1-\alpha_2}$ are smallest for PSZ1~G186.98+38.66.
Our estimates of $\tilde\alpha$  for PSZ1~G109.88+27.94 and PSZ1~G139.61+24.20, however, might be biased low or high, since we have not corrected for the effect of $c^\mathrm{mPSF}$.

\begin{figure}
	\includegraphics[width=\columnwidth]{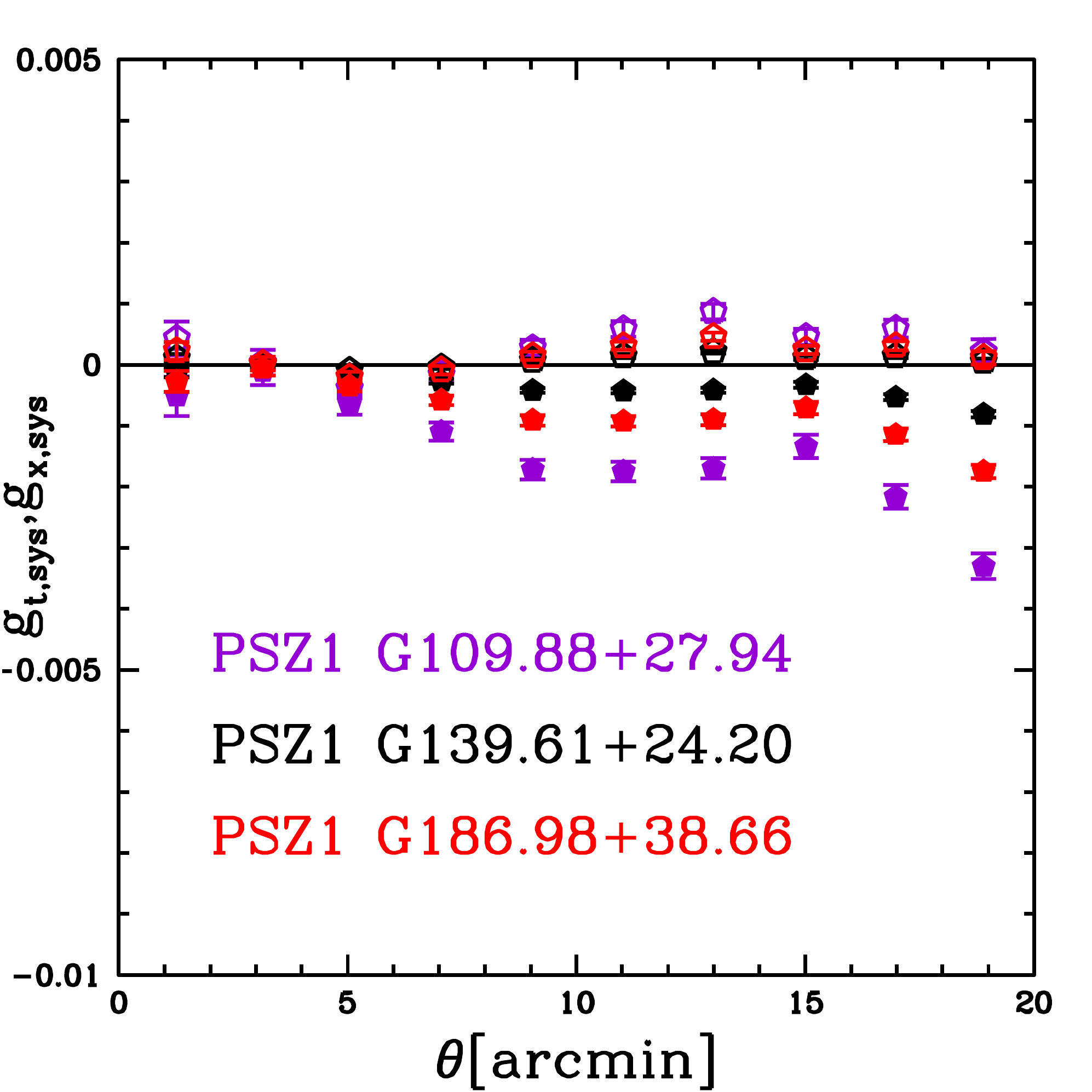} 
    \caption{Our model for the systematic tangential (filled symbols) and cross (open symbols) shear in the cluster fields as a function of distance from the cluster centre. A negative systematic shear means that the observed reduced shear in the cluster fields will be biased low. We use our calculated values of $\langle e_\mathrm{t}^\mathrm{sys} \rangle$ (and $\langle e_\mathrm{x}^\mathrm{sys} \rangle$) to apply a leakage correction to the measured weak lensing shear signal.}
    \label{fig:etsys}
\end{figure}

We can now use our estimates of $\tilde{\alpha}$ to measure the systematic tangential shear in radial bins of $\boldsymbol\theta$
\begin{equation}
\left\langle e_\mathrm{t}^\mathrm{sys} (\boldsymbol\theta)\right\rangle = \tilde{\alpha} \left\langle e_\mathrm{t}^\mathrm{PSF} (\boldsymbol\theta) \right\rangle \ ,
  \label{etsys}
  \end{equation}
where $\boldsymbol\theta$ is the distance of the galaxy from the cluster centre (cf.~Fig.~\ref{fig:etsys}).
We find that $\left\langle e_\mathrm{t}^\mathrm{sys}\right\rangle$ is negative and that the effect is larger in the outskirts of the stacks. While the overall shape of the systematic tangential shear seems to be the same in all cluster fields, PSZ1~G109.88+27.94 possesses the largest additive shear bias with \ ${e_\mathrm{t}^\mathrm{sys} = -3.6\pm0.2 \times 10^{-3}}$ at a distance of 18.9~arcmin from the centre of the cluster. This shows that the tangential alignment of the PSF shapes is the same in all stacks. 

We use 10 radial bins in the distance range of ${0'<\boldsymbol\theta<20'}$ to measure the systematic tangential shear and correct for the effect of leakage on the galaxy shapes by subtracting $\left\langle e_\mathrm{t}^\mathrm{sys}\right\rangle$ from $\left\langle e_\mathrm{t}^\mathrm{obs}\right\rangle$.
Accordingly, we can define the systematic cross shear for all cluster fields and correct the measured B-modes.
We also apply a leakage correction to the data in our weak lensing analysis. 

There is no impact of a model error for $\tilde\alpha = \mathrm{const.}$ on our estimates of the position dependent ${e_\mathrm{t}^\mathrm{sys}}$. If we assume a false systematic tangential shear profile, we bias our NFW cluster mass estimate of PSZ1~G109.88+27.94 by 2~per~cent and those of PSZ1~G139.61+24.20 and PSZ1~G186.98+38.66 by 1~per~cent. 
Compared to the mean ellipticity bias and in some fields compared to the PSF~model~bias discussed in Sections~\ref{sec:meane}-\ref{sec:modelbias}, this \emph{PSF leakage calibration bias} $\sigma_\mathrm{e_{t}^{sys}}$ is subdominant.

\subsubsection{Star-galaxy cross-correlation}\label{sec:stargal}

\begin{figure}
   	{\includegraphics[width=\columnwidth]{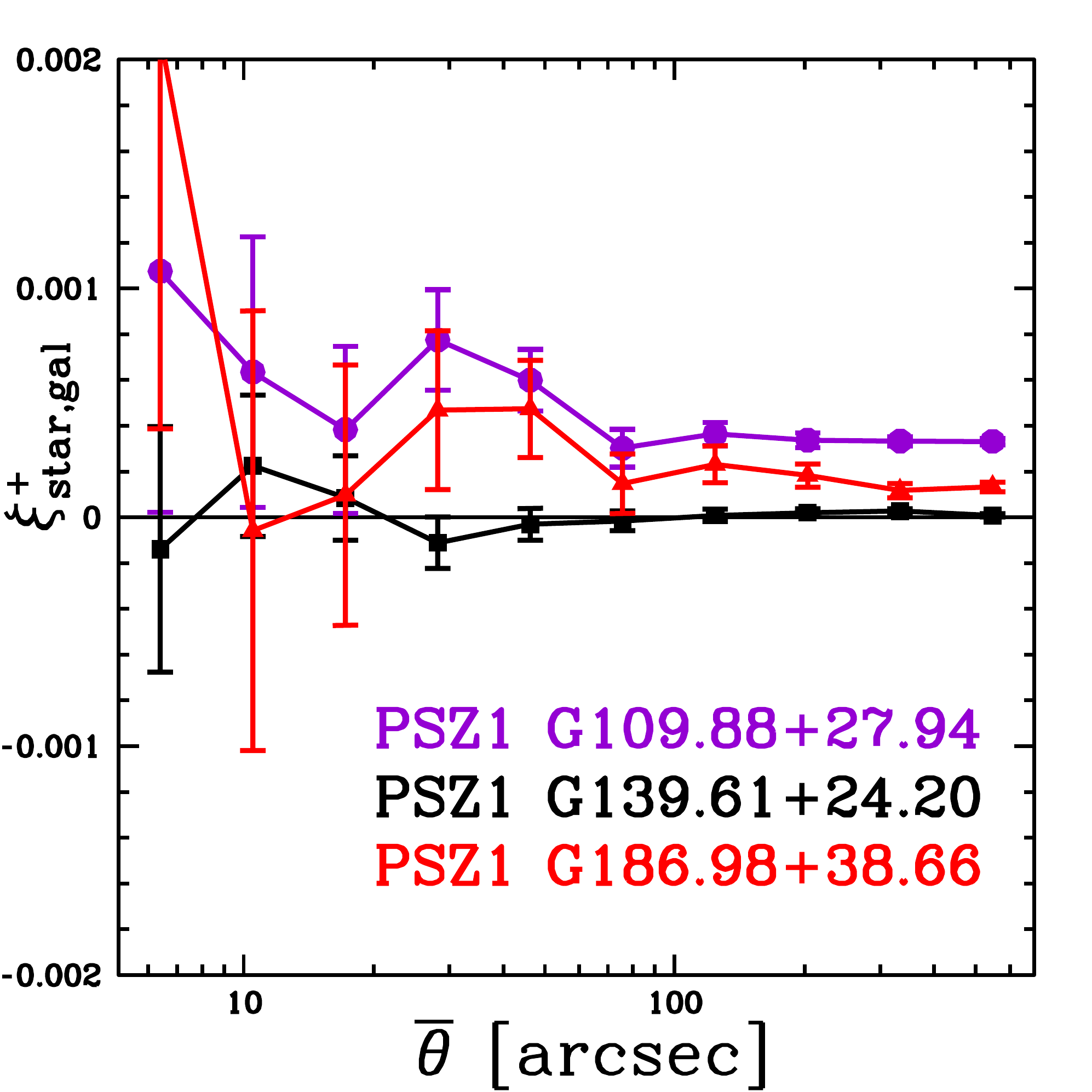}}
\caption{The cross-correlation function of the ellipticities of field stars and galaxies as a function of angular 
separation for PSZ1~G109.88+27.94 (violet circles), PSZ1~G139.61+24.20 (black squares) 
and PSZ1~G186.98+38.66 (red triangles). The star-galaxy cross-correlation shows the same behavior as $\xi^+_{\mathrm{PSF,gal}}$ (cf.~Fig.~\ref{fig:xipg}).}
       \label{fig:stargal}
\end{figure}

Finally, we consider the cross-correlation functions of the stellar and the galaxy ellipticities 
\begin{equation}
{\xi^\pm_{\star,\mathrm{gal}}} = \left\langle 
\boldsymbol{e}^\star
*
 \boldsymbol\epsilon^\mathrm{gal}
\right\rangle
 \pm 
 \left\langle
 \boldsymbol\epsilon^\mathrm{gal}
*
\boldsymbol{e}^\star
\right\rangle.
\label{FCORRELATION}
\end{equation} 
An obviously different behaviour of $\xi^\pm_{\star,\mathrm{gal}}$ compared to $\xi^\pm_{\mathrm{PSF,gal}}$ would indicate PSF model bias. Consequently, we use the star-galaxy~cross-correlation function as a consistency check to test whether our leakage model in Section~\ref{sec:leakage} sufficiently describes the data.
We calculate ${\xi^\pm_{\star,\mathrm{gal}}(\boldsymbol\theta)}$ 
and present the results in Fig.~\ref{fig:stargal}. 
As expected, the star-galaxy cross-correlation function $\xi^+_{\star,\mathrm{gal}}$ is consistent with $\xi^+_{\mathrm{PSF,gal}}$ (Fig.~\ref{fig:xipg}) but with larger errors.  
This is also the case for $\xi^-_{\star,\mathrm{gal}}$ and $\xi^-_\mathrm{PSF,gal}$, with $\xi^-_\mathrm{PSF,gal}$.
On the smallest scales with ${\bar{\theta} \lesssim 30\ \mathrm{arcsec}}$, we cannot measure the star-galaxy cross-correlation well 
and the errorbars in Fig.~\ref{fig:stargal} are accordingly large. 
$\xi^+_{\star,\mathrm{gal}}$ and $\xi^+_{\mathrm{PSF,gal}}$ show that stellar and galaxy ellipticities and also the ellipticities of the modeled PSF and the measured galaxies ellipticities are not, or only minimally correlated on any scale for  
PSZ1~G139.61+24.20. The other two clusters of our sample, however, show a small correlation between galaxy shapes and the shapes of stars (and PSF model). As discussed in the previous section, this is due to systematics in the PSF modeling and can be accounted for by applying a leakage correction if the leakage factor $\tilde\alpha$ is known. 

\section{Background sample selection}\label{sec:four:photoz}

\begin{figure}
	\includegraphics[width=\columnwidth]{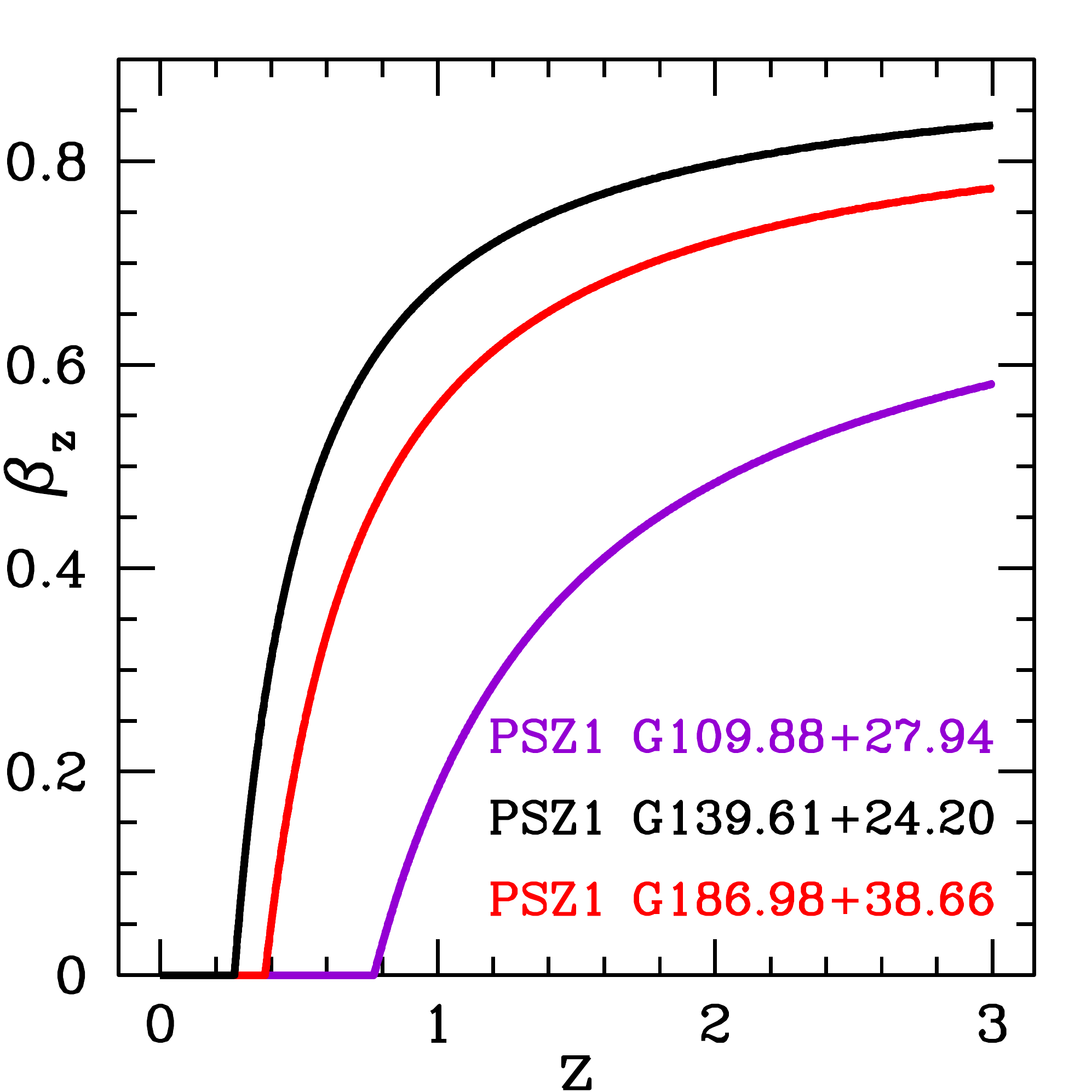} 
    \caption{The lensing strength $\beta_\mathrm{z}(z) = D_\mathrm{ds}/D_\mathrm{s}$ for the WWL pathfinder sample. 
   $\beta_\mathrm{z}$ is zero for $z \leq z_\mathrm{d}$ and reaches an asymptotic value at $\beta_\mathrm{z}<1$.
   The cluster redshifts are $z_\mathrm{d}=0.267,0.3774,0.77$ for PSZ1~G139.61+24.20, PSZ1~G186.98+38.66 and PSG1~G109.88+27.94, respectively.}
    \label{fig:beta}
\end{figure}

This section describes our method to select a sample of lensed background galaxies for weak lensing analysis from our three-band WWFI photometry. 
The lensing signal scales as 
\begin{equation}
{\beta_\mathrm{z}}(z_\mathrm{d},z_\mathrm{s})\ D_\mathrm{d}(z_\mathrm{d}) 
\mathrel{\mathop:}=  D_\mathrm{d}
\begin{cases}
    \frac{D_\mathrm{ds}}{D_\mathrm{s}} \ & z_\mathrm{s} \leq z_\mathrm{d}\ ,\\
    0 & \mathrm{otherwise}\ .
\end{cases}
\label{eq:beta}\end{equation} 
We know the spectroscopic cluster redshift $z_\mathrm{d}$, so the lensing strength ${\beta_\mathrm{z}}$ 
is simply a function of source redshift $z_\mathrm{s}$. For foreground galaxies and cluster members, $\beta_\mathrm{z}$ is zero. It rises steeply with increasing source redshift until it reaches an asymptotic value, ${\beta_\mathrm{z}} < 1$, as $z_\mathrm{s}\to\infty$ (cf.~Fig.~\ref{fig:beta}). 

In weak lensing surveys, photometric redshifts are often calculated using template fitting 
algorithms \citep[e.g.][]{brimioulle2008photometric} or machine learning approaches \citep[e.g.][]{rau2015accurate}. 
However, photometric redshifts cannot be well constrained using either method, if only three-band photometric information is available.  
Due to non-linear error propagation from equation~(\ref{eq:beta}), imprecise redshift point estimates bias ${\beta_\mathrm{z}}$ estimates and 
make a clean background sample selection impossible. 
We use the probabilistic method of \citet{gruen2014weak} to determine source redshift probability distributions 
for the galaxies, despite the limited photometric information. 
The position of a galaxy in magnitude space can be compared to that of reference galaxies with accurate redshift information.   
Using this approach, we consider the full colour-magnitude distribution of the reference galaxies empirically and gain information on $p(z)$ of each source galaxy (cf.~Section~\ref{sec:3band}). 

\subsection{Photometric redshifts for EGS}\label{sec:photoz}

\begin{figure}
  \includegraphics[width=2.9in]{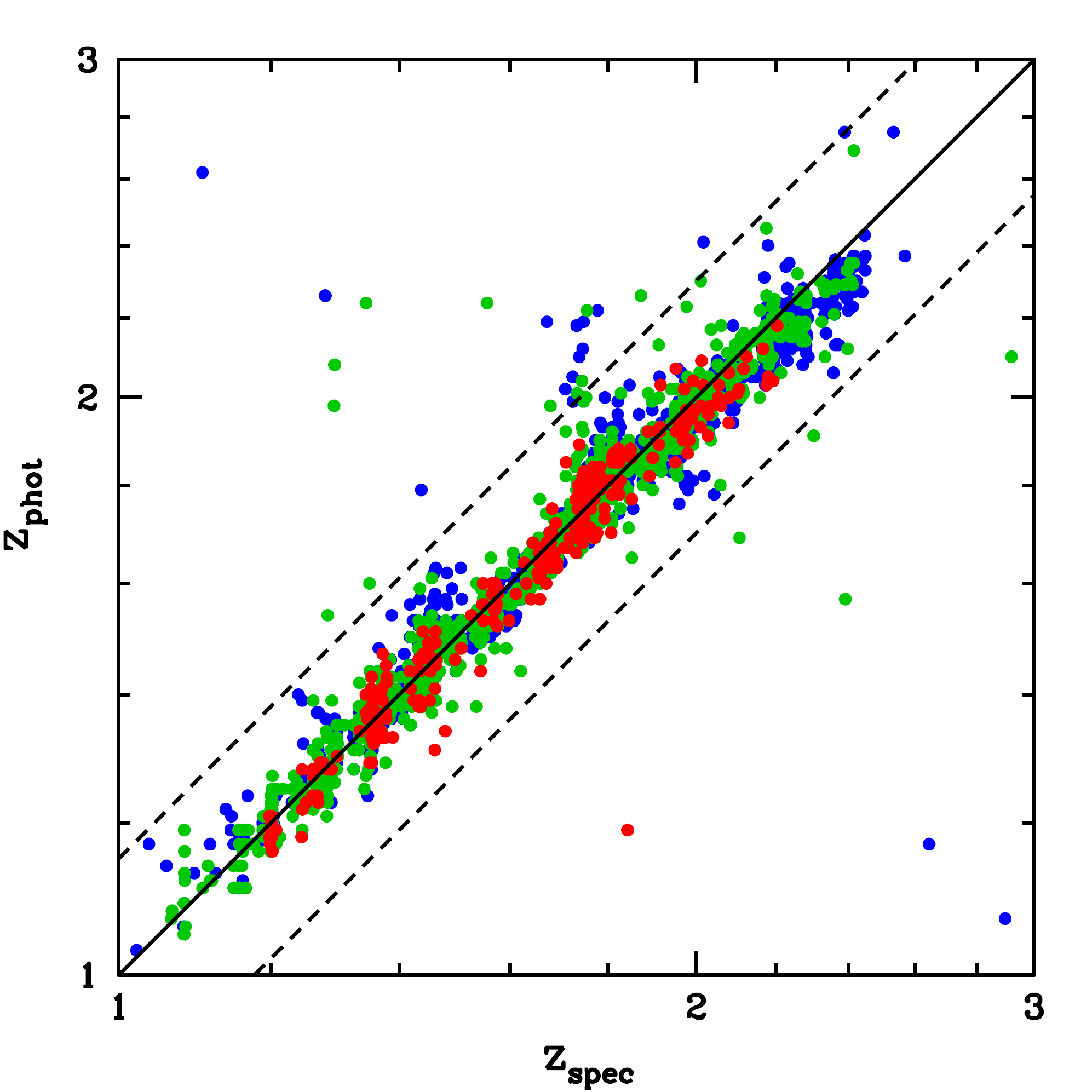}
  \caption[]{Comparison of photometric and spectroscopic redshifts for W-EGS. The color code highlights galaxies of different spectral types. Early type galaxies are colored in red, star-forming galaxies in blue and spirals in green.}
\label{fig:zzplot}\end{figure}

As a first step, we obtain precise photometric redshifts for the galaxies in CFHTLS-D3.
As mentioned in Section~\ref{sec:EGS}, we have eight-band $u^*,g',r',i',z',J,H,K$ 
photometric data at our disposal \citep{davis2007all,bielby2012wircam}.
We detect sources on the deepest ($r$ band) W-EGS co-added image, extract fluxes, magnitudes and errors from the CFHTLS-D3 pointing and create a photometric catalogue as described in Section~\ref{sec:photometry}.
We use the photometric template-fitting \texttt{Photo-{Z}} code of 
\cite{bender2001fors} and follow the approach of \cite{brimioulle2013dark}.
The overlap of the field with the Deep Extragalactic Evolutionary Probe-2 (DEEP2) survey \citep{newman2013deep2} allows for a spectroscopic calibration and validation of our achieved photometric redshift estimates. 
The spectroscopic sample has a limiting apparent magnitude of $R_\mathrm{AB} = 24.1$ and contains $\sim2\%$ of all galaxies with photometric information up to  $z_\mathrm{spec} \approx 2$. 
We obtain a photometric redshift accuracy of 
${\sigma_{\Delta z/(1+z_\mathrm{spec})} = 0.026}$ and 
an outlier rate of
${\eta = 1.5}$~per~cent (cf.~Fig.~\ref{fig:zzplot}).  

We match our CFHTLS-D3 photometric redshifts to the sources detected on our stacked W-EGS images. We obtain 
a catalogue, which contains photometric information using WWFI $g,r,i$ filter bands and reliable 
CFHTLS $u^*,g',r',i',z',J,H,K$ photometric redshift estimates. This data set is used as a reference field and connects the redshift of a galaxy to its WWFI magnitudes/colors.

\subsection{Using three band photometry to estimate galaxy redshifts}\label{sec:3band}

We now use the three-band WWFI photometry of our cluster fields and our redshift catalogue for W-EGS to estimate the $\beta_\mathrm{z}$ of each background galaxy.
Our background selection procedure is based on the fact that galaxies with similar magnitudes
belong to the
same distribution of morphological classes and redshifts.
Here, we only give a brief overview, as the method is described in detail in \citet{gruen2014weak}. 

Our aim is to get an estimate on the lensing strengths $\beta(z)$ of each galaxy in the cluster fields.
For each cluster field, we have a catalogue of $i$ galaxies with a set of apparent magnitudes 
${\bf{m}}^\mathrm{cl}_i = \{ m^\mathrm{cl}_{i,j} \}$, where ${j = g,r,i}$. 
For W-EGS, we have a similar catalogue with ${{\bf{m}}^\mathrm{egs}_{k} = \{ m^\mathrm{egs}_{k,j} \}}$, (${k >} i$).
We apply a cut on the flux radius of the reference galaxies that corresponds to the cut we make on the galaxies in our shape catalogues.    
Galaxies are considered to be comparable, if 
${\lvert \Delta{\bf{m}}\rvert~=~\lvert {\bf{m}}^\mathrm{cl}_i~-~{\bf{m}}^\mathrm{egs}_k \rvert~\leq~0.1.}$ 
In this way, a reference sample of galaxies $k'$ can be assigned to each galaxy in the cluster fields.
The redshift probability distribution $p(z \mid \bf{m}^\mathrm{cl})$ is then given by the distribution of the reference redshifts $p(z \mid \bf{m}^\mathrm{egs})$.
Thanks to the overlap of W-EGS and CFHTLS~D3, we have been able to assign a reliable photometric redshift estimate to each source in this reference field (Section~\ref{sec:photoz}).
The lensing strength for a given galaxy $i$ can then be estimated to be
\begin{equation}{\beta_\mathrm{z}}({\bf{m}}^\mathrm{cl}_i) = \langle{\beta_\mathrm{z}}({\bf{m}}^\mathrm{egs}_{k'})\rangle_{k'} .\end{equation} 

In order for the source catalogues not to be contaminated by cluster member galaxies, we apply a cluster member correction \citep[cf.][their Section~3.1.3]{gruen2014weak}. Naturally, the excess of cluster members is a function of position, with the maximum found in the cluster center. 
If we were to neglect this effect, the estimated ${\beta_\mathrm{z}}$ would be affected in such a way that ${\beta_\mathrm{z}}$ near the 
cluster center would be overestimated the strongest.

\begin{figure}
	\centering
	\includegraphics[width = 2.9in]{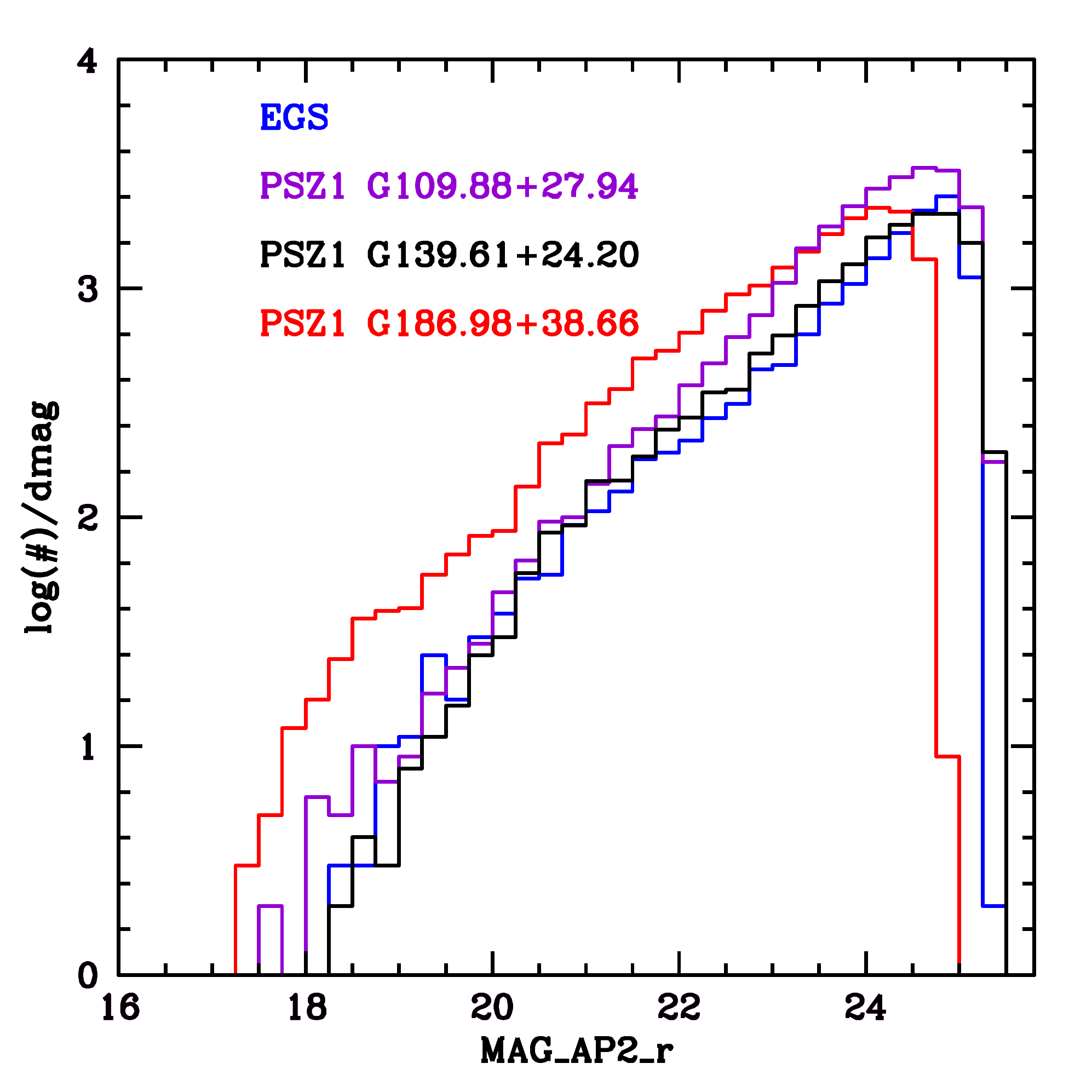}
	
    \caption{Distribution of $1''$ aperture $r$-band magnitudes of galaxies with  $\mathrm{S/N}>10$. The blue histogram shows the distribution of apparent brightnesses of W-EGS $r$-band stacked images. The limiting $r$-band magnitude is largest for W-EGS (blue), PSZ1~G109.88+27.94 (purple) and PSZ1~G139.61+24.20 (black) and is smallest for PSZ1~G186.98+38.66 (red).} 
    \label{fig:depth}
\end{figure}

W-EGS has to contain a sufficient amount of galaxies with similar photometric properties as those of objects in the cluster fields. 
Fig.~\ref{fig:depth} shows the distribution of the $1''$ aperture limiting 
$r$-band magnitudes for each field, for which $\mathrm{S/N}>10$. PSZ1~G186.98+38.66 is the shallowest cluster field in the lensing band with an $r$-band limiting magnitude of $\sim25$~mag. The $r$-band stack of W-EGS is as deep as the other cluster fields. With $\sim25.5$~mag in $g$ and $\sim24.8$~mag in $i$, it is deeper than the cluster fields by $\leq0.5$~mag.


Table~\ref{tab:galaxies} provides basic properties of the background sample with successful shape measurement. Compared to \citet{gruen2014weak}, we have higher number densities of background galaxies at comparable redshifts, which is due to the fact that our stacks are slightly deeper than theirs. Furthermore, they require a higher minimum lensing strength $\beta_\mathrm{min}$ for their source galaxies. We have found that a cut on the lensing strength of $\beta_\mathrm{min}=0.05$ is conservative enough to remove the remaining contamination by cluster members from the source catalogues but is still low enough that not too many background galaxies are lost for the analysis. 
\begin{table*}
	\centering
	\caption{Basic properties of the background samples. Colums from left to right: Object name, threshold $\beta_\mathrm{min}$, corresponding threshold redshift $z(\beta_\mathrm{min})$, mean lensing strength $\langle\beta\rangle$, the corresponding redshift $z(\langle\beta\rangle)$, total number and number densitiy of background galaxies (i.e. with $\langle\beta\rangle {>\beta_\mathrm{min}}$) and at a distance $r > 500$~\~kpc from the cluster centre, and the fraction of masked area in the images.}
	\label{tab:galaxies}
	\begin{tabular}{lccccccr} 
		\hline
		Object &$\beta_\mathrm{min}$&$z(\beta_\mathrm{min})$& $\langle\beta\rangle$&$z(\langle\beta\rangle)$& $N^\mathrm{KSB}_\mathrm{bg}$ & $n^\mathrm{KSB}_\mathrm{bg}$ [arcmin$^{-2}$]&$A_\mathrm{masked}/A$\\
		\hline
				\hline
		PSZ1~G109.88+27.94 &	0.05	&	0.82	&	0.08	&	0.85		&	6992	&	9 &	0.114 \\
		PSZ1~G139.61+24.20 &	0.05	&	0.28	&	0.41	&	0.48		&	7293	&	9  & 0.118\\
		PSZ1~G186.98+38.66 &	0.05	&	0.40	&	0.27	&	0.54		&	9739	&	12   & 0.112 \\
		\hline
                      
	\end{tabular}
\end{table*}

\subsection{Photometric redshift uncertainty}\label{sec:cosmicvar}

Due to the proportionality of the lensing signal and ${\beta_\mathrm{z}}$, biases in our catalogue of lensing strengths have a direct impact on the estimated cluster masses. In the case that the lensing strength is over-/under- estimated for a galaxy, its lensing signal will be under-/over- estimated. 
Since W-EGS is comparatively small, we expect the dominant bias to arise from cosmic variance. As we only have this single reference field, we use the findings of \citet[][]{gruen2017selection} to estimate the cosmic variance in W-EGS.

\citet[][]{gruen2017selection} use an approach that is quite similar to our own method. Instead of collecting all galaxies with a comparable set of magnitudes, they build a colour-magnitude decision tree based on a set of reference fields and using different sets of photometric bands. They split the multi-band colour-magnitude space into $l$ boxes $B_l$ in such a way that each source $m$ in the lensing sample falls in one such colour-magnitude boxes. Its p(z) is then described by the observed distribution of photometric redshifts of the reference galaxies in the box. The lensing weight is then given by the mean of all reference galaxy redshifts in $B_l$, i.e. ${\beta_\mathrm{z}}_m = \langle{\beta_\mathrm{z}}\rangle_{B_l}$. 
As \citet[][]{gruen2017selection} use the full set of the CFHT~Deep~fields\footnote[2]{\url{http://www.cfht.hawaii.edu/Science/CFHLS/ cfhtlsdeepwidefields.html}}, their catalogue of reference galaxies is much larger than ours and contains galaxies from four different pointings.
\citet[][]{gruen2017selection} test the impact of different types of biases on the lensing strengths and find cosmic variance to be the dominant bias in data sets comparable to our own.  
They calculate the cosmic variance $\sigma_{\beta/B_i}$ in their catalogue as the mean of the lensing strengths in each colour-magnitude box over their four reference fields using a jackknife approach. 
They find that this bias depends strongly on the set of filters and decreases with increasing number of bands.
For a combination of $gri$ band data, they estimate the cosmic variance at redshifts $z=0.267,0.3774,0.77$ to be equal to $\sigma \approx 0.006,0.009,0.023$ \citep[cf.][their~Fig.~B1]{gruen2017selection}.
As the cosmic variance scales as $\sqrt{4/N_\mathrm{fields}}$, where $N_\mathrm{fields}$ is the number of different pointings, the bias on our lensing strengths due to cosmic variance should be less than $2.4\%,3.6\%$ and $9.2\%$ for PSZ1~G139.61+24.20, PSZ1~G186.98+38.66 and PSZ1~G109.88+27.94, respectively.

\section{Weak lensing analysis}\label{sec:five:wlanalysis}

The surface mass density of a foreground object acting as a gravitational lens determines the 
distortion of the background images. The mean tangential component of the shear on a circle $\left\langle\gamma_\mathrm{t}(\boldsymbol\theta)\right\rangle$ is given by the convergence $\kappa$ inside the circle subtracted by the convergence at its edge,
\begin{equation}\left\langle\gamma_\mathrm{t}(\boldsymbol\theta)\right\rangle = \left\langle{\kappa(<\boldsymbol\theta)}\right\rangle - \left\langle\kappa (\boldsymbol\theta)\right\rangle.
\label{eq:gammat}\end{equation}
The reduced shear is given by $g=\gamma/(1-\kappa)$.
Equation~(\ref{eq:gammat}) implies that the measurement of the tangential shear averaged over all galaxies in a circular annulus 
around the cluster centre can be used to directly estimate the azimuthally averaged mass profile 
of a cluster of galaxies. Yet,  in order to retain the full information on the surface mass density, 
we need more information about the system than just the reduced shear alone. 
If only the gravitational shear is measured, one cannot distinguish between models for $\kappa$ that are similar except for an additional sheet of mass.
A possible way to break this \textit{mass sheet degeneracy} is to assume a functional form of $\kappa(\boldsymbol\theta)$.
In order to constrain weak lensing cluster masses, we consider two model profiles: the Singular Isothermal Sphere (SIS) and the NFW profile.

\subsection{Fit of a Singular Isothermal Sphere}

The simplest model to describe the density profile of a cluster of galaxies is the Singular Isothermal Sphere  \citep[e.g.][]{binney1998galactic}. In this model, the galaxy cluster is thought of as a spherically symmetric, self-gravitating ideal gas cloud consisting of collisionless particles.      
The surface mass density at a distance $r$ from the cluster centre with a constant velocity dispersion $\sigma_\mathrm{v}$ is given by
\begin{equation} 
\Sigma(r) =  \frac{\sigma_\mathrm{v}^2}{2\pi G} \int \limits_{-\infty}^{\infty} \frac{\mathrm{{d}z}}{r^2 + z^2} = \frac{\sigma_\mathrm{v}^2}{2G r},
\end{equation}
where $G$ is the gravitational constant.
Using the definition of the convergence $\kappa$, it can be shown that
\label{sigma}
\begin{equation} 
\kappa(r) = \frac{\sigma_\mathrm{v}^2}{2G\Sigma_\mathrm{crit}}\frac{1}{r} = {2\pi}{D_\mathrm{L}\beta_\mathrm{z}}\left(\frac{\sigma_\mathrm{v}}{c}\right)^2 \frac{1}{r} = \gamma(r).
\label{gammasis}
\end{equation}
We average the reduced tangential shear in eight radial distance bins from $\theta = 2-20\ \mathrm{arcmin}$ and fit equation~(\ref{gammasis}) to the signal. We use the average lensing strength of the sample $\left\langle\beta_\mathrm{z}\right\rangle$ in this one-parameter fit and constrain the velocity dispersion of the WWL galaxy clusters. 
We exclude galaxies, for which $\theta = r/D_\mathrm{L} \leq 2\ \mathrm{arcmin}$ and apply a cut on the distance fraction with $\beta_\mathrm{z, min} = 0.05$. 
The mass of an SIS out to a radius $r$ can then be calculated as 
\begin{equation} 
M_\mathrm{SIS}(r) = \frac{2\sigma_\mathrm{v}^2}{G}r.
\label{Msis}
\end{equation} 
\subsection{Significance map}\label{sec:significance}

We estimate the significance of the detected tangential alignment of background galaxies by using the aperture mass statistic \citep{schneider1996detection}.
The aperture mass significance allows us to visualize the two-dimensional weak lensing signal and helps to identify neighbouring mass distributions acting as lenses themselves.
We calculate the aperture mass as the weighted sum over all background galaxies in a circular aperture
\begin{equation} M_\mathrm{ap}\left(\lvert\boldsymbol\theta\rvert\right) = \sum_iw\left(\lvert\boldsymbol\theta -\boldsymbol\theta_i\rvert\right) 
g _\mathrm{i,\mathrm{t}}.
\label{Map}
\end{equation} 
Here, $g_\mathrm{i,\mathrm{t}}$ denotes the reduced tangential shear of a galaxy $i$ at a position $\boldsymbol\theta$ with respect to the centre of mass. 
The uncertainty of this aperture mass is then given by 
\begin{equation} \sigma_\mathrm{m_\mathrm{ap}}\left(\lvert\boldsymbol\theta\rvert\right) = 
\sqrt{\frac{1}{2}\sum_iw^2\left(\lvert\boldsymbol\theta -\boldsymbol\theta_i\rvert\right)\lvert g _\mathrm{i,\mathrm{t}}\rvert^2}. 
\label{sigmaMap}
\end{equation} 
In our analysis, we use a Gaussian weight function 
\begin{equation} w\left(\lvert\boldsymbol\theta\rvert\right)\propto\left\{%
\begin{array}{ll}
\exp\left[-\lvert\boldsymbol\theta\rvert^2/(2\sigma_\mathrm{w}^2)\right] & \lvert\boldsymbol\theta\rvert < 3\sigma_\mathrm{w},\\
0 & \textrm{otherwise}
\end{array}%
\right.
\label{WMap}
 \end{equation} 
and choose $\sigma_\mathrm{w} = 3\ \mathrm{arcmin}$ as the width of the aperture.
The ratio of the aperture mass and its uncertainty gives the significance $M_\mathrm{ap}/\sigma_\mathrm{m_\mathrm{ap}}$, 
which is positive, where a tangential shear signal of an overdensity has been detected. It is strongest at the centre of a cluster of galaxies and 
grows weaker with the distance to the core of the dark matter halo.

In order to check for systematics, we define the cross aperture in analogy to equation~(\ref{Map}) by replacing the tangential shear with 
the cross component $g_\mathrm{x}$.

\subsection{NFW model}\label{sec:nfw}

We further use the two-parametric density profile of a dark matter halo 
\citep{navarro1996structure, navarro1997universal} to perform a likelihood analysis.
Numerical simulations \citep{navarro1995simulations} have shown that the 3-dimensional density of dark matter as a function of the radius $r$ is best described by  
\begin{equation}\rho_\mathrm{NFW}(r)=\frac{\rho_0}{{r}/{r_\mathrm{s}}(1+{r}/{r_\mathrm{s}})^2},\label{NFWdensity}\end{equation}
where the scale radius $r_\mathrm{s}$ and $\rho_0$ are parameters describing the density profile of the individual dark matter halo. 
The dimensionless concentration parameter is related to the virial radius and the scale radius via $c=r_{vir}/{r_\mathrm{s}}$. 
We use the virial radius $r_\mathrm{200m}$ at which the enclosed average density 
$\bar{\rho}$ reaches a value that is 200 times that of the mean matter density for our fit. In order to compare our results to
literature, we also express this quantity in terms of the critical density $r_\mathrm{200c}$ (and accordingly $r_\mathrm{500c}$).

The virial mass $M_\mathrm{200m}$ and concentration parameter $c_\mathrm{200m}$ can be used to predict the shear signal
$\gamma^\mathrm{NFW}_\infty(\boldsymbol\theta)$ and convergence $\kappa^\mathrm{NFW}_\infty(\boldsymbol\theta)$ a dark matter halo would cause for a 
galaxy at infinite redshift 
at a given position $\boldsymbol\theta$ \citep[cf.][]{bartelmann1996arcs, wright2000gravitational}. 
We use our distance fraction estimates from Section~\ref{sec:four:photoz} to compute the theoretical value of the reduced shear as described in \cite{seitz1997steps}
\begin{equation}
\left\langle g^\mathrm{NFW}(\boldsymbol\theta)\right\rangle = \frac{\left\langle\beta_\mathrm{z}\right\rangle\gamma^\mathrm{NFW}_\infty}
{1 - \frac{\left\langle\beta_\mathrm{z}^2\right\rangle}{\left\langle\beta_\mathrm{z}\right\rangle}\kappa^\mathrm{NFW}_\infty}.
\label{eq:gfit}
\end{equation}
Assuming that the convergence is negligibly small, this relation is reduced to $\left\langle g^\mathrm{NFW}(\boldsymbol\theta)\right\rangle = {\left\langle\beta_\mathrm{z}\right\rangle\gamma^\mathrm{NFW}_\infty}$, which gives consistent results.
Equation~(\ref{eq:gfit})
includes a correction for the linear response of the reduced shear to the dispersion of $\beta_\mathrm{z}$. 
We fit the theoretical reduced shear of the observed galaxies to their tangential shear profiles 
(cf. equation~\ref{eq:gtgx}) using the minimum $\chi^2$ method of \cite{avni1976energy}. 

The log-likelihood is proportional to $\chi^2(M,c)$,
 \begin{equation}\mathcal{L}(M,c \mid e^\mathrm{data}_{i,j}) \propto \mathrm{e}^{- \frac{\chi^2(M,c)}{2}},\label{likelihood}\end{equation} and depends on mass and concentration of the cluster of galaxies.   
The $\chi^2$ can easily be calculated as 
\begin{equation}\chi^2=\sum\limits_{i,j}\frac{(e^\mathrm{NFW}_{i,j}-e^\mathrm{data}_{i,j})^2}{\sigma_{i,j}^2+\sigma_\mathrm{int}^2}\label{chisq}\end{equation} 
for a set of given parameters $M_\mathrm{200}$ and $c_\mathrm{200}$. Here, $e^\mathrm{data}_{i,j}$ describes the $i^{th}$ component of the 
ellipticity of the $j^{th}$ galaxy in the data set. The predicted ellipticities are expressed by $e^\mathrm{NFW}_{i,j}$. 
The uncertainty $\sigma_{i,j}$ of the shear is given by the quadratic average of the ellipticity errors. 
The intrinsic ellipticity dispersion is set to $\sigma_\mathrm{int} = 0.25$.

The maximization of the likelihood $\mathcal{L}$ is equivalent to the minimization of equation~(\ref{chisq}). 
We fit $c$ and $M$ simultaneously and calculate the confidence intervals according to \cite{avni1976energy}. 

The two parameters of the fit are, in fact, correlated. The dependency of concentration at a given cluster redshift $z_\mathrm{cl}$ 
on the halo mass has been investigated using simulations.
We adopt the mass-concentration relation of \cite{duffy2008dark} 
\begin{equation}c(M,z_\mathrm{cl}) = A \left(\frac{M}{M_\mathrm{pivot}}\right)^B\left(1+z_\mathrm{cl}\right)^C,\label{eq:cduffy}\end{equation} 
where we use $M_\mathrm{pivot} = 2\cdot10^{12}\ \mathrm{h^{-1} M_\odot}$, $A=10.14$, $B=-0.081$ and $C=-1.01$, as expected for dark 
matter halos at intermediate redshifts. Uncertainties for the fitting parameters $A$, $B$ and $C$ have been provided in \cite{duffy2008dark} 
and are found to be small enough to not affect our measurement. The impact of imperfect simulations or the assumption of different cosmological 
parameters in the derivation of $c(M,z_\mathrm{cl})$ have been discussed in \cite{hoekstra2012canadian} and are found to be negligible.
   
Assuming a lognormal distribution of concentrations with $\sigma_{\log c} = 0.18$ \citep{bullock2001profiles}, we use equation~(\ref{eq:cduffy}) to define 
a concentration prior
\begin{equation} P_\mathrm{c}(c,M,z_\mathrm{cl}) = \exp\left[-\frac{\Big(\log\big(c\big) - \log\big(c(M,z_\mathrm{cl})\big)\Big)^2}{2\sigma_{\log c}^2}\right] , 
\label{eq:duffyprior}\end{equation}  
where $\log$ denotes the logarithm to base 10.
Multiplying the likelihood with the concentration term $P_\mathrm{c}(c,M,z_\mathrm{cl})$, we are able to reduce the uncertainty on $c$ (and $M$). 
We find maximum-likelihood solutions for virial mass $M_\mathrm{200m}$ and concentration parameter $c_\mathrm{200m}$ and determine projected 
and combined confidence regions in the parameter space of the model.

\section{Results}\label{sec:six:results}

In this section, we present our results of the weak lensing analysis of three 
\emph{Planck} SZ clusters of galaxies. For each object, we provide:

\begin{enumerate}
\item A color image, a description of the visual appearance of the cluster and, if available, results from previous studies.
\item The mass significance map, calculated using equation~(\ref{Map}-\ref{WMap}).
By applying a cut on ${\beta^\mathrm{min}_\mathrm{z} = 0.05}$, we try to optimize the lensing signal. 
If this threshold would have been chosen lower, the small lensing signal of objects would increase the shape noise, 
as the scatter of intrinsic ellipticities is large compared to the weak lensing shear. 
At the same time, a threshold too high would remove too many lensed sources which, too, would increase the noise. 
For each of the clusters, we show the significance of non-zero aperture masses as a function of position in the field.   
\item The tangential shear profile and the results of a fit of an SIS model.
The virial masses are computed as
    \begin{equation}
  \begin{split}
M^\mathrm{SIS}_\mathrm{200c} &= \frac{800\pi}{3} \rho_\mathrm{crit}(z) \left[\frac{\sqrt{2}\sigma^2}{10H(z_\mathrm{cl})}\right]^3,\\
M^\mathrm{SIS}_\mathrm{200m} &= \frac{800\pi}{3} \rho_{m}(z) \left[\frac{\sqrt{2}\sigma^2}{10H_0\sqrt{\Omega_{m,0}(1+z_\mathrm{cl})^3}}\right]^3,
 \end{split}
  \end{equation}
  where we have used equation~(\ref{Msis}) and the definitions of the critical and the mean matter density.

\item The results of our NFW fit.
In order to stay in the weak lensing regime and reduce the likelihood to include cluster member galaxies in the background sample, we exclude  galaxies at a distance $r <  500\ \mathrm{kpc}$ from the fitting procedure. We do not  
exclude galaxies in the outer regions of the field, as that would remove more sources from the analysis, which would result in an increase of the shot noise.

Whenever we expect line-of-sight structures to be present in the cluster fields, we perform multiple halo fits to find constraints on the masses of the groups. In order to reduce the amount of fitting parameters, we fix the concentrations of the halos using the mass-concentration relation of \citet{duffy2008dark}.

\end{enumerate}

\subsection{PSZ1~G109.88+27.94}\label{sec:109}

\begin{figure}
	\includegraphics[width=\columnwidth]{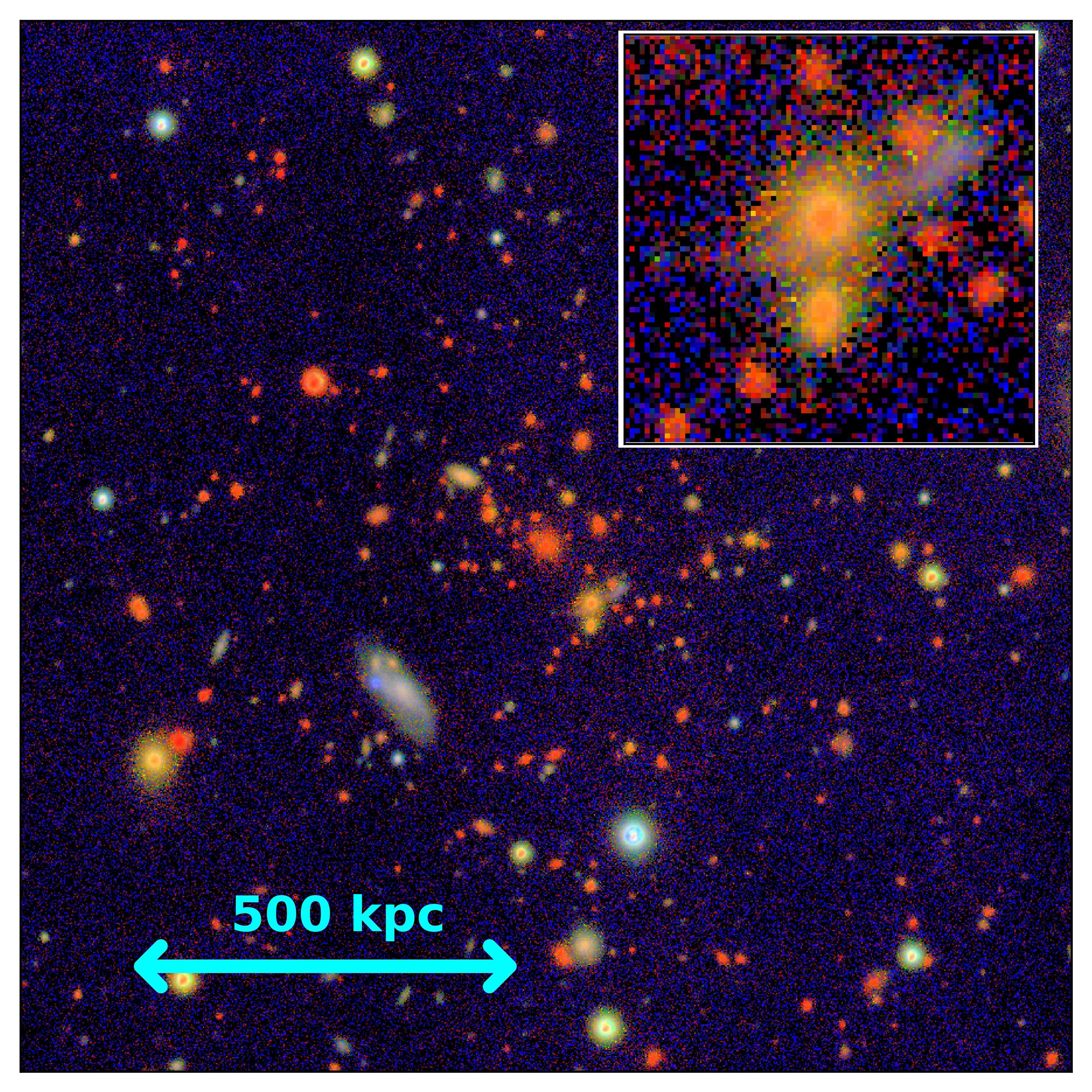}  
    \caption{RGB image using WWFI $g$, $r$ and $i$~band stacks. The dispayed $3\times 3$~arcmin$^2$
region is centred on the brightest cluster galaxy of PSZ1~G109.88+27.94. The cyan line indicates a distance of 500~kpc at the cluster redshift. This corresponds to the physical distance of the radius, that has been excluded from the NFW analysis described in the subsection below. 
The BCG has a photometric redshift of $z=0.77$. Two blue arc candidates can be seen towards the south-west of the BCG. A close-up of the strong lensing feature candidates near the two yellow foreground galaxies is shown in the upper right corner of the image.
North-east is towards the upper left in the image.}
    \label{fig:psz109visual}
\end{figure}

PSZ1~G109.88+27.94 is an SZ selected cluster candidate that has been discovered by the \emph{Planck} satellite with a relatively low S/N 
of $5.3$ ($5.8$) \citep{ade2014a,ade2015a}. 
A group of elliptical galaxies is clearly visible in the $gri$-image we show in Fig.~\ref{fig:psz109visual}. 
While the SZ signal is centred on the brightest cluster galaxy, the X-ray centroid is shifted to the south-west relative of the cluster
\footnote[3]{For more details on the clusters, SZ and X-ray footprints, see \url{http://szcluster-db.ias.u-psud.fr/sitools/client-user/SZCLUSTER_DATABASE/project-index.html}.}.
We cannot verify the spectroscopic redshift of $z=0.4$ given in the \emph{Planck} catalogues \citep{ade2014b}. 
Thanks to the coverage of the field by SDSS DR-14 data, we use their photometric redshift estimate of the BCG (Brightest Cluster Galaxy) for our analysis. A comparison of the $r-i$ colours of our WWFI observations to those of the reference galaxies in W-EGS support the assumption that the cluster is at a redshift of $z_\mathrm{phot} = 0.77\pm0.04$.

Our deep colour image reveals a set of two arc candidates towards the south of the BCG. The arcs 
seem to lie on opposite sides of the critical line and are distorted by two nearby yellow galaxies in the foreground of the cluster.
We use their projected distance from the cluster centre to get a rough estimate of the Einstein radius. ${\theta_\mathrm{E}\approx14''}$ from the BCG, which corresponds to a physical distance of  ${\sim100\ \mathrm{kpc}}$ at the cluster redshift. 

\citet{wen2015calibration} have identified two groups in the field. Fg1 (Foreground group 1) has a photometric redshift of $z_\mathrm{phot} =0.22$, Fg2 (Foreground group 2) is at $z_\mathrm{phot} =0.29$, 

\subsubsection{Significance map}				

\begin{figure}
	\includegraphics[width=\columnwidth]{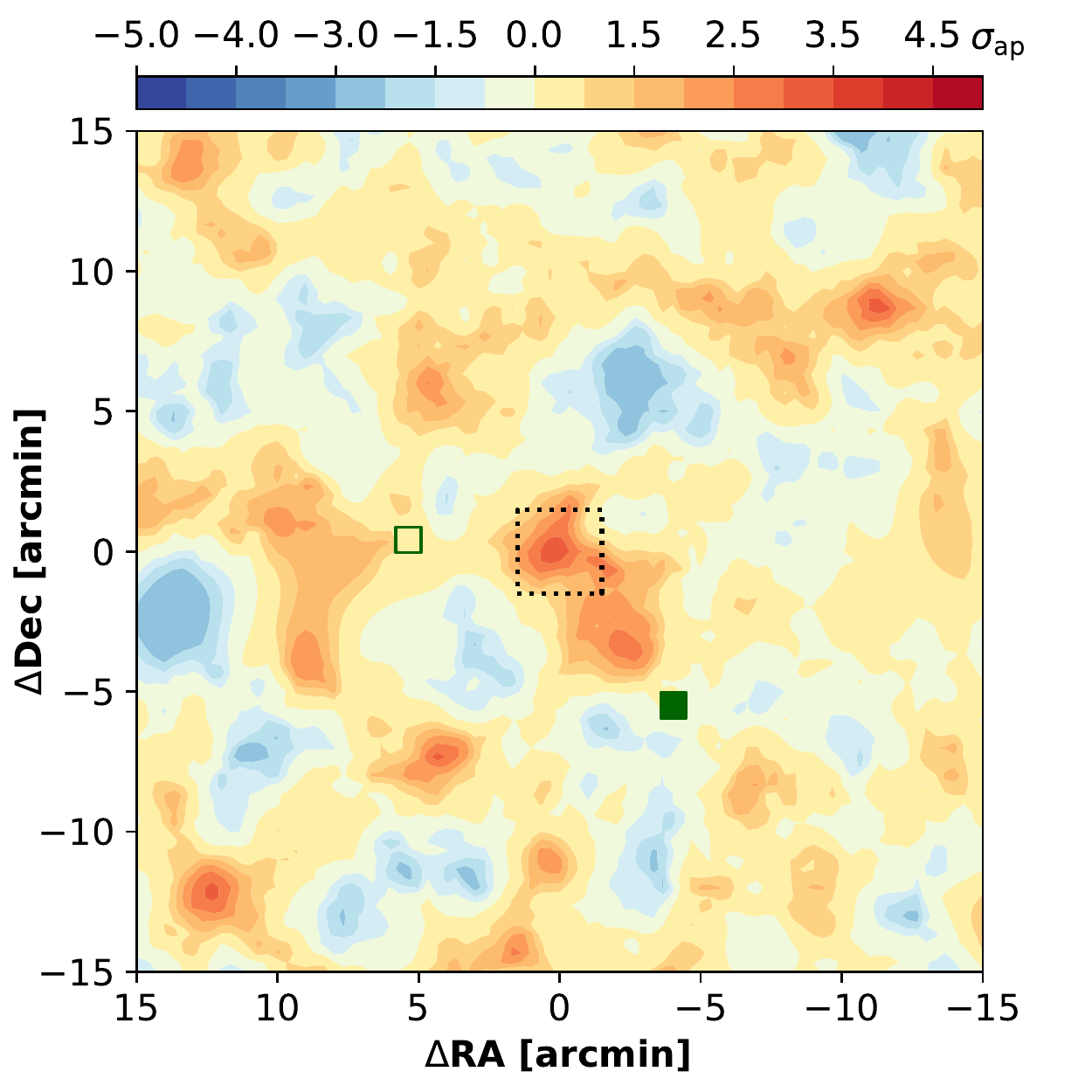}
    \caption{Aperture mass significance map of PSZ1~G109.88+27.94 centred on the brightest
cluster galaxy. 
The background color represents the significance calculated according to equation~(\ref{Map}).
Galaxies with $\beta_\mathrm{z} < 0.05$ have been excluded from the analysis.
The dotted box encompasses the inner $3\times 3$~arcmin$^2$ region of the cluster, for which we present a colour image in Fig.~\ref{fig:psz109visual}.
The positions of the foreground groups Fg1 and Fg2 are marked by the open and solid green square symbols.}
    \label{fig:psz109significance}
\end{figure}

The significance map for PSZ1~G109.88+27.94 is shown in Fig.~\ref{fig:psz109significance}. 
The cluster has been detected with an aperture mass significance of $3.5\sigma$. The $\sigma_\mathrm{ap}$-peak is centred on the BCG. There is another $3\sigma$ that seems to be correlated to the central peak at $(\Delta RA, \Delta Dec)\sim(-2,-4)$. This peak could be due to the imprint of Fg2 on the background galaxy shapes. It does not coincide with the BCG of Fg2 but is shifted by $2.7'$ towards the center of the main cluster.
The aperture mass peaks with $| \Delta RA |> 10'$ are close to the image boarder and are thus likely due to noise. 

\subsubsection{Tangential shear profile and fit of an SIS}			

\begin{figure}
	\includegraphics[width=\columnwidth]{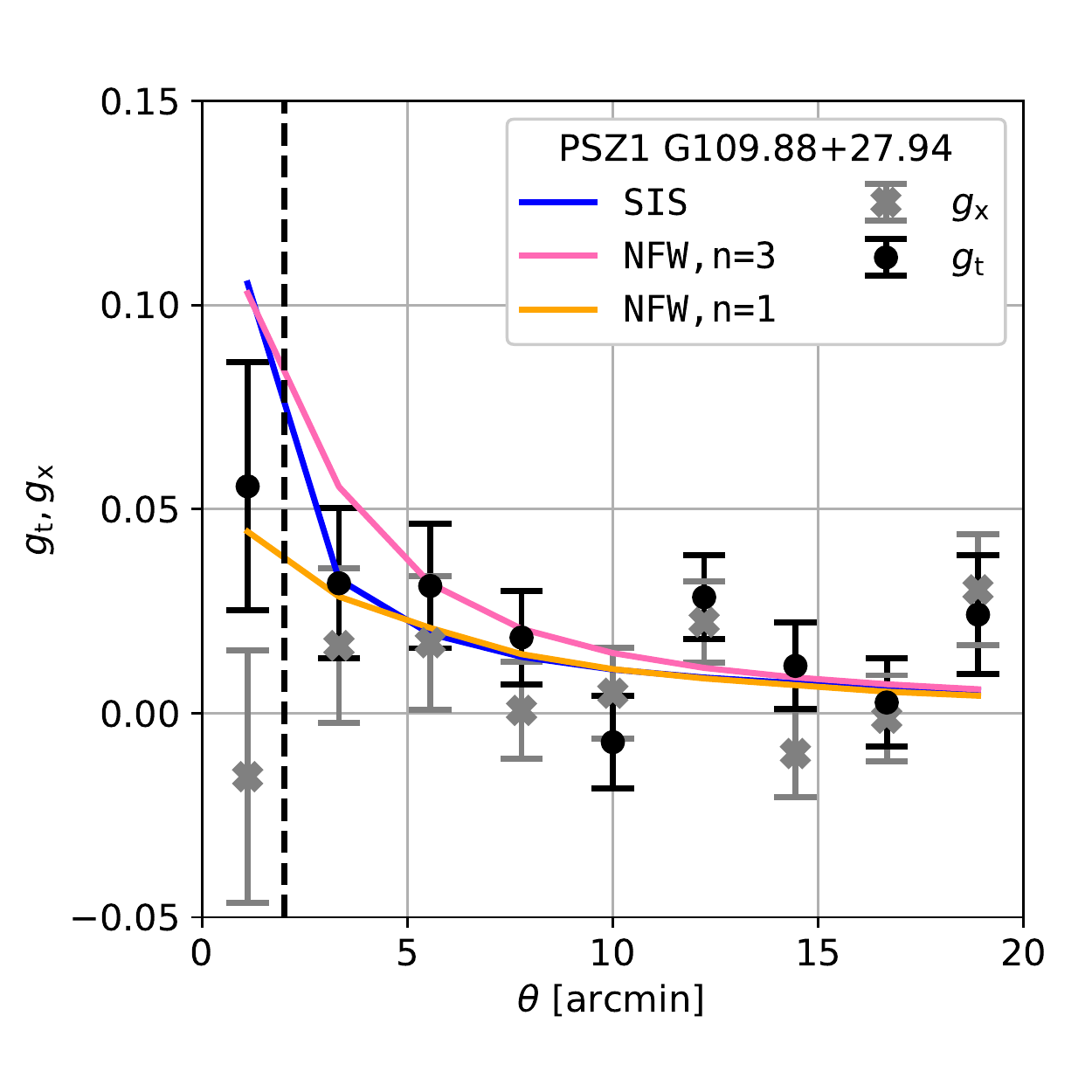}  
    \caption{Tangential alignment of PSZ1~G109.88+27.94. The black (grey) circles (crosses) show the tangential (cross) reduced shear $g_\mathrm{t}$ ($g_\mathrm{x}$). 
The blue line shows the fit of
an SIS density profile to the binned reduced shear $g_\mathrm{t}$. An inner region with a radius of $2'$ (dashed line) has been excluded
from the analysis. The cross shear is mostly consistent
with zero. A leakage correction has been applied to the shear (cf.~Section~\ref{sec:leakage}).
The orange and pink lines show the theoretical tangential shear profile of one and three dark matter NFW halos with centres and redshifts as used in the one-, two- and three-halo NFW fits (cf.~Section~\ref{sec:nfw109}) and the best fitting masses (and concentration parameters).
} 
    \label{fig:psz109tangential}
\end{figure}

The tangential alignment of background galaxies around PSZ1~G109.88+27.94 is shown in
Fig.~\ref{fig:psz109tangential}. 
We have applied a leakage correction to the data as explained in Section~\ref{sec:leakage}. As the systematic tangential shear increases with increasing distance from the centre of the image, neglecting this effect would cause $g_\mathrm{t}$ to be biased low for large $\theta$.

The fit of an SIS profile to the measured tangential shear implies a velocity dispersion of 
${\sigma_\mathrm{v} = (1800 \pm 200)\ \mathrm{km s^{-1}}}$ (blue line). The mass is estimated to be
$M^\mathrm{SIS}_\mathrm{200m} = 44^{+16}_{-13}\times10^{14}~\mathrm{M_\odot}$.  

\subsubsection{NFW fit}\label{sec:nfw109}					

\begin{figure}
			\includegraphics[width=\columnwidth]{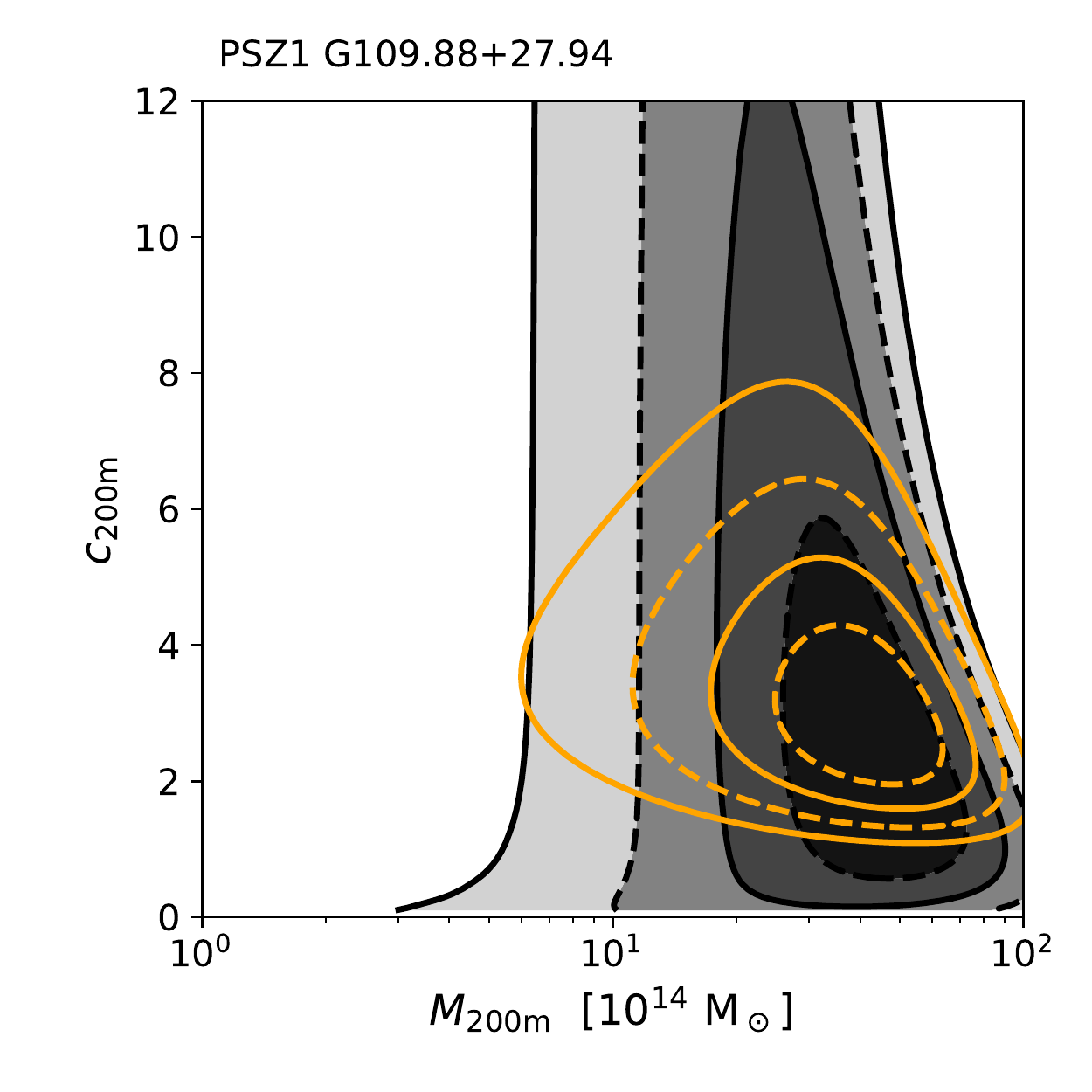} 
  \caption{
	We perform a maximum likelihood estimation assuming an NFW profile of the main cluster component in the field of PSZ1~G109.88+27.94.
	We present the likelihood contours of the fit of an NFW shear profile with virial mass $M_\mathrm{200m}$ and concentration $c_\mathrm{200m}$ to the data.
The solid contours show the combined 1 and 2$\sigma$ confidence regions of the two-parametric fit, whereas the dashed lines mark the projected confidence intervals.
The orange contours have been obtained using the concentration prior of \citet{bullock2001profiles} and \citet{duffy2008dark}, whereas the black lines show the NFW fit without any further assumptions about $c_\mathrm{200m}$.
    }
    \label{fig:psz109nfw}
\end{figure}

We perform four different NFW fits. In our first two-parameter fit, we assume a single dark matter halo at $z=0.77$ and fit for the virial mass $M_\mathrm{200m}$ and concentration parameter
$c_\mathrm{200m}$. 
While the
cluster mass is estimated to be ${M^\mathrm{NFW}_\mathrm{200m} = 46^{+27}_{-19}\times10^{14}~\mathrm{M_\odot}}$ with concentration ${c^\mathrm{NFW}_\mathrm{200m} = 2.0^{+3.8}_{-1.5}}$ (Fig.~\ref{fig:psz109nfw}, black contours). The use of a concentration prior reduces the uncertainty on $c_\mathrm{200m}$ and we obtain 
${M^\mathrm{NFW,p}_\mathrm{200m} = 42^{+21}_{-17}\times10^{14}~\mathrm{M_\odot}}$
and ${c^\mathrm{NFW,p}_\mathrm{200m} = 2.9^{+1.4}_{-0.9}}$. This fit yields $\chi^2_\mathrm{min} = 1.3$ (Fig.~\ref{fig:psz109nfw}, orange contours).

We try to find the masses of the two group candidates identified by \citet{wen2015calibration} by assuming the mass-concentration relation of \citet{duffy2008dark} and performing a three parameter fit of the halo masses. 
We find ${M^\mathrm{NFW,p}_\mathrm{200m} = \left(40^{+10}_{-15}, 4.0^{+2.3}_{-2.3}, 4.0^{+2.3}_{-2.3}\right) \times 10^{14}\ \mathrm{M_\odot}}$ for the main component, Fg1 and Fg2 respectively. 
Our three-halo fit shows the cluster at $z=0.77$ is clearly the dominant mass component in the field. 
Our analysis suggests that Fg1 and Fg2 have a comparable mass but we can only claim a $1\sigma$ detection of these group candidates. The $\chi^2_\mathrm{min}$ is approximately 1.3. The tangential shear signal predicted by our best fitting three-halo model is shown in pink in (Fig.~\ref{fig:psz109tangential}).

\subsection{PSZ1~G139.61+24.20}

\begin{figure}
	\includegraphics[width=\columnwidth]{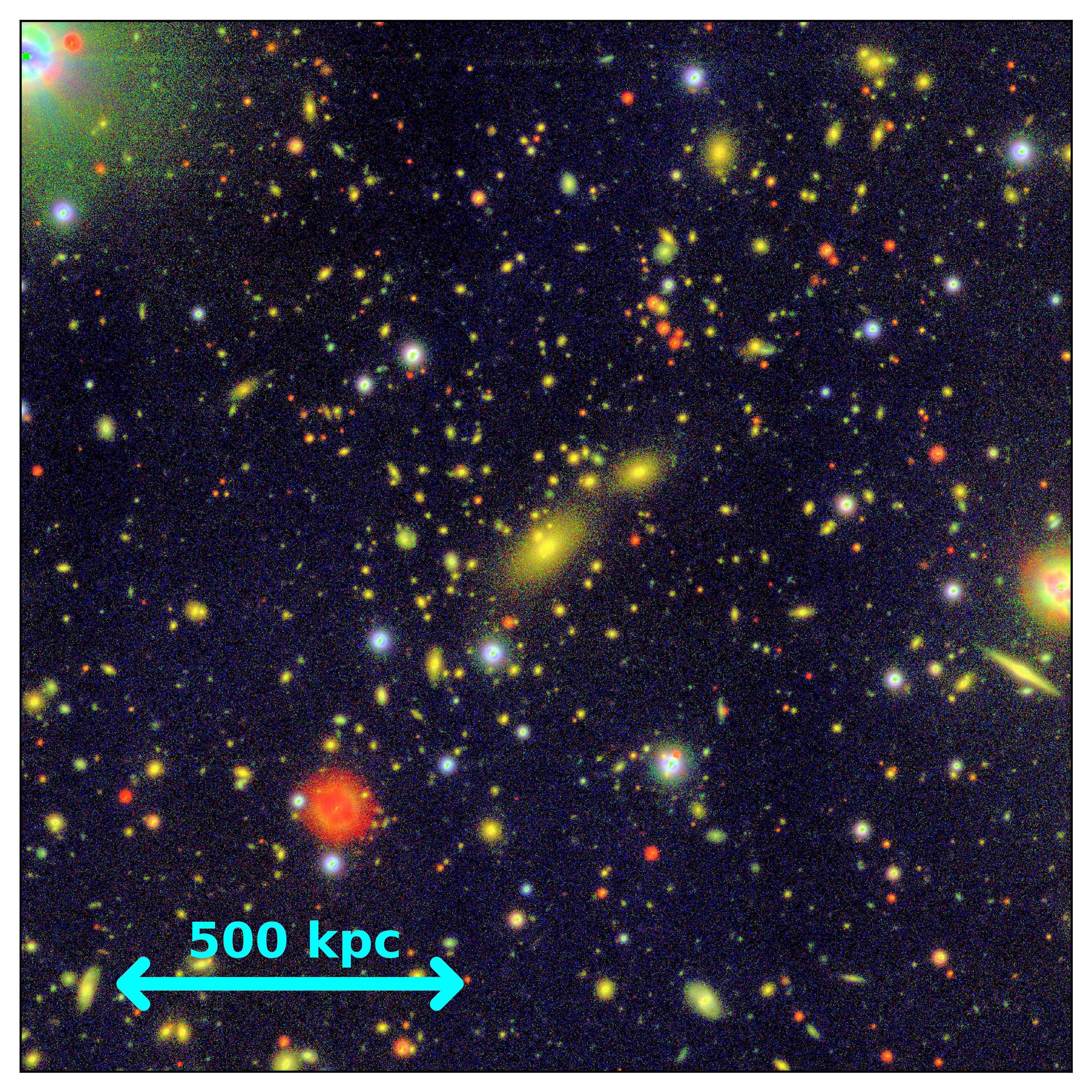}  
    \caption{RGB image using WWFI $g$, $r$ and $i$~band stacks. The dispayed $6\times 6$~arcmin$^2$
region is centred on the brightest cluster galaxy of PSZ1~G139.61+24.20. The cyan line
indicates a distance of 500~kpc at the cluster redshift $z_\mathrm{cl} = 0.267$. This corresponds to the
physical distance of the radius, that has been excluded from the NFW analysis described
in the subsection below. North-east is towards the upper left in the image.}
    \label{fig:psz139visual}
\end{figure}

PSZ1~G139.61+24.20 is at a spectroscopic redshift of $z_\mathrm{cl} = 0.267$ \citep{ade2015b}.
Fig.~\ref{fig:psz139visual} shows the central region of the cluster, which coincides with both, the SZ footprint and the X-ray centroid. 
We have chosen the diffuse brightest
cluster galaxy as the centre of the dark matter halo of PSZ1~G139.61+24.20 in our analysis.

\citet{giacintucci2017occurrence} detect a radio minihalo in the core of PSZ1~G139.61+24.20, which implies that there is a diffuse radio source in the centre of the cluster. We confirm PSZ1~G139.61+24.20 to be a massive cluster of galaxies and present the first weak lensing mass estimate for this object.

\subsubsection{Significance map}\label{sec:139sig}		

\begin{figure}
	\includegraphics[width=\columnwidth]{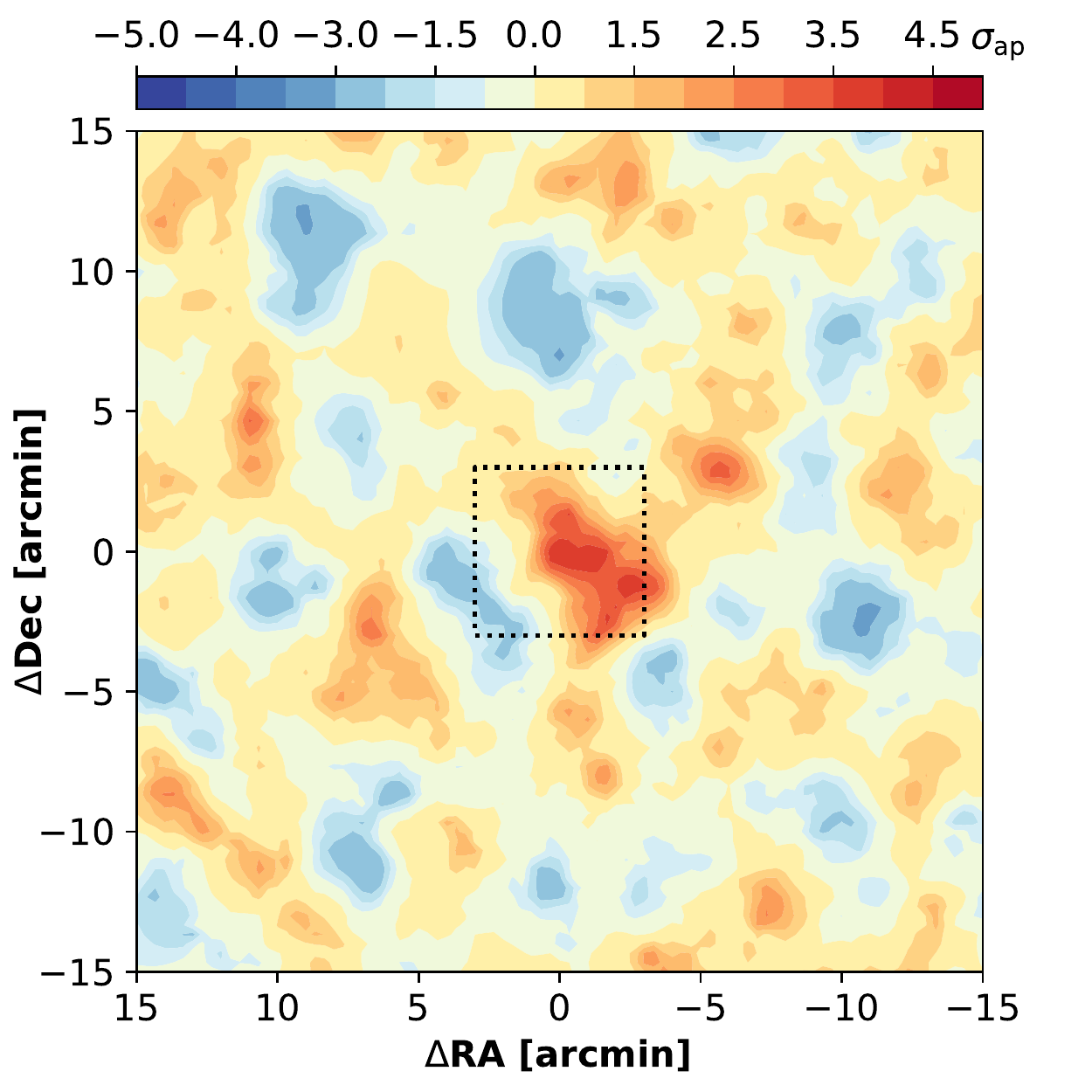}
    \caption{Aperture mass significance map of PSZ1~G139.61+24.20 centred on the brightest
cluster galaxy. 
The background color represents the significance calculated according to equation~(\ref{Map}).
Galaxies with $\beta_\mathrm{z} < 0.05$ have been excluded from the analysis.
The dotted box encompasses the inner $6\times 6$~arcmin$^2$ region of the cluster, for which we present a colour image in Fig.~\ref{fig:psz139visual}.
}
    \label{fig:psz139significance}
\end{figure}

The significance map (Fig.~\ref{fig:psz139significance}) shows a peak at the position of the BCG at a significance of $\sim4\sigma$.
The SZ footprint of the cluster shows the same orientation as our 2d-lensing signal.
Another small $3.5\sigma$~peak can be seen towards the North-East of the cluster but is likely due to noise.

\subsubsection{Tangential shear profile and fit of an SIS}		

\begin{figure}
	\includegraphics[width=\columnwidth]{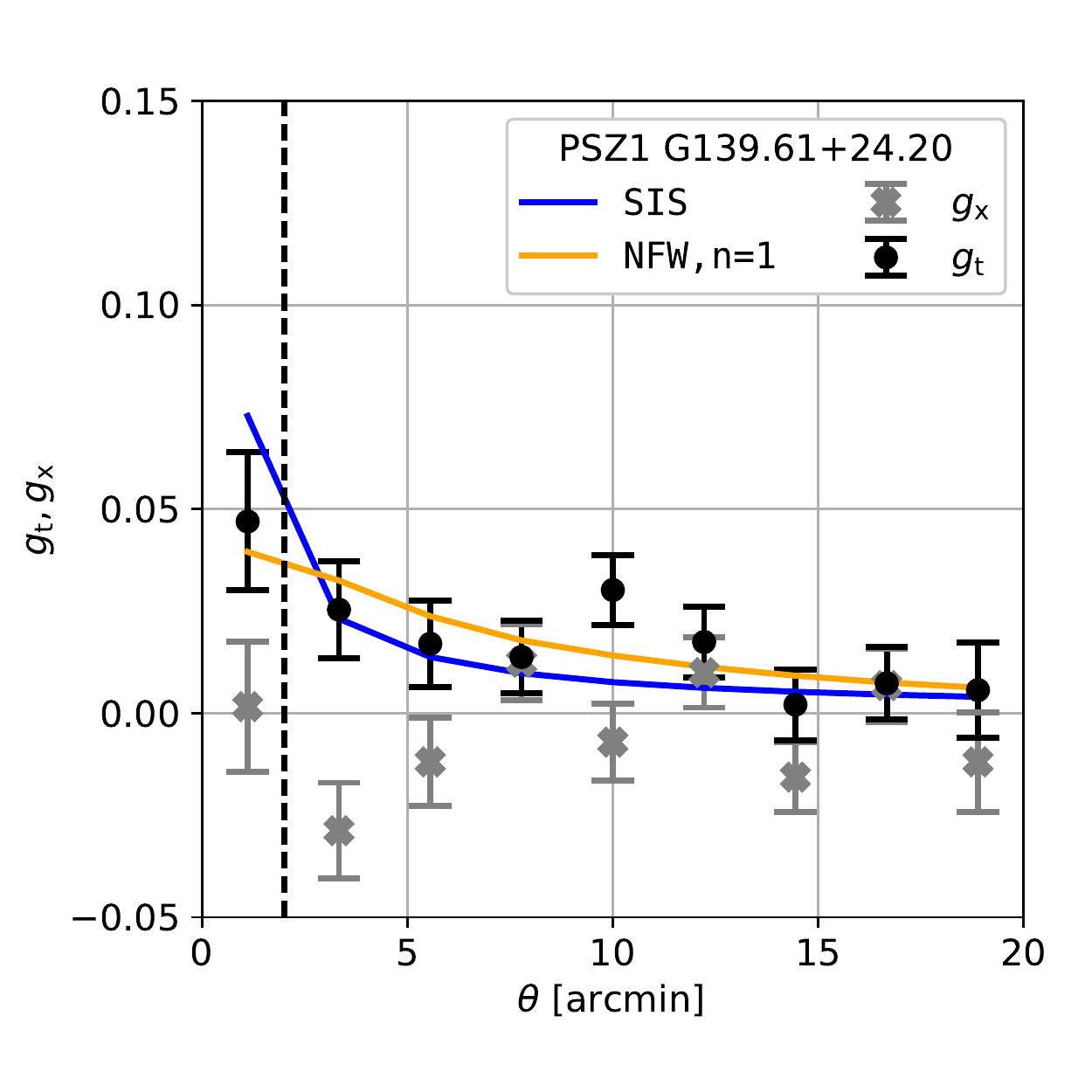}  
    \caption{Tangential alignment of PSZ1~G139.61+24.20. The black (grey) circles (crosses) show the tangential (cross) reduced shear $g_\mathrm{t}$ ($g_\mathrm{x}$). 
The blue line shows the fit of
an SIS density profile to the binned reduced shear $g_\mathrm{t}$. An inner region with a radius of $2'$ (dashed line) has been excluded
from the analysis. The cross shear is mostly consistent
with zero. A leakage correction has been applied to the shear (cf.~Section~\ref{sec:leakage}). The orange line shows the best fitting model of our two-parameter NFW fit (cf.~Section~\ref{sec:nfw139}).}
    \label{fig:psz139tangential}
\end{figure}

The measured shear profile of PSZ1~G139.61+24.20 is shown in Fig.~\ref{fig:psz139tangential}. 
The velocity dispersion is estimated to be ${\sigma_\mathrm{v} = (800 \pm 100)\ \mathrm{km s^{-1}}}$, which corresponds to a mass of
$M^\mathrm{SIS}_\mathrm{200m} = 6.3^{+2.7}_{-2.1}\times 10^{14}\mathrm{M_\odot}$.
The cross component of the shear is mostly consistent with zero, except for the second bin, in which we measure a negative $g_\mathrm{x}$ at a $\sim 2\sigma$ level.

\subsubsection{NFW fit}\label{sec:nfw139}			

\begin{figure}
		\includegraphics[width=\columnwidth]{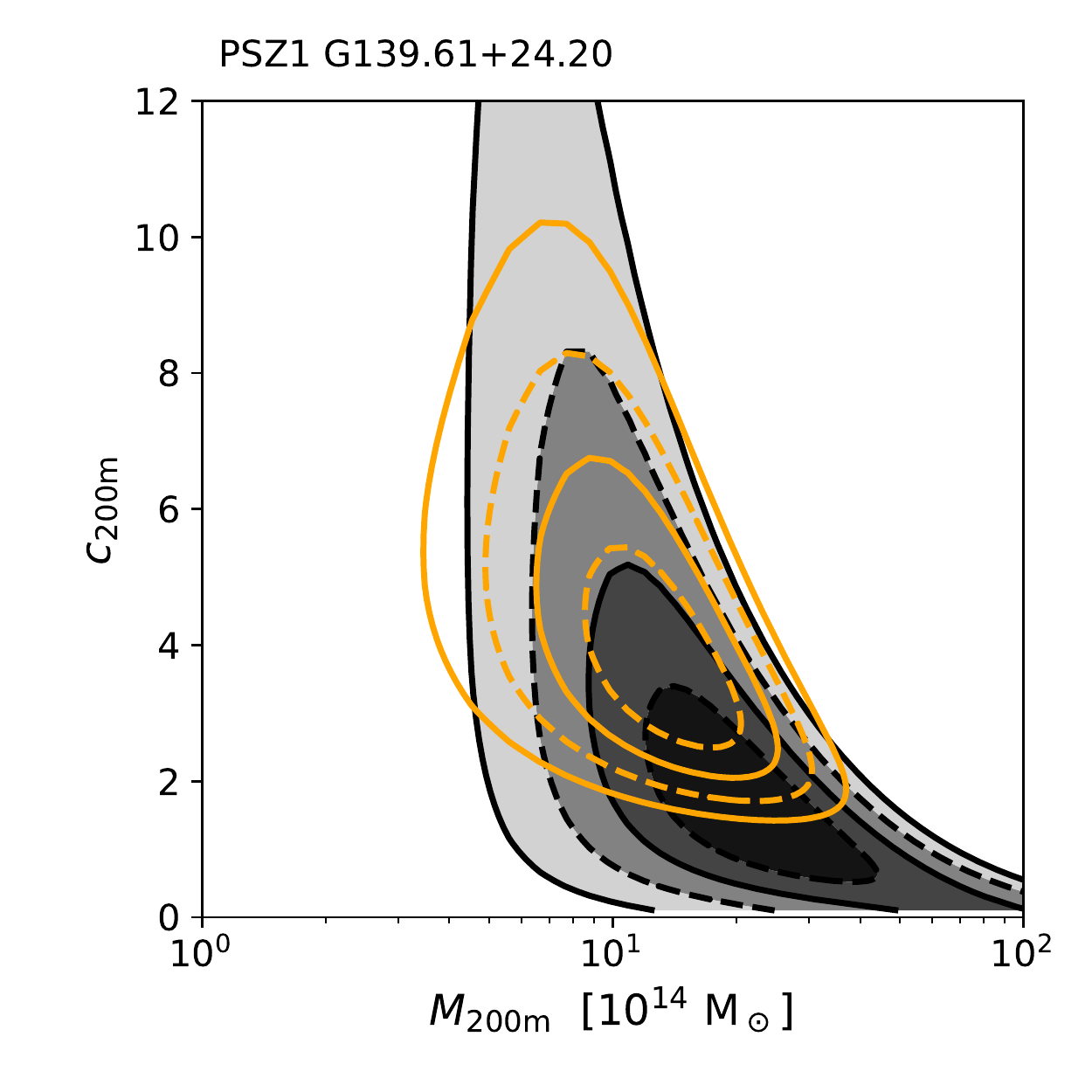} 

    \caption{
    We perform a maximum likelihood estimation assuming an NFW profile of the main cluster component in the field of PSZ1~G139.61+24.20.
    We present the likelihood contours of the fit of an NFW shear profile with virial mass $M_\mathrm{200m}$ and concentration $c_\mathrm{200m}$ to the data.
The solid contours show the combined 1 and 2$\sigma$ confidence regions of the two-parametric fit, whereas the dashed lines mark the projected confidence intervals.
The orange contours have been obtained using the concentration prior of \citet{bullock2001profiles} and \citet{duffy2008dark}, whereas the black lines show the NFW fit without any further assumptions about $c_\mathrm{200m}$.
    }
    \label{fig:psz139nfw}
\end{figure}

Without making any assumptions about the concentration of the cluster, the two-parametric
NFW fit (Fig.~\ref{fig:psz139nfw}) yields {$M^\mathrm{NFW}_\mathrm{200m} = 22^{+22}_{-10}\times10^{14} \mathrm{M_\odot}$}.
Again, the concentration that we get from this fit is low with ${c_\mathrm{200m} = 1.6^{+1.9}_{-1.0}}$.
Upper limits of the projected intervals of 68\%, 90\% and 99\% confidence are indicated by the grey lines.
Using the concentration prior of \citet{bullock2001profiles} and \citet{duffy2008dark}
gives ${M^\mathrm{NFW,p}_\mathrm{200m} = 13.5^{+6.9}_{-4.8}\times10^{14} \mathrm{M_\odot}}$, a concentration parameter of
${c^{NFW,p}_{200m} = 3.7^{+1.8}_{-1.2}}$ and $\chi^2_\mathrm{min} = 1.4$. 

\subsection{PSZ1~G186.98+38.66}

PSZ1~G186.98+38.66 has an assigned spectroscopic redshift of $z_\mathrm{cl} = 0.378$ \citep{piffaretti2011vizier} in the \emph{Planck} catalogues. 
The field contains a massive cluster of galaxies at ${\mathrm{(RA,Dec)}= (08:50:11.2,+36:04:21)}$ that has first been visually identified by \citet{1961Z} and is commonly referred to as \emph{Zwicky}~1953, ZwCl~0847+3617, RXC~J0850.2+3603, or MACS~J0850.1+3604.
The cluster shows a prominent SZ imprint and X-ray signal which are aligned with the optical cluster centre. 

According to the high concentration of luminous red galaxies (LRGs) at a redshift of about 0.35-0.4, this cluster field belongs to the 200 most massive lines of sight in the SDSS \citep{wong2013new}. \citet{ammons2013mapping} have studied this field with Hectospec at the MMT telescope on Mt. Hopkins, Arizona.
Using their spectroscopic catalogues, they have identified two groups in the field, which are traced by LRGs.
They find that the pointing is dominated by a massive cluster at redshift $z=0.3774$ with a velocity dispersion of $\sigma = 1300\ \mathrm{km s^{-1}}$. A second, smaller group at redshift $z=0.2713$ has a velocity dispersion of $\sigma = 300\ \mathrm{km s^{-1}}$. There is another LRG at a spectroscopic redshift of $z=0.563$ in the field. \citet{ammons2013mapping} do not find evidence for the presence of a third group at this redshift, though this could be due to the limited depth of their spectroscopic sample.

\citet{ammons2013mapping}  also use six-band $B,V,R_c,I_c,i',z'$ Subaru/Suprime-Cam data to search for strong lensing features. They have found a candidate multiply-imaged source at a photometric redshift of $z = 5.03^{+0.21}_{-0.17}$. This source galaxy has been used by  \citet{wong2013new} to perform a joint weak and strong lensing analysis of the field using the same data as \citet{ammons2013mapping}.  

\begin{figure}
	\includegraphics[width=\columnwidth]{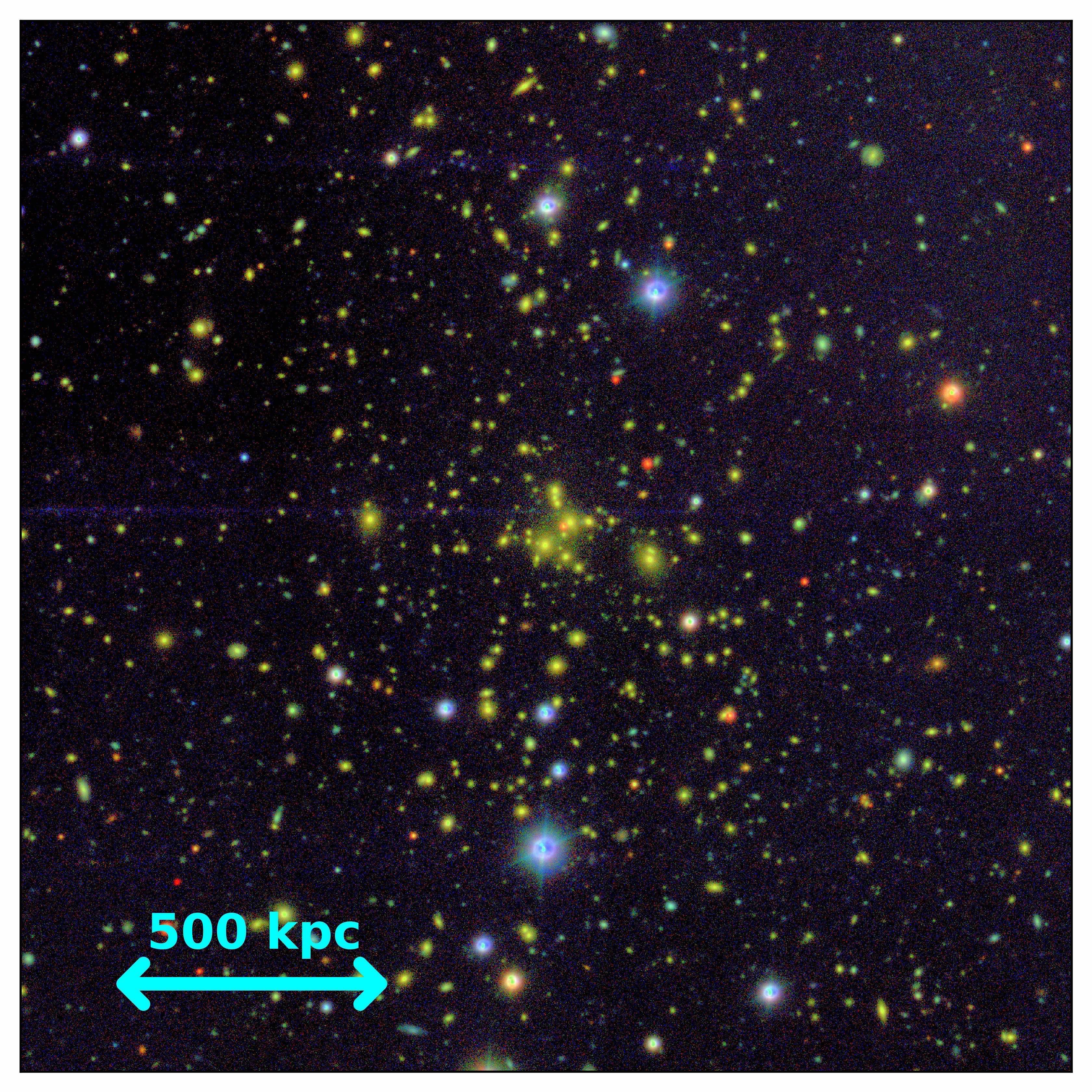}  
    \caption{RGB image using WWFI $g$, $r$ and $i$~band stacks. The dispayed $6\times 6$~arcmin$^2$ region 
      region is centred on the BCG of PSZ1~G186.98+38.66. The cyan line indicates a distance of 500~kpc at the cluster redshift. 
      This corresponds to the physical distance of the radius, that has been excluded from the NFW analysis
      described in the subsection below. 
      The cluster has a group redshift of $0.3774$ \citep{ammons2013mapping}.
      North-east is towards the upper left in the image.}
    \label{fig:psz186visual}
\end{figure}

We choose the centre of the cluster at $z=0.3774$ to be the brightest cluster galaxy (cf.~Fig.~\ref{fig:psz186visual}). 
The field is very crowded with yellow cluster member galaxies.

PSZ1~G186.98+38.66 is part of the WtG sample of 51 massive galaxy clusters \citep{applegate2014weighing}.
They only consider one halo az $z=0.378$ to find a virial mass of $M(<1.5\mathrm{Mpc}) = (15.8\pm 2.6)\times 10^{14}\ \mathrm{M_\odot}$ using their photometric redshift estimates
$M(<1.5\mathrm{Mpc}) = (16.9\pm 3.2)\times 10^{14}\ \mathrm{M_\odot}$ by applying a colour-cut.

\subsubsection{Significance map}				

\begin{figure}
	\includegraphics[width=\columnwidth]{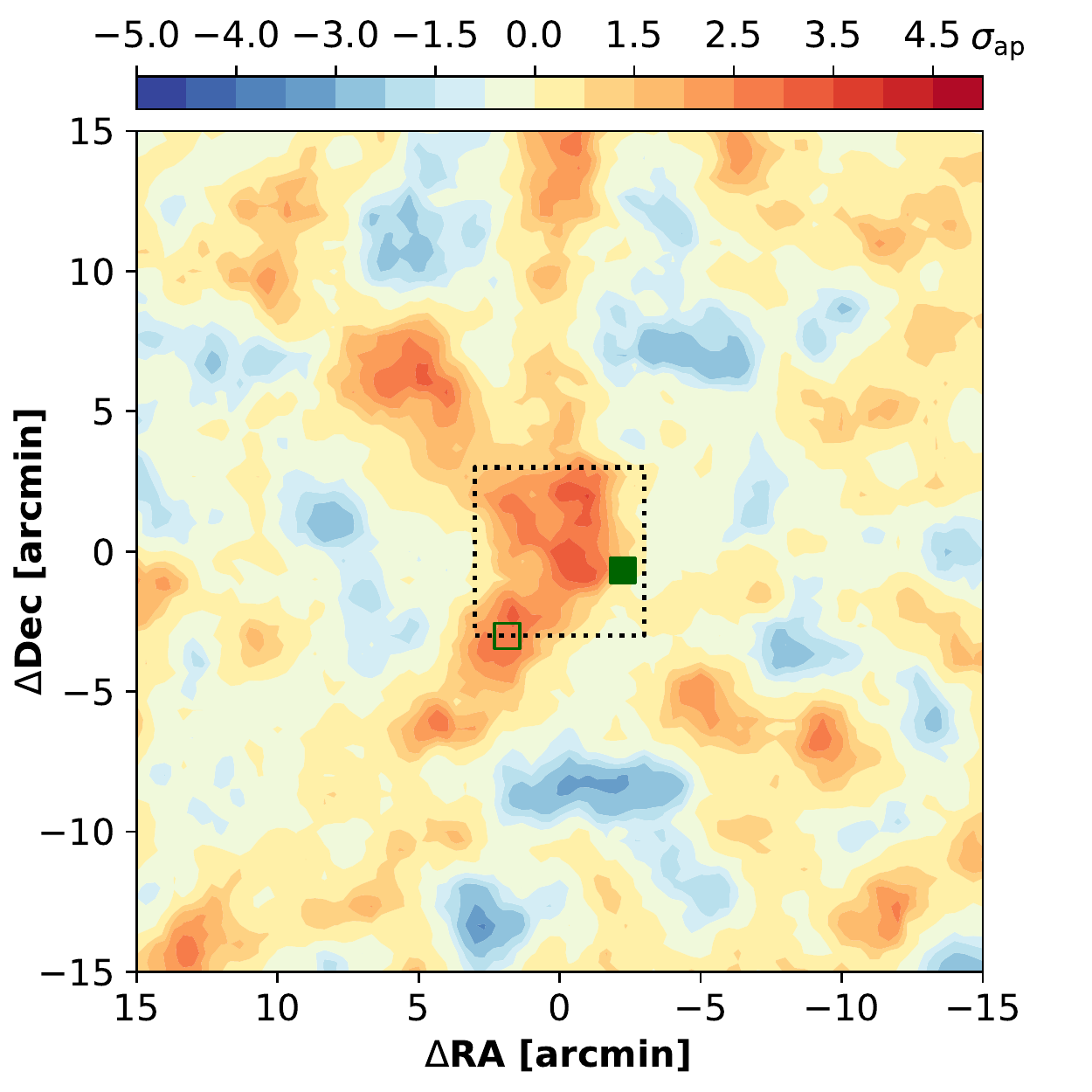}
    \caption{PSZ1~G186.98+38.66 centred on the BCG. The background color represents the significance calculated according to equation~(\ref{Map}). Galaxies with $\beta_\mathrm{z} < 0.05$ have been excluded from the analysis.
The dotted box encompasses the inner $6\times 6$~arcmin$^2$ region of the cluster, for which we present a colour image in Fig.~\ref{fig:psz186visual}. The position of the foreground group is marked by the open green square symbol. The solid green symbol is at the position of the LRG at $z=0.563$.
}
    \label{fig:psz186significance}
\end{figure}

The significance map of PSZ1~G186.98+38.66 (Fig.~\ref{fig:psz186significance}) is centred on the cluster
position. 
The main component of the cluster at $z=0.3774$ is detected at a significance of $4\sigma$.
The foreground group (Fg) leaves an imprint on the 2d-lensing signal at a significance of $3.5\sigma$ at $(\Delta RA, \Delta Dec) = (2,-2.5)$. The centroid position of this group is indicated by the open green symbol. The projected distance of the LRG at $z=0.563$ to the main halo centre is very small. 
There is another $3.5\sigma$ peak of the aperture mass but we cannot find a corresponding overdensity of red galaxies at the designated position of the background group candidate (Bg). 

\subsubsection{Tangential shear profile and fit of an SIS}			

\begin{figure}
	\includegraphics[width=\columnwidth]{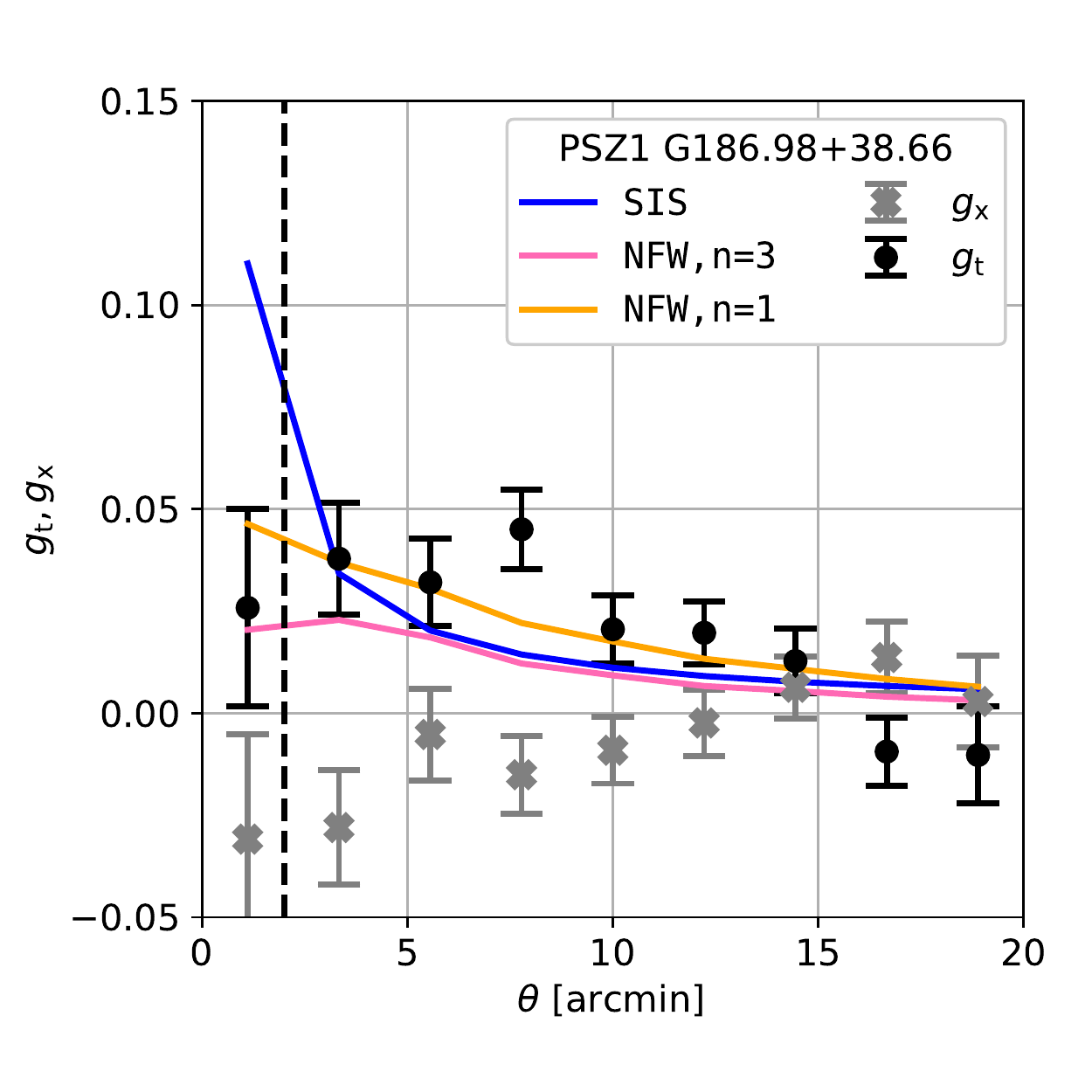}  
    \caption{Tangential alignment of PSZ1~G186.98+38.66. The black (grey) circles (crosses) show the tangential (cross) reduced shear $g_\mathrm{t}$ ($g_\mathrm{x}$). 
The blue line shows the fit of
an SIS density profile to the binned reduced shear $g_\mathrm{t}$. An inner region with a radius of $2'$ (dashed line) has been excluded
from the analysis. The cross shear is mostly consistent
with zero. A leakage correction has been applied to the shear (cf.~Section~\ref{sec:leakage}). The orange and pink lines show the theoretical tangential shear profile of one and three dark matter NFW halos with centres and redshifts as used in the one-, two- and three-halo NFW fits (cf.~Section~\ref{sec:nfw186}) and the best fitting masses (and concentration parameters).
}
    \label{fig:psz186tangential}
\end{figure}

Fig.~\ref{fig:psz186tangential} shows the tangential shear profile of 
PSZ1~G186.98+38.66. 
Despite the leakage correction applied to the data, the measured tangential shear signal is still slightly negative for $\gtrsim 16'$.

We fit an SIS profile to our measurement and find a velocity dispersion of ${\sigma_\mathrm{v} = (1100\pm200)\ \mathrm{km s^{-1}}}$, which corresponds to a mass of ${M^\mathrm{SIS}_\mathrm{200m}=  14.5^{+9.4}_{- 6.5}\times10^{14}\mathrm{M_\odot}.}$
The tangential shear signal in the cluster centre is lowered for the three-halo NFW model.
This could explain the low tangential shear signal in the first bin,
as opposed to the SIS model which predicts a much larger value of $g_\mathrm{t}$ at this radius. 

\subsubsection{NFW fit}\label{sec:nfw186}			

\begin{figure}
		\includegraphics[width=\columnwidth]{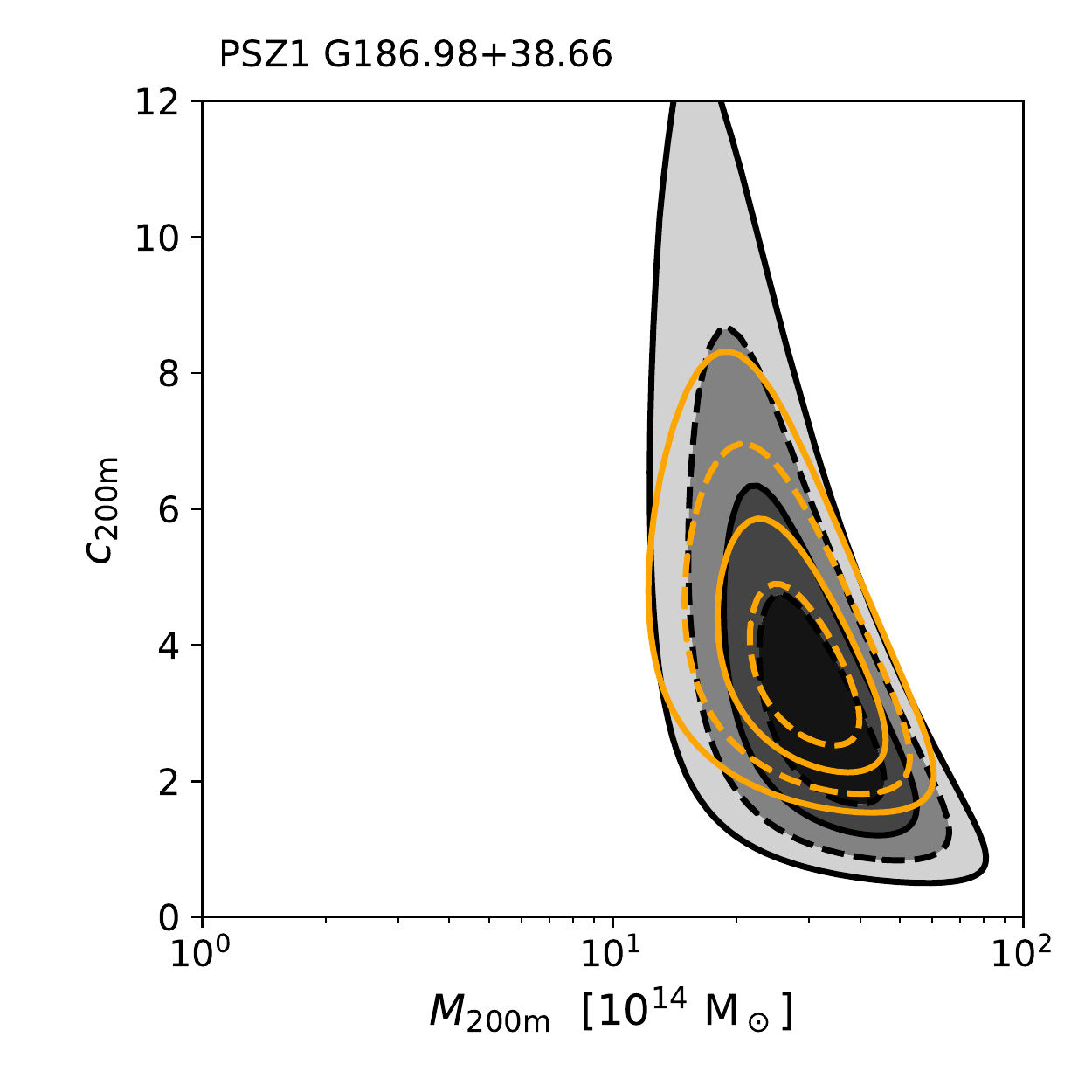} 
    \caption{
    We perform a maximum likelihood estimation assuming an NFW profile of the main cluster component in the field of PSZ1~G186.98+38.66.
    We present the likelihood contours of the fit of an NFW shear profile with virial mass $M_\mathrm{200m}$ and concentration $c_\mathrm{200m}$ to the data.
The solid contours show the combined 1 and 2$\sigma$ confidence regions of the two-parametric fit, whereas the dashed lines mark the projected confidence intervals.
The orange contours have been obtained using the concentration prior of \citet{bullock2001profiles} and \citet{duffy2008dark}, whereas the black lines show the NFW fit without any further assumptions about $c_\mathrm{200m}$.
}
    \label{fig:psz186nfw}
\end{figure}

Fitting an NFW density profile to the measurement gives a mass of
${M^\mathrm{NFW}_\mathrm{200m} = 32^{+13}_{-20}\times10^{14} \mathrm{M_\odot}}$
 with a concentration of 
${c^\mathrm{NFW}_\mathrm{200m} = 2.8^{+1.9}_{-1.2}}$ and a minimum $\chi^2$ of $\chi_\mathrm{min}^2 = 1.3$. Using a prior
on the concentration, we find 
$M^\mathrm{NFW,p}_\mathrm{200m} = 30^{+10}_{-8}\times10^{14} \mathrm{M_\odot}$
and 
$c^\mathrm{NFW,p}_\mathrm{200m} = 3.5^{+1.4}_{-1.0}$, where $\chi_\mathrm{min}^2$ does not change significantly (cf.~Fig~\ref{fig:psz186nfw}). 

The three-halo fit reveals the foreground group to have a rather low mass of ${M^\mathrm{NFW}_\mathrm{200m}\lesssim2.5^{+2.5}_{-1.9}\times10^{14} \mathrm{M_\odot}}$ to which we are not sensitive in our weak lensing study. The fit prefers a mass of ${M^\mathrm{NFW}_\mathrm{200m} = 2.5^{+5.4}_{-2.5}\times10^{14} \mathrm{M_\odot}}$ for the background cluster candidate. The presence of the two groups implies a lower mass of  ${M^\mathrm{NFW}_\mathrm{200m} = 20^{+5}_{-4}\times10^{14} \mathrm{M_\odot}}$ compared to the one-halo fit for the main cluster component. The minimum $\chi^2$ of this fit is slightly smaller compared to the one-halo NFW model with  $\chi_\mathrm{min}^2 = 1.2$.
We show the best fit model tangential shear in Fig.~\ref{fig:psz186tangential} in pink.

\section{Discussion and comparison to literature}\label{sec:seven:compare}

\subsection{Comparison to SZ mass estimates}			

Having determined weak lensing masses for the WWL pathfinder sample, 
we compare our estimates to the SZ masses reported by \cite{ade2014b,ade2015b} and \cite{ade2015c}. 
We discuss the results of this work and suggest strategies to improve the weak lensing analysis of future WWL precision measurements.

Mass estimates based on {\it{Planck}} measurements of the Compton parameter $Y$ can be obtained via
\begin{equation}D_A^2 Y = \frac{\sigma_\mathrm{T}}{m_\mathrm{e} c^2}\int P_\mathrm{e} \mathrm{d}V ,
\label{DY}\end{equation}
where $\sigma_\mathrm{T}$ is the Thompson scattering cross section, $m_\mathrm{e}$ is the electron mass and $c$ is the speed of light.
As the volume integral of the electron pressure $P_e$ equals the thermal energy of the electron gas, the Compton parameter is very closely correlated with both the temperature and the mass of the gas and thus, as both of these properties depend on the cluster mass, with the total mass $M_\mathrm{tot}$.  

The two \textit{Planck} SZ cluster catalogues offer mass estimates within a radius of $R_{500c}$, where
\begin{equation}
M_{500c} = 500 \times \frac{4\pi \rho_\mathrm{crit}}{3} R^3_{500c}.
\end{equation}

The angular size $\theta_{500}$\footnote[4]{$\theta_{500}$ denotes the angular size corresponding to the physical scale of the virial radius $R_{500}$.} 
is the aperture used to extract the integrated Compton parameter $Y_{500}$. Consequently, there is a degeneracy between cluster size and SZ flux, 
which has to be broken in order to derive the cluster mass \citep{ade2011planck}. This is accomplished by using the $M_{500}$-$D^2_\mathrm{A}Y_{500}$ 
relation between $\theta_{500}$ and $Y_{500}$ \citep{arnaud2010universal}.

The catalogue of \cite{ade2014b} (hereafter PSZ1) includes $1\ 227$ SZ selected clusters of galaxies, some of which have not yet 
been verified. PSZ1 has been created using the first 15.5 months of \textit{Planck} observations. $M^\mathrm{Yz}_\mathrm{500}$ is 
based on X-ray calibrated scaling relations under the assumption of a flat universe with $h = 0.7,\ \Omega_\mathrm{m} = 0.3$ and $\Omega_\mathrm{\Lambda} = 0.7$ \citep{ade2015c}.
\citet[][hereafter PSZ2]{ade2015b} derive their mass proxy $M_\mathrm{SZ}$ in a similar fashion. 

We convert the weak lensing masses $M^\mathrm{NFW}_\mathrm{200m}$ to $M_\mathrm{500c}$ and find an overall agreement with the SZ masses 
within the errors (cf.~Table~\ref{tab:masses}) for the single-halo fits of PSZ1~G139.61+24.20 and PSZ1~G186.98+38.66. In the case of PSZ1~G109.88+27.94, the SZ mass is severely underestimated by PSZ1 and PSZ2, as a wrong redshift of $z=0.4$ was assumed in their analysis.
\begin{table*}
	\centering
        \caption[Mass proxies for the cluster sample]{\emph{Planck} name, line-of-sight structure, right ascension, declination and the redshifts we used in our weak lensing analysis. Our weak lensing mass estimates $M^\mathrm{WL}_\mathrm{500c}$ from the NFW fits assuming one and three dark matter halos along the line-of-sight and the SZ mass proxies of PSZ1 and PSZ2, where the masses are given in units of $10^{14}\ \mathrm{M_\odot}$.}
        \begin{tabular}{ l c c c c c c c c c c c r}
          \hline
          Field name & LSS &RA (J2000) & Dec (J2000) &  $z_\mathrm{analysis}$ &$M^\mathrm{WL,1}_\mathrm{200m}$&$M^\mathrm{WL,1}_\mathrm{500c}$&$M^\mathrm{WL,3}_\mathrm{200c}$&$M^\mathrm{Yz}_\mathrm{500}$&$M^\mathrm{SZ}_\mathrm{500}$\\
          \hline
                    \hline
          PSZ1~G109.88+27.94 & Main & 18:23:23.0 & +78:23:13 & 0.77 & $42^{+27}_{-19}$ & $21^{+16}_{-10}$& $40^{+10}_{-8}$& $5.6^{+0.6}_{-0.7}$ & $5.2^{+0.4}_{-0.5}$\\
                    &Fg1& 18:25:09.4 & +78:23:37& 0.22& - &- & $4.0^{+2.3}_{-2.4}$&-  & -\\
           & Fg2 &18:22:02.8 & +78:17:43 & 0.29 &-&-& $4.0^{+2.3}_{-2.4}$& -&-\\
          \hline
           PSZ1~G139.61+24.20 &  &6:21:48.9 & +74:42:04& 0.267 & $13.5^{+6.9}_{-4.8}$ & $6.0^{+4.7}_{-2.9}$&-& $7.1^{+0.6}_{-0.6}$ & $7.6^{+0.5}_{-0.5}$\\
                    \hline
           PSZ1~G186.98+38.66 & Main &8:50:07.9 & +36:04:13& 0.3774 & $30^{+10}_{-8}$ & $13.7^{+7.5}_{-5.1}$& $20.0^{+5.2}_{-4.1}$&$7.4^{+0.7}_{-0.8}$ & $6.8^{+0.5}_{-0.5}$\\

           & Fg &8:50:17.1 & +36:01:13 & 0.2713 &  - &-&$2.5^{+2.5}_{-1.9}$&-&-\\

           & Bg &8:49:56.8 &  +36:03:33& 0.563 &-&- & $2.5^{+5.4}_{-2.5}$& -&-\\
          \hline
        \end{tabular}
\label{tab:masses}
\end{table*}

{Compared to the \textit{Planck} mass estimates, the WWFI weak lensing masses $M^\mathrm{WL}_\mathrm{500c}$ have very large uncertainties. 
The errors of the SZ masses presented in Table~\ref{tab:masses} are purely measurement uncertainties. Neither intrinsic, statistical, nor systematic 
errors on the scaling relations have been taken into account. An intrinsic scatter of the SZ mass arises from a scatter in the SZ signal at fixed halo mass. 
\citet{ade2014a,ade2015a} have found a value of $\sigma_\mathrm{\ln Y} = 0.127\pm 0.023$ for their data.    
As the weak lensing shear is sensitive to all of the matter along the line of sight to the cluster, a scatter into lensing mass has to be considered as well. 

For the estimation of the SZ cluster masses, only one halo at the designated cluster redshift was assumed. If we compare the results of our one-halo NFW masses $M^\mathrm{WL,1}_\mathrm{500c}$ to the results from \emph{Planck}, we find that they agree within the errors for  PSZ1~G139.61+24.20. 
The $\sim1.3\sigma$ discrepancy between $M^\mathrm{WL,1}_\mathrm{500c}$ and $M^\mathrm{Yz}_\mathrm{500}$ (/$M^\mathrm{SZ}_\mathrm{500}$) implies a larger cluster mass for 
PSZ1~G186.98+38.66 than the SZ signal suggests.

\subsection{Comparison to weak lensing mass estimates} 	

As mentioned before, only for PSZ1~G186.98+38.66 previous weak and strong lensing mass estimates do exist.
The first weak lensing mass estimate from \citet{applegate2014weighing} predicts a total mass of $M(<1.5\mathrm{Mpc}) = (16.9\pm 3.2)\times 10^{14}\ \mathrm{M_\odot}$ 
in a sphere at redshift 0.378 with radius $1.5\ $~Mpc from the cluster centre. We conclude, that our weak lensing mass estimates for this field agree well with the findings of \citet{applegate2014weighing}.

\subsection{Comparison to dynamical mass estimates}		

Finally, the velocity dispersions of PSZ1~G139.61+24.20 and PSZ1~G186.98+38.66 are known from literature. 
\citet{amodeo2017calibrating} have identified 
20 cluster member galaxies of PSZ1~G139.61+24.20 and 41 cluster cluster members for PSZ1~G186.98+38.66 using spectra obtained with 
the GMOS multi-object spectrograph at Gemini observatory. 
They 
measure a velocity dispersion of ${\sigma_{200c} = 1052^{+390}_{-273}\ \mathrm{km s^{-1}}}$ for PSZ1~G139.61+24.20 and
${\sigma_{200c} = 1432^{+200}_{-166}\ \mathrm{km s^{-1}}}$ for PSZ1~G186.98+38.66.
Our SIS fits suggest velocity dispersions of ${\sigma_\mathrm{v} = 800\pm100\ \mathrm{km s^{-1}}}$ for PSZ1~G139.61+24.20 and ${\sigma_\mathrm{v} = 1100\pm200\ \mathrm{km s^{-1}}}$ for PSZ1~G186.98+38.66 
Even with this crude model of the density profiles of the clusters, our ${\sigma_\mathrm{v}}$ estimates are in agreement with the measurements of \citet{amodeo2017calibrating}.

\citet{ammons2013mapping} have used a dynamical model to constrain the NFW mass and concentration of PSZ1~G186.98+38.66 from more than 500 galaxy spectra. They find a ${\sigma_\mathrm{los}=1300^{+60}_{-60}\ \mathrm{km s^{-1}}}$ for the main cluster component at $z=0.3774$ and ${\sigma_\mathrm{los}=300^{+110}_{-90}\ \mathrm{km s^{-1}}}$ for the group at $z=0.3774$, They estimate the virial cluster mass to be equal to ${M_{200} = 32^{+3.1}_{-2.9}\times 10^{14}\ \mathrm{M_\odot}}$ and the mass of the smaller group to be equal to ${M_{200} = 0.6^{+0.4}_{-0.4}\times 10^{14}\ \mathrm{M_\odot}}$.
Their mass estimate for the components of PSZ1~G186.98+38.66 agrees well with our results from the multiple halo fits. Though we cannot constrain the mass of the foreground group well, their prediction is well below our upper mass limit of ${M_{200} \lesssim 0.8\times 10^{14}\ \mathrm{M_\odot}}$

\subsection{Impact of biases on our mass estimates}

\textit{On average}, weak lensing masses should be unbiased (${M^\mathrm{true} \approx M^\mathrm{WL}}$).
We have carefully investigated sources of multiplicative and additive bias that could affect our weak lensing mass estimates. We present an overview of the bias budget in Table~\ref{tab:systematics}.
\begin{table*}
	\centering
        \caption[]{Bias budget for our NFW cluster mass estimates. 
         We present the total (statistical) systematic uncertainty on $M_\mathrm{200m}$ due to systematics in the shapes and lensing strength estimates.}
        \begin{tabular}{ l c c r}
          \hline
          {\bf{Bias}}		&					&				&\\
          	 		& PSZ1~G109.88+27.94	& PSZ1~G139.61+24.20  	& PSZ1~G186.98+38.66 \\
			&$z = 0.77$&$z = 0.267$&$z=0.3774$\\
                    \hline
                    \hline    
            
         {\bf{Statistical uncertainties}} & & & \\
         Mean ellipticity mass bias $\sigma_\mathrm{{c}^{\langle\epsilon\rangle}}$ (Section~\ref{sec:meane})& $0.09$ & $0.07$ & $0.03$\\
         PSF leakage calibration bias $\sigma_\mathrm{e_{t}^{sys}}$ (Section~\ref{sec:leakage1})& $0.002$&  $0.001$&  $0.001$\\
\hline
		       {\bf{Total }} & {\bf{$<9.0\%$}}& {\bf{$<7.0\%$}}&{\bf{$<3.0\%$}}\\ 
\\
          {\bf{Systematic uncertainties}} & & & \\
         Residual multiplicative shape bias $\sigma_\mathrm{m}$ (Section~\ref{sec:noisebias})& $0.07$ & $0.07$ & $0.07$\\
         PSF model bias $\sigma_{{c}^\mathrm{mPSF}}$ (Section~\ref{sec:modelbias})& $0.085$ & $0.035$ & $0.005$\\
                  Cosmic variance $\sigma_\beta/\beta$ (Section~\ref{sec:cosmicvar}) & 0.04 & 0.03 & 0.03\\
	\hline
		       {\bf{Total }} & {\bf{$<12\%$}}& {\bf{$<8.4\%$}}&{\bf{$<7.6\%$}}\\ 
	        	\hline
		\hline
		 {\bf{Total bias}} & {\bf{$<15\%$}}& {\bf{$<11\%$}}&{\bf{$<8.1\%$}}\\ 
		\hline
	\end{tabular}
	
\label{tab:systematics}
\end{table*}

While we do correct for multiplicative shape bias $m$ by applying a S/N-calibration (cf.~Section~\ref{sec:noisebias}), we do not take a dependency of $m$ on the galaxy profiles or on the distributions of sizes and ellipticities into account.
Moreover, the simulations used to constrain the multiplicative shape bias might not be a good enough replication of our WWL~data. A better calibration of $m$ for future WWL projects will be necessary. 
We estimate the residual multiplicative shear bias to be less than 5~per~cent.

Usually, additive shear biases are neglected in cluster weak lensing studies, since the shear is measured in circular apertures and additive offsets cancel out. 
However, since our masks are not radially symmetric, additive shape biases can still affect the measured tangential shear signal. 
We consider three different types of additive shape bias. The mean ellipticity of our galaxies is not zero but equal to 
${c^\mathrm{\langle\epsilon\rangle}_{1,2} = (-2.8,-2.4)\pm(1.7,1.7)\times10^{-3}}$. 
As it depends on the random orientation of the applied masks, this constant additive shape bias causes a statistical uncertainty on the cluster mass.
The PSF~model~bias is different in each cluster field. We use the \emph{Rowe}~statistics to constrain an upper limit of ${c^\mathrm{mPSF} \lesssim 4,3,1\times10^{-3}}$ for this bias in the field of PSZ1~G109.88+27.94, PSZ1~G139.61+24.20 and PSZ1~G186.98+38.66, respectively. 
The PSF model bias may be correlated between different pointings, in which the cluster is always near the centre.  It will not decrease by taking more data.
The third source of additive systematics in the shape catalogues is PSF~leakage, which arises when the deconvolution of the PSF from the source images is not perfect. This type of additive shape bias is not spatially constant over the fields but should be the same for all observed fields.
PSF~leakage can be approximated as a linear dependency of the galaxy shapes on the PSF~ellipticity, i.e. $\alpha e^\mathrm{PSF}$. We calibrate the tangential shear signal by modeling the systematic tangential shear signal $e_\mathrm{t}^\mathrm{sys}$ as a function of distance from the cluster centre. We estimate the remaining PSF leakage calibration bias on the cluster mass to be less than 2~per~cent.
The PSF leakage calibration bias $\sigma_\mathrm{e_t^{sys}}$ has been estimated using the statistical uncertainty of our leakage correction and will decrease with increasing cluster sample size. 

We have also considered biases in our background sample selection.
The photometric calibration is very precise with negligible errors on the zero-point and the photometric redshifts of the reference galaxies. 
The only significant contribution to the systematics budget comes from the cosmic variance and depends only on the cluster redshift. 
It is smallest for PSZ1~G139.61+24.20 at $z=0.267$ and largest for PSZ1~G109.88+27.94 at $z=0.77$ and is equal to ${\sigma_\beta = 0.023,0.006,0.009}$ for PSZ1~G109.88+27.94, PSZ1~G139.61+24.20 and PSZ1~G186.98+38.66, respectively.
For clusters with high redshifts, such as PSZ1~G109.88+27.94, our applied background sample selection technique does not perform well using only $gri$ photometry.

We give upper limits of the total bias on the cluster mass: $15\%$ for PSZ1~G109.88+27.94, $11\%$ for PSZ1~G139.61+24.20 and $8.1\%$ for PSZ1~G186.98+38.66. Note that some biases are expected to cancel each other out, so the true mass bias might be significantly smaller than the values given in Table~\ref{tab:systematics}.
The statistical uncertainties are almost of the same order as the systematical uncertainties and will decrease with increasing cluster sample size and the observation of additional reference fields.
The residual multiplicative shear bias is the dominant source of systematics. An accurate calibration of $m$, or a shape measurement technique that performs better than \texttt{KSB+}, will be needed in order to bring this bias down in future studies. 
The large scatter of the PSF~model~bias upper limit shows that the performance of our PSF modeling technique is field dependent but performs well with $\sigma_{\boldsymbol{c}^\mathrm{mPSF}} < 0.5\%$ in the case of PSZ1~G186.98+38.66. 

\section{Conclusions}\label{sec:eight:conclude}

We present the results of the first cluster weak lensing study using only data obtained at the Wendelstein Observatory in Bavaria, Germany. Our pathfinder sample consists of three massive SZ-selected clusters of galaxies.
We obtain shape catalogues using an implementation of the \texttt{KSB}~code and determine lensing strengths from our deep $g,r,i$~band photometric data by following the approach of \citet{gruen2014weak}.

We carefully test for the impact of biases on our cluster mass estimates and present a simple method to approximate and correct for PSF~leakage in weak lensing data where the large statistical uncertainty makes a precise estimation of the additive shear bias difficult.
We use the mass-concentration relation of \citet{duffy2008dark} as a prior and perform an NFW likelihood analysis to estimate the mass and concentration of the objects. 

We present the first weak lensing mass estimates for the massive SZ-selected galaxy clusters PSZ1~G109.88+27.94 and PSZ1~G139.61+24.20. We correct the redshift estimate from PSZ1 and PSZ2 for PSZ1~G109.88+27.94 from $z=0.4$ to $z_\mathrm{phot} = 0.77$.

A two-parameter NFW-fit for the mass and concentration of a single dark matter halo in the field of PSZ1~G186.98+38.66 yields results, which are consistent with the weak lensing mass constraints from WtG.
Assuming the presence of a foreground group at ${z=0.2713}$ in the field of PSZ1~G186.98+38.66 we try to constrain its mass but are not sensitive to such low mass halos. We cannot exclude the existence of the background group candidate at a redshift of ${z=0.563}$ in the field of PSZ1~G186.98+38.66, which would lower the mass of the main cluster significantly. 
Our findings confirm that the presence of line of sight structures can have a significant impact on recovered weak lensing cluster masses.

We conclude that we can use multi-band WWFI data to perform independent weak lensing studies of good quality for small samples of individual clusters. We plan to further improve our analysis and target a new sample of relaxed clusters of galaxies and individual galaxy clusters we deem worthy to be studied in more detail.

\section*{Acknowledgements}

This paper contains data obtained at the Wendelstein Observatory of the Ludwig-Maximilians University Munich. 

This work was supported by SFB-Transregio 33 'The Dark Universe' by the Deutsche Forschungsgemeinschaft (DFG). We also acknowledge the support by the DFG Cluster of Excellence "Origin and Structure of the Universe". 

Support for DG was provided by NASA
through Einstein Postdoctoral Fellowship grant number PF5-160138
awarded by the Chandra X-ray Center, which is operated by the
Smithsonian Astrophysical Observatory for NASA under contract
NAS8-03060.



\bibliographystyle{mnras}
\bibliography{literature} 

\begin{thebibliography}{}
\makeatletter
\relax
\def\mn@urlcharsother{\let\do\@makeother \do\$\do\&\do\#\do\^\do\_\do\%\do\~}
\def\mn@doi{\begingroup\mn@urlcharsother \@ifnextchar [ {\mn@doi@}
  {\mn@doi@[]}}
\def\mn@doi@[#1]#2{\def\@tempa{#1}\ifx\@tempa\@empty \href
  {http://dx.doi.org/#2} {doi:#2}\else \href {http://dx.doi.org/#2} {#1}\fi
  \endgroup}
\def\mn@eprint#1#2{\mn@eprint@#1:#2::\@nil}
\def\mn@eprint@arXiv#1{\href {http://arxiv.org/abs/#1} {{\tt arXiv:#1}}}
\def\mn@eprint@dblp#1{\href {http://dblp.uni-trier.de/rec/bibtex/#1.xml}
  {dblp:#1}}
\def\mn@eprint@#1:#2:#3:#4\@nil{\def\@tempa {#1}\def\@tempb {#2}\def\@tempc
  {#3}\ifx \@tempc \@empty \let \@tempc \@tempb \let \@tempb \@tempa \fi \ifx
  \@tempb \@empty \def\@tempb {arXiv}\fi \@ifundefined
  {mn@eprint@\@tempb}{\@tempb:\@tempc}{\expandafter \expandafter \csname
  mn@eprint@\@tempb\endcsname \expandafter{\@tempc}}}

\bibitem[\protect\citeauthoryear{Abolfathi et~al.,}{Abolfathi
  et~al.}{2017}]{abolfathi2017fourteenth}
Abolfathi B.,  et~al., 2017, arXiv preprint arXiv:1707.09322

\bibitem[\protect\citeauthoryear{Allen, Evrard  \& Mantz}{Allen
  et~al.}{2011}]{allen2011cosmological}
Allen S.~W.,  Evrard A.~E.,   Mantz A.~B.,  2011, Annual Review of Astronomy
  and Astrophysics, 49, 409

\bibitem[\protect\citeauthoryear{Ammons, Wong, Zabludoff  \& Keeton}{Ammons
  et~al.}{2013}]{ammons2013mapping}
Ammons S.~M.,  Wong K.~C.,  Zabludoff A.~I.,   Keeton C.~R.,  2013, The
  Astrophysical Journal, 781, 2

\bibitem[\protect\citeauthoryear{Amodeo et~al.,}{Amodeo
  et~al.}{2017}]{amodeo2017calibrating}
Amodeo S.,  et~al., 2017, arXiv preprint arXiv:1704.07891

\bibitem[\protect\citeauthoryear{Applegate et~al.,}{Applegate
  et~al.}{2014}]{applegate2014weighing}
Applegate D.~E.,  et~al., 2014, Monthly Notices of the Royal Astronomical
  Society, 439, 48

\bibitem[\protect\citeauthoryear{Arnaud, Pratt, Piffaretti, B{\"o}hringer,
  Croston  \& Pointecouteau}{Arnaud et~al.}{2010}]{arnaud2010universal}
Arnaud M.,  Pratt G.,  Piffaretti R.,  B{\"o}hringer H.,  Croston J.,
  Pointecouteau E.,  2010, Astronomy \& Astrophysics, 517, A92

\bibitem[\protect\citeauthoryear{Avni}{Avni}{1976}]{avni1976energy}
Avni Y.,  1976, The Astrophysical Journal, 210, 642

\bibitem[\protect\citeauthoryear{Barmby, Huang, Ashby, Eisenhardt, Fazio,
  Willner  \& Wright}{Barmby et~al.}{2008}]{barmby2008catalog}
Barmby P.,  Huang J.-S.,  Ashby M.,  Eisenhardt P.,  Fazio G.,  Willner S.,
  Wright E.,  2008, The Astrophysical Journal Supplement Series, 177, 431

\bibitem[\protect\citeauthoryear{Bartelmann}{Bartelmann}{1996}]{bartelmann1996arcs}
Bartelmann M.,  1996, arXiv preprint astro-ph/9602053

\bibitem[\protect\citeauthoryear{Bender, Appenzeller, B{\"o}hm  et~al.}{Bender
  et~al.}{2001}]{bender2001fors}
Bender R.,  Appenzeller I.,  B{\"o}hm A.,   et~al., 2001, in Deep Fields, ed.
  S. Cristiani, A. Renzini, \& RE Williams, ESO astrophysics symposia
  (Springer).

\bibitem[\protect\citeauthoryear{Bertin}{Bertin}{2006}]{bertin2006automatic}
Bertin E.,  2006, in Astronomical Data Analysis Software and Systems XV. p.~112

\bibitem[\protect\citeauthoryear{Bertin}{Bertin}{2011}]{bertin2011automated}
Bertin E.,  2011, in Astronomical Data Analysis Software and Systems XX. p.~435

\bibitem[\protect\citeauthoryear{Bertin \& Arnouts}{Bertin \&
  Arnouts}{1996}]{bertin1996sextractor}
Bertin E.,  Arnouts S.,  1996, Astronomy and Astrophysics Supplement Series,
  117, 393

\bibitem[\protect\citeauthoryear{Bertin, Mellier, Radovich, Missonnier, Didelon
   \& Morin}{Bertin et~al.}{2002}]{bertin2002terapix}
Bertin E.,  Mellier Y.,  Radovich M.,  Missonnier G.,  Didelon P.,   Morin B.,
  2002, in Astronomical Data Analysis Software and Systems XI. p.~228

\bibitem[\protect\citeauthoryear{Bielby et~al.,}{Bielby
  et~al.}{2012}]{bielby2012wircam}
Bielby R.,  et~al., 2012, Astronomy \& Astrophysics, 545, A23

\bibitem[\protect\citeauthoryear{Binney \& Tremaine}{Binney \&
  Tremaine}{1998}]{binney1998galactic}
Binney J.,  Tremaine S.,  1998, Princeton Series in Astrophysics,(Princeton
  University Press, Princeton, NJ, 1987).[Google Books].(Cited on page 36.)

\bibitem[\protect\citeauthoryear{Brimioulle, Lerchster, Seitz, Bender  \&
  Snigula}{Brimioulle et~al.}{2008}]{brimioulle2008photometric}
Brimioulle F.,  Lerchster M.,  Seitz S.,  Bender R.,   Snigula J.,  2008, arXiv
  preprint arXiv:0811.3211

\bibitem[\protect\citeauthoryear{Brimioulle, Seitz, Lerchster, Bender  \&
  Snigula}{Brimioulle et~al.}{2013}]{brimioulle2013dark}
Brimioulle F.,  Seitz S.,  Lerchster M.,  Bender R.,   Snigula J.,  2013,
  Monthly Notices of the Royal Astronomical Society, 432, 1046

\bibitem[\protect\citeauthoryear{Bullock, Kolatt, Sigad, Somerville, Kravtsov,
  Klypin, Primack  \& Dekel}{Bullock et~al.}{2001}]{bullock2001profiles}
Bullock J.~S.,  Kolatt T.~S.,  Sigad Y.,  Somerville R.~S.,  Kravtsov A.~V.,
  Klypin A.~A.,  Primack J.~R.,   Dekel A.,  2001, Monthly Notices of the Royal
  Astronomical Society, 321, 559

\bibitem[\protect\citeauthoryear{Davis et~al.,}{Davis
  et~al.}{2007}]{davis2007all}
Davis M.,  et~al., 2007, The Astrophysical Journal Letters, 660, L1

\bibitem[\protect\citeauthoryear{Duffy, Schaye, Kay  \& Dalla~Vecchia}{Duffy
  et~al.}{2008}]{duffy2008dark}
Duffy A.~R.,  Schaye J.,  Kay S.~T.,   Dalla~Vecchia C.,  2008, Monthly Notices
  of the Royal Astronomical Society: Letters, 390, L64

\bibitem[\protect\citeauthoryear{Flewelling et~al.,}{Flewelling
  et~al.}{2016}]{flewelling2016pan}
Flewelling H.,  et~al., 2016, arXiv preprint arXiv:1612.05243

\bibitem[\protect\citeauthoryear{Giacintucci, Markevitch, Cassano, Venturi,
  Clarke  \& Brunetti}{Giacintucci et~al.}{2017}]{giacintucci2017occurrence}
Giacintucci S.,  Markevitch M.,  Cassano R.,  Venturi T.,  Clarke T.~E.,
  Brunetti G.,  2017, The Astrophysical Journal, 841, 71

\bibitem[\protect\citeauthoryear{G\"ossl \& Riffeser}{G\"ossl \&
  Riffeser}{2002}]{goessl2002image}
G\"ossl C.~A.,  Riffeser A.,  2002, Astronomy \& Astrophysics, 381, 1095

\bibitem[\protect\citeauthoryear{Gruen \& Brimioulle}{Gruen \&
  Brimioulle}{2017}]{gruen2017selection}
Gruen D.,  Brimioulle F.,  2017, Monthly Notices of the Royal Astronomical
  Society, 468, 769

\bibitem[\protect\citeauthoryear{Gruen et~al.,}{Gruen
  et~al.}{2013}]{gruen2013weak}
Gruen D.,  et~al., 2013, Monthly Notices of the Royal Astronomical Society, p.
  stt566

\bibitem[\protect\citeauthoryear{Gruen et~al.,}{Gruen
  et~al.}{2014}]{gruen2014weak}
Gruen D.,  et~al., 2014, Monthly Notices of the Royal Astronomical Society,
  442, 1507

\bibitem[\protect\citeauthoryear{Heymans et~al.,}{Heymans
  et~al.}{2006}]{heymans2006shear}
Heymans C.,  et~al., 2006, Monthly Notices of the Royal Astronomical Society,
  368, 1323

\bibitem[\protect\citeauthoryear{Heymans et~al.,}{Heymans
  et~al.}{2012}]{heymans2012cfhtlens}
Heymans C.,  et~al., 2012, Monthly Notices of the Royal Astronomical Society,
  427, 146

\bibitem[\protect\citeauthoryear{Hoekstra, Franx, Kuijken  \& Squires}{Hoekstra
  et~al.}{1998}]{hoekstra1998weak}
Hoekstra H.,  Franx M.,  Kuijken K.,   Squires G.,  1998, The Astrophysical
  Journal, 504, 636

\bibitem[\protect\citeauthoryear{Hoekstra, Mahdavi, Babul  \&
  Bildfell}{Hoekstra et~al.}{2012}]{hoekstra2012canadian}
Hoekstra H.,  Mahdavi A.,  Babul A.,   Bildfell C.,  2012, Monthly Notices of
  the Royal Astronomical Society, 427, 1298

\bibitem[\protect\citeauthoryear{Hopp et~al.,}{Hopp
  et~al.}{2008}]{hopp2008improving}
Hopp U.,  et~al., 2008, in SPIE Astronomical Telescopes+ Instrumentation. pp
  70161T--70161T

\bibitem[\protect\citeauthoryear{Hopp, Bender, Grupp, Goessl, Lang-Bardl,
  Mitsch, Riffeser  \& Ageorges}{Hopp et~al.}{2014}]{hopp2014commissioning}
Hopp U.,  Bender R.,  Grupp F.,  Goessl C.,  Lang-Bardl F.,  Mitsch W.,
  Riffeser A.,   Ageorges N.,  2014, in SPIE Astronomical Telescopes+
  Instrumentation. pp 91452D--91452D

\bibitem[\protect\citeauthoryear{Jarvis et~al.,}{Jarvis
  et~al.}{2016}]{jarvis2015science}
Jarvis M.,  et~al., 2016, Monthly Notices of the Royal Astronomical Society,
  460, 2245

\bibitem[\protect\citeauthoryear{Joachimi et~al.,}{Joachimi
  et~al.}{2015}]{joachimi2015galaxy}
Joachimi B.,  et~al., 2015, Space Science Reviews, 193, 1

\bibitem[\protect\citeauthoryear{Kaiser, Squires  \& Broadhurst}{Kaiser
  et~al.}{1994}]{kaiser1994method}
Kaiser N.,  Squires G.,   Broadhurst T.,  1994, arXiv preprint astro-ph/9411005

\bibitem[\protect\citeauthoryear{Kosyra, G{\"o}ssl, Hopp, Lang-Bardl, Riffeser,
  Bender  \& Seitz}{Kosyra et~al.}{2014}]{kosyra201464}
Kosyra R.,  G{\"o}ssl C.,  Hopp U.,  Lang-Bardl F.,  Riffeser A.,  Bender R.,
  Seitz S.,  2014, Experimental Astronomy, 38, 213

\bibitem[\protect\citeauthoryear{Luppino \& Kaiser}{Luppino \&
  Kaiser}{1997}]{luppino1997detection}
Luppino G.,  Kaiser N.,  1997, The Astrophysical Journal, 475, 20

\bibitem[\protect\citeauthoryear{{Mantz} et~al.,}{{Mantz}
  et~al.}{2016}]{mantz2016weighing}
{Mantz} A.~B.,  et~al., 2016, preprint, \href
  {http://adsabs.harvard.edu/abs/2016arXiv160603407M} {} (\mn@eprint {arXiv}
  {1606.03407})

\bibitem[\protect\citeauthoryear{Marrone et~al.,}{Marrone
  et~al.}{2012}]{marrone2012locuss}
Marrone D.~P.,  et~al., 2012, The Astrophysical Journal, 754, 119

\bibitem[\protect\citeauthoryear{Melchior et~al.,}{Melchior
  et~al.}{2017}]{melchior2017weak}
Melchior P.,  et~al., 2017, Monthly Notices of the Royal Astronomical Society,
  469, 4899

\bibitem[\protect\citeauthoryear{Navarro \& White}{Navarro \&
  White}{1996}]{navarro1996structure}
Navarro J.,  White S.~D.,  1996, in SYMPOSIUM-INTERNATIONAL ASTRONOMICAL UNION.
  pp 255--258

\bibitem[\protect\citeauthoryear{Navarro, Frenk  \& White}{Navarro
  et~al.}{1995}]{navarro1995simulations}
Navarro J.~F.,  Frenk C.~S.,   White S.~D.,  1995, Monthly Notices of the Royal
  Astronomical Society, 275, 720

\bibitem[\protect\citeauthoryear{Navarro, Frenk  \& White}{Navarro
  et~al.}{1997}]{navarro1997universal}
Navarro J.~F.,  Frenk C.~S.,   White S.~D.,  1997, The Astrophysical Journal,
  490, 493

\bibitem[\protect\citeauthoryear{Newman et~al.,}{Newman
  et~al.}{2013}]{newman2013deep2}
Newman J.~A.,  et~al., 2013, The Astrophysical Journal Supplement Series, 208,
  5

\bibitem[\protect\citeauthoryear{Pickles}{Pickles}{1998}]{pickles1998stellar}
Pickles A.,  1998, Publications of the Astronomical Society of the Pacific,
  110, 863

\bibitem[\protect\citeauthoryear{Piffaretti, Arnaud, Pratt, Pointecouteau  \&
  Melin}{Piffaretti et~al.}{2011}]{piffaretti2011vizier}
Piffaretti R.,  Arnaud M.,  Pratt G.,  Pointecouteau E.,   Melin J.-B.,  2011,
  VizieR Online Data Catalog, 353, 40109

\bibitem[\protect\citeauthoryear{{Planck Collaboration} et~al.,}{{Planck
  Collaboration} et~al.}{2011}]{ade2011planck}
{Planck Collaboration} et~al., 2011, Astronomy \& Astrophysics, 536, A8

\bibitem[\protect\citeauthoryear{{Planck Collaboration} et~al.,}{{Planck
  Collaboration} et~al.}{2014a}]{ade2014a}
{Planck Collaboration} et~al., 2014a, Astronomy \& Astrophysics, 571, A20

\bibitem[\protect\citeauthoryear{{Planck Collaboration} et~al.,}{{Planck
  Collaboration} et~al.}{2014b}]{ade2014b}
{Planck Collaboration} et~al., 2014b, Astronomy \& Astrophysics, 571, A29

\bibitem[\protect\citeauthoryear{{Planck Collaboration} et~al.,}{{Planck
  Collaboration} et~al.}{2015a}]{ade2015a}
{Planck Collaboration} et~al., 2015a, preprint, \href
  {http://adsabs.harvard.edu/abs/2015arXiv150201597P} {} (\mn@eprint {arXiv}
  {1502.01597})

\bibitem[\protect\citeauthoryear{{Planck Collaboration} et~al.,}{{Planck
  Collaboration} et~al.}{2015b}]{ade2015b}
{Planck Collaboration} et~al., 2015b, Astronomy \& Astrophysics, 582, A29

\bibitem[\protect\citeauthoryear{{Planck Collaboration} et~al.,}{{Planck
  Collaboration} et~al.}{2015c}]{ade2015c}
{Planck Collaboration} et~al., 2015c, A\&A, submitted, arXiv, 1502

\bibitem[\protect\citeauthoryear{Rau, Seitz, Brimioulle, Frank, Friedrich,
  Gruen  \& Hoyle}{Rau et~al.}{2015}]{rau2015accurate}
Rau M.~M.,  Seitz S.,  Brimioulle F.,  Frank E.,  Friedrich O.,  Gruen D.,
  Hoyle B.,  2015, Monthly Notices of the Royal Astronomical Society, 452, 3710

\bibitem[\protect\citeauthoryear{Rowe}{Rowe}{2010}]{rowe2010improving}
Rowe B.,  2010, Monthly Notices of the Royal Astronomical Society, 404, 350

\bibitem[\protect\citeauthoryear{Schneider}{Schneider}{1996}]{schneider1996detection}
Schneider P.,  1996, Monthly Notices of the Royal Astronomical Society, 283,
  837

\bibitem[\protect\citeauthoryear{Seitz \& Schneider}{Seitz \&
  Schneider}{1997}]{seitz1997steps}
Seitz C.,  Schneider P.,  1997, Astronomy and Astrophysics, 318, 687

\bibitem[\protect\citeauthoryear{Wen \& Han}{Wen \&
  Han}{2015}]{wen2015calibration}
Wen Z.,  Han J.,  2015, The Astrophysical Journal, 807, 178

\bibitem[\protect\citeauthoryear{Wong, Zabludoff, Ammons, Keeton, Hogg  \&
  Gonzalez}{Wong et~al.}{2013}]{wong2013new}
Wong K.~C.,  Zabludoff A.~I.,  Ammons S.~M.,  Keeton C.~R.,  Hogg D.~W.,
  Gonzalez A.~H.,  2013, The Astrophysical Journal, 769, 52

\bibitem[\protect\citeauthoryear{Wright \& Brainerd}{Wright \&
  Brainerd}{2000}]{wright2000gravitational}
Wright C.~O.,  Brainerd T.~G.,  2000, The Astrophysical Journal, 534, 34

\bibitem[\protect\citeauthoryear{Zuntz et~al.,}{Zuntz
  et~al.}{2017}]{zuntz2017dark}
Zuntz J.,  et~al., 2017, arXiv preprint arXiv:1708.01533

\bibitem[\protect\citeauthoryear{{Zwicky}, {Herzog}, {Wild}, {Karpowicz}  \&
  {Kowal}}{{Zwicky} et~al.}{1961}]{1961Z}
{Zwicky} F.,  {Herzog} E.,  {Wild} P.,  {Karpowicz} M.,   {Kowal} C.~T.,  1961,
  {Catalogue of galaxies and of clusters of galaxies, Vol. I}

\bibitem[\protect\citeauthoryear{von~der Linden et~al.,}{von~der Linden
  et~al.}{2014}]{von2014weighing}
von~der Linden A.,  et~al., 2014, Monthly Notices of the Royal Astronomical
  Society, 439, 2

\makeatother
\end{thebibliography}



%


\bsp	
\label{lastpage}
\end{document}